\documentclass[longauth,structabstract]{aa}  
\usepackage{graphicx}
\usepackage{natbib} 
\usepackage{longtable,lscape}
\bibpunct{(}{)}{;}{a}{}{,} 

\usepackage{txfonts}
\def\l{~$\lambda$}
\def\ll{~$\lambda\lambda$}

\def\halph{H$\alpha$}

\def\ha{H\,{\sc i}}

\def\he{He}
\def\hea{He\,{\sc i}}
\def\heb{He\,{\sc ii}}
\def\nc{N\,{\sc iii}}

\def\ob{O\,{\sc ii}}

\def\kms{km\,s$^{-1}$}

\def\msun{M$_{\odot}$}
\def\msol{M$_{\odot}$}

\def\p{$\pi$}
\def\k{$\kappa$}

\def\fbin{$f_\mathrm{bin}$}
\def\s{$\sigma$}

\def\chisq{$\chi^2$}
\def\vsini{$v \sin i$}
\def\snr{$S/N$}

\def\fw{{\sc fastwind}}

\def\aap{A\&A}

\def\apjs{ApJS}
\def\apjl{ApJ}

\def\na{{\it n/a}}

\defcitealias{ETHB11}{Paper~I}
\defcitealias{TES11}{Paper~II}
\defcitealias{BVH11}{Paper~III}
\defcitealias{BBE11}{Paper~IV}
\defcitealias{BTT09}{BTT09}

\begin{document}
   \title{The VLT-FLAMES Tarantula Survey\thanks{Based on observations collected at the European Southern Observatory under program ID 182.D-0222}}

   \subtitle{VIII. Multiplicity properties of the O-type star population}

   \author{
     H. Sana      \inst{1}
     \and
     A. de Koter  \inst{1,2}
     \and
     S.E. de Mink \inst{3,4}\thanks{Hubble Fellow}
     \and
     P.R. Dunstall  \inst{5}
     \and
     C.J. Evans   \inst{6}
     \and
     V. H\'enault-Brunet  \inst{7}
     \and
     J. Ma\'{i}z Apell\'aniz \inst{8}
     \and \\
     O.H. Ram\'{i}rez-Agudelo   \inst{1}
     \and
     W.D. Taylor  \inst{7}
     \and
     N.R. Walborn   \inst{3}
     \and
     J.S. Clark     \inst{9}
     \and
     P.A. Crowther  \inst{10}
     \and
     A. Herrero   \inst{11,12}
     \and
     M. Gieles    \inst{13}
     \and\\
     N. Langer    \inst{14}
     \and
     D.J. Lennon    \inst{15,3}
     \and
     J.S. Vink      \inst{16}
   }

\institute{
          Astronomical Institute Anton Pannekoek, 
          Amsterdam University,  
          Science Park 904, 1098~XH, 
          Amsterdam, The Netherlands\\
          \email{h.sana@uva.nl}
\and 
           Utrecht University,
           Princetonplein 5, 3584CC,
           Utrecht, The Netherlands
\and 
           Space Telescope Science Institute,
           3700 San Martin Drive,
           Baltimore,
           MD 21218,
           USA 
\and 
           Johns Hopkins University,
           11100 Johns Hopkins Rd Laurel,
           Baltimore, MD, USA
\and 
           Astrophysics Research Centre, School of Mathematics and Physics, 
           Queen’s University Belfast, 
           Belfast BT7 1NN, UK
\and 
           UK Astronomy Technology Centre,
           Royal Observatory Edinburgh,
           Blackford Hill, Edinburgh, EH9 3HJ, UK
\and 
           Scottish Universities Physics Alliance (SUPA),
           Institute for Astronomy,
           University of Edinburgh,
           Royal Observatory Edinburgh,
           Blackford Hill, Edinburgh, EH9 3HJ, UK
\and 
           Instituto de Astrof\'{i}sica de Andaluc\'ia-CSIC,
           Glorieta de la Astronom\'ia s/n,
           E-18008 Granada, Spain
\and 
           Department of Physics and Astronomy,
           The Open University, Walton Hall, Milton Keynes, MK7 6AA, UK
\and 
           Dept. of Physics \& Astronomy,
           Hounsfield Road,
           University of Sheffield, S3 7RH, UK
\and 
           Instituto de Astrof\'{i}sica de Canarias, 
           C/ V\'{i}a L\'{a}ctea s/n, E-38200 La Laguna, Tenerife, Spain
\and 
           Departamento de Astrof\'{i}sica, 
           Universidad de La Laguna, 
           Avda. Astrof\'{i}sico Francisco S\'{a}nchez s/n, 
           E-38071 La Laguna, Tenerife, Spain
\and 
           Institute of Astronomy, University of Cambridge,
           Madingley Road, Cambridge, CB3 0HA, UK
\and 
           Argelander-Institut f\"ur Astronomie, 
           Universit\"at Bonn, 
           Auf dem H\"ugel 71, 
           53121 Bonn, Germany
\and 
           ESA/STScI,
           3700 San Martin Drive,
           Baltimore, MD 21218, USA
\and 
           Armagh Observatory, College Hill,
           Armagh, BT61 9DG, Northern Ireland, UK
}
   \date{Received May 17, 2012; accepted September 18, 2012}

 
  \abstract
   {The Tarantula Nebula in the Large Magellanic Cloud is our closest view of a starburst region and is the ideal environment to investigate important questions regarding the formation, evolution and final fate of the most massive stars.  }
   {We analyze the multiplicity properties of the massive O-type star population observed through multi-epoch spectroscopy in the framework of the VLT-FLAMES Tarantula Survey. With 360 O-type stars, this is the largest homogeneous sample of massive stars analyzed to date.}
   {We use multi-epoch spectroscopy and variability analysis to identify spectroscopic binaries. We also use a Monte-Carlo method to correct for observational biases. By modeling simultaneously the observed binary fraction, the distributions of the amplitudes of the radial velocity variations and the distribution of the time scales of these variations, we derive the intrinsic current binary fraction and period and mass-ratio distributions. }
   {We  observe a spectroscopic binary fraction of $0.35\pm0.03$, which corresponds to the fraction of objects displaying statistically significant radial velocity variations with an amplitude of at least 20~\kms. We compute the intrinsic binary fraction to be $0.51\pm0.04$. We adopt power-laws to describe the  intrinsic period and mass-ratio distributions:  $f(\log_{10} P/\mathrm{d})\sim (\log_{10} P/\mathrm{d})^\pi$ (with $\log_{10} P/\mathrm{d}$ in the range 0.15-3.5) and $f(q)\sim q^\kappa$ with $0.1 \leq q=M_2/M_1 \leq1.0$. The power-law indexes  that best reproduce the observed quantities are $\pi=-0.45\pm0.30$ and $\kappa=-1.0\pm0.4$. The period distribution that we obtain thus favours shorter period systems compared to an \"Opik law ($\pi=0$). The  mass ratio distribution is slightly skewed towards low mass ratio systems but remains incompatible with a random sampling of a classical mass function ($\kappa=-2.35$). The binary fraction seems mostly uniform across the field of view and independent of the spectral types and luminosity classes. The binary fraction in the outer region of the field of view ($r>7.8\arcmin$, i.e.\ $\approx$117~pc) and among the O9.7~I/II objects are however significantly lower than expected from statistical fluctuations. The observed and intrinsic binary fractions are also lower for the faintest objects in our sample ($K_\mathrm{s}>15.5$~mag), which  results from observational effects and the fact that our O star sample is not magnitude-limited but is defined by a spectral-type cutoff. We also conclude that magnitude-limited investigations are biased towards larger binary fractions.}
   {Using the multiplicity properties of the O stars in the Tarantula region and simple evolutionary considerations, we estimate that over 50\%\ of the current O star population will exchange mass with its companion within a binary system. This shows that binary interaction is greatly affecting the evolution and fate of massive stars, and must be taken into account to correctly interpret unresolved populations of massive stars. }


   \maketitle
%

\section{Introduction}

Massive binaries are spectacular systems that may have been among the first objects that formed in the early universe \citep{TAOS09,SGB12}. Because the binary fraction among massive stars is 
high \citep{MHG09} and as a large fraction of them are part of short period systems \citep{SaE11}, an accurate knowledge of their binary properties is crucial to understand the role of massive stars as a population \citep{LCY08, dMCL09}. On the one hand, the fraction of binaries and their orbital parameter distributions are the tracers of massive star formation and of the early dynamical evolution of their host star cluster. On the other hand, they determine the frequency of evolved massive binary systems, including high-mass X-ray and double compact binaries \citep{SBB08}, Type Ib/c supernovae \citep{YWL10}, short and long-duration gamma-ray bursts \citep{PIJ10}. 

Among the various multiplicity properties, the most investigated one is the observed fraction of spectroscopic binaries, that sets a lower limit on the true binary fraction. Based on an extensive compilation of the literature, \citet{MHG09} showed that 51\%\ of the Galactic O-type stars investigated up to 2008 by multi-epoch spectroscopy are spectroscopic binaries. This fraction is 56\%\ for objects in clusters and OB associations. The spectroscopic survey of Galactic O and WN stars \citep{BGA10} obtains a similar fraction: 60\%\ of the  240 southern stars investigated indeed display significant radial velocity (RV) variations with an amplitude larger than 10~\kms. Finally, studies of individual young open clusters, such as  IC 1805 \citep{DBRM06,HGB06}, IC 1848 \citep{HGB06},  IC 2944 \citep{SJG11}, NGC 2244 \citep{MNR09}, NGC 6231 \citep{SGN08}, NGC 6611 \citep{SGE09} and Tr 16 \citep{RNF09} reveal observed binary fractions ranging between 30 and 60\%. These cluster to cluster variations are so far compatible with random fluctuations expected from the size of the samples \citep{SaE11}.

The  intrinsic period and mass-ratio distributions of the O-type binaries have remained poorly constrained until recently. In a pioneering work, \citet{KoF07} attempted to constrain these distributions using a sample of 900 RV observations of 32 O- and 88 B-type stars in Cyg~OB2 and a Monte-Carlo method to correct for observational biases. Their solution was unfortunately degenerate leading the authors to fix the period distribution to an \"Opik law (i.e., a flat distribution in the logarithm of the separation or, equivalently, of the period). They find a very high intrinsic binary fraction of 80 to 100\%, although the range of separations considered implies an extrapolation of their results over one to two orders of magnitude outside the sensitivity domain of their observations (see also Sect.~\ref{sect: MW}). Based on a smaller sample of 13 O+OB and 37 B+B eclipsing binaries \citep{HHH03,HHH05}, \citet{PiS06} suggested the presence of a population of equal-mass (`twin') binaries albeit \citet{Luc06} argued  that the considered sample was too small to draw statistically significant conclusions on the presence of a peak close to unity in the mass-ratio distribution.

\citet{SdMdK12} used a Monte-Carlo method to retrieve the intrinsic multiplicity properties of   O-type stars in  nearby Galactic open clusters. Based on asample of 71 O-type objects, they showed that the period distribution does not follow the widely used \"Opik law but is overabundant in short period systems and that the mass-ratio distribution is flat in the range $0.1< M_2/M_1 <1$, ruling out both the presence of a twin population of equal mass binaries and random pairing from a classical initial mass function as a process to associate components in a binary. These results have strong implications for the evolution of massive stars as over two-thirds of the stars born as O star will, during their lifetime, interact with a nearby companion in a binary system. 

The clusters considered by \citet{SdMdK12} have however relatively low masses, in the range 1,000-5,000~M$_\odot$. In view of the much more energetic environment provided by $10^5-10^6$~M$_\odot$ clusters, containing up to thousands of massive stars, it is currently unknown whether the binary fraction and the parameter distributions would remain unaffected. Addressing this question  has implications in the context of super-star-clusters and cluster complexes observed well beyond the Local Group \citep[e.g.][]{GSPZ10}.

The Tarantula Nebula (or 30 Doradus) in the LMC is one of the largest concentrations of massive stars in the Local Group \citep{WaB97} with a total mass of $\sim$5.5$ \times 10^{4}$~\msol\ \citep{AZM09}. Combined with its youth \citep[the central cluster R136 is approximately 1 to 2 Myr;][]{dKHH98} and the presence of several sub-populations, indicative of violent star forming activity in the past
tens of millions of years, this complex system is our closest view of a starburst-like region in the Local Universe and an ideal laboratory to investigate the characteristics of massive stars.

By targeting $\sim$800 massive stars the VLT-FLAMES Tarantula Survey \citep[VFTS; ][\citetalias{ETHB11}]{ETHB11} provides a near-complete census of the 30 Doradus region. The science drivers of this program
are presented in e.g. \citet{ETS11} and \citet{dKSE11}. The study of the multiplicity properties of this region is an important component of VFTS and to this end a multi-epoch observing strategy has been implemented to identify binaries with periods $\lesssim 10^3$ days (see Sect.~\ref{sect: bias}), i.e.\ the systems in which the components are expected to interact through Roche-lobe overflow. Early serendipitous findings \citep{TES11, DDE11} clearly emphasize the importance of binaries and the large potential these systems have for our understanding of binary properties and evolution.

\begin{figure}[t!]
  \centering
  \includegraphics[width=\columnwidth]{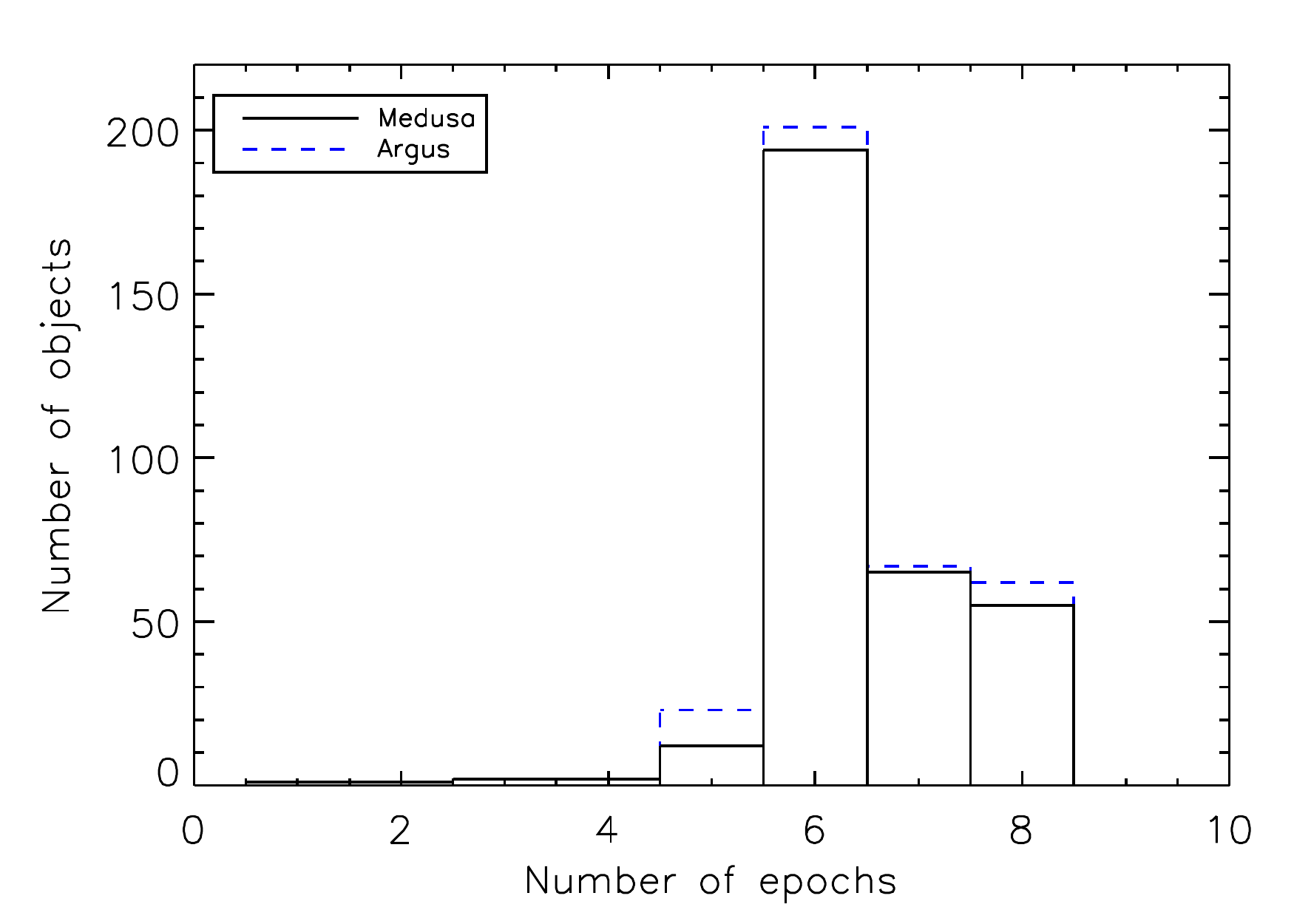}
  \includegraphics[width=\columnwidth]{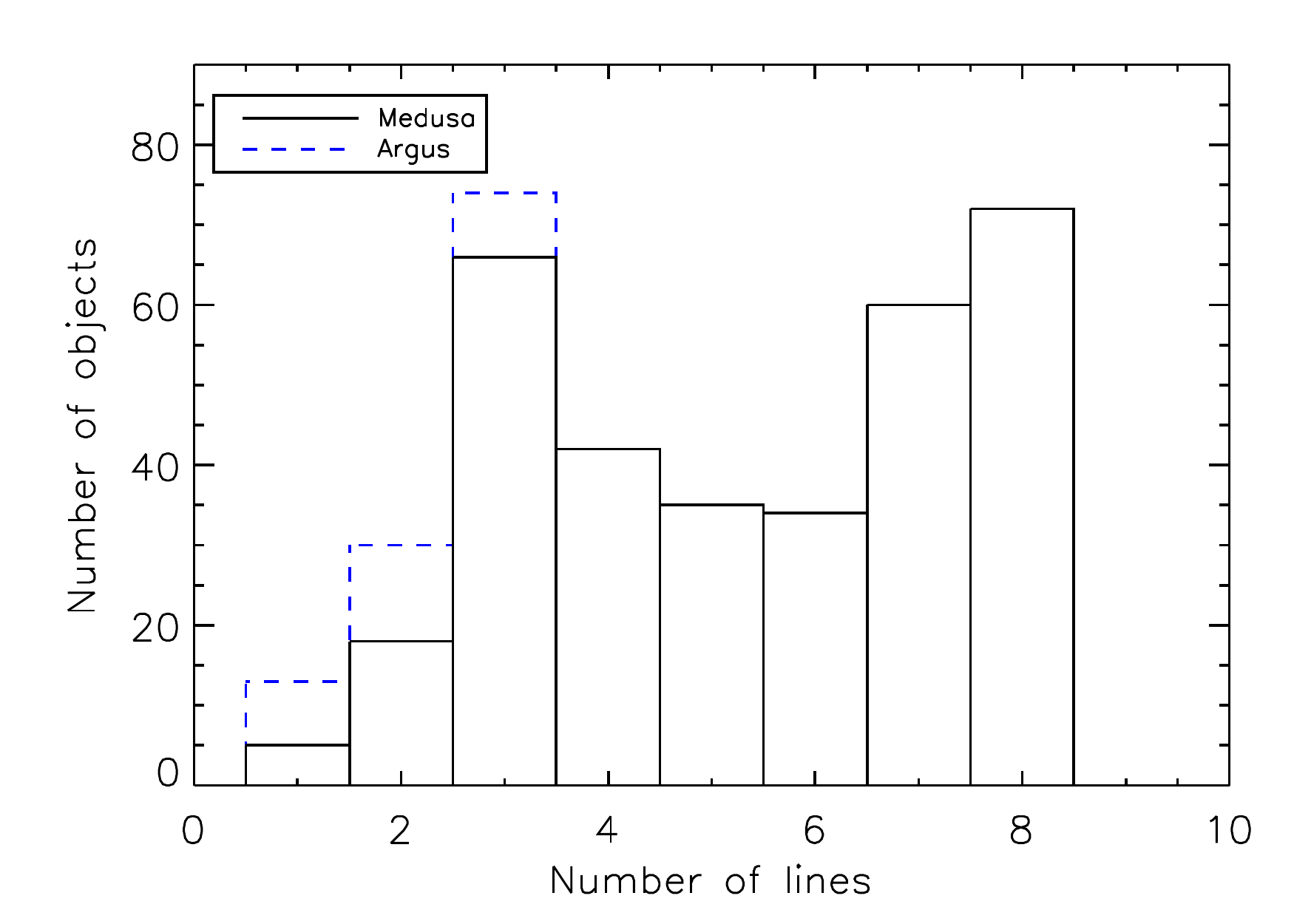}
  \includegraphics[width=\columnwidth]{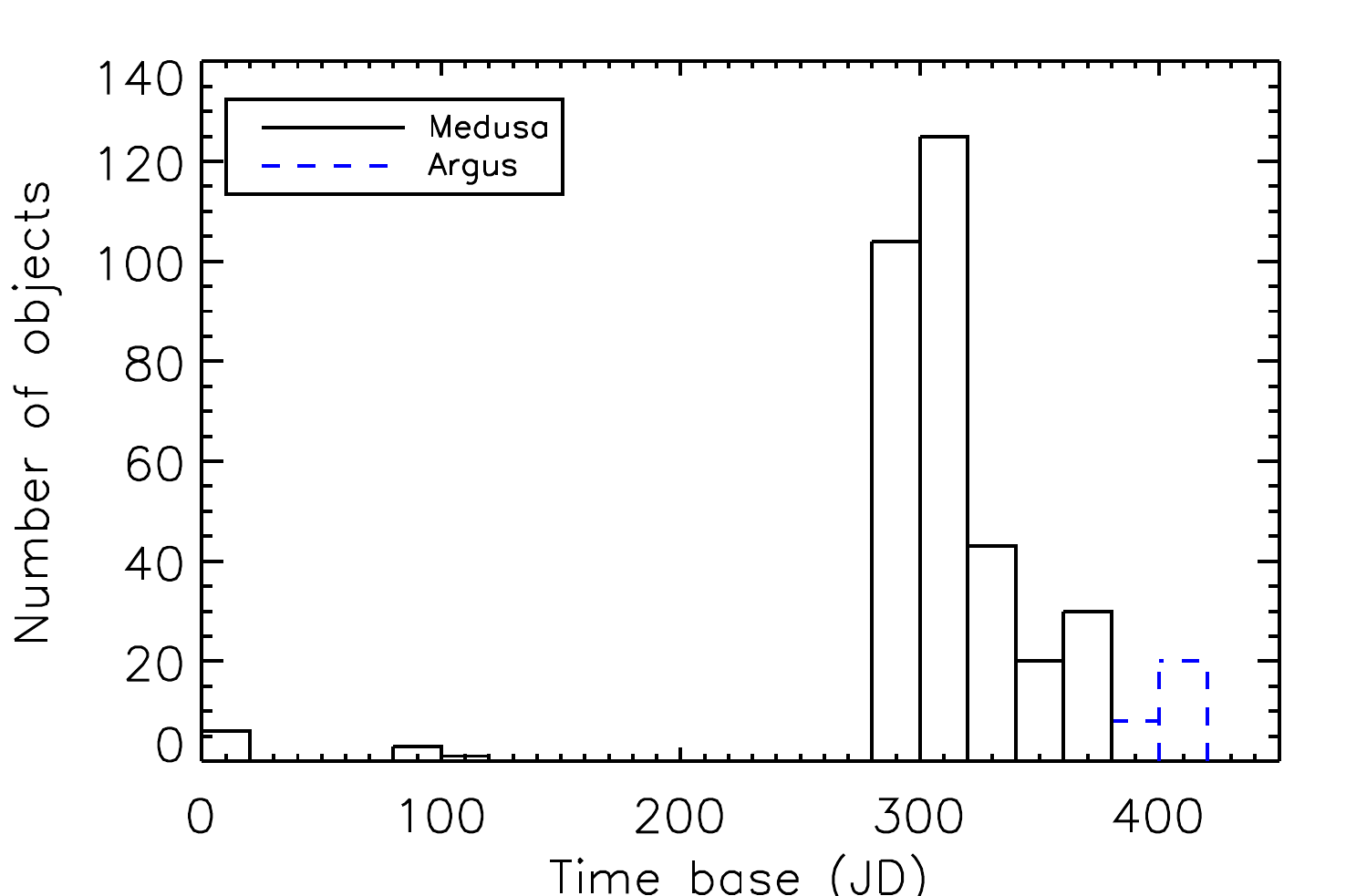}
  \caption{ Top panel:   Number of RV epochs obtained for each object. One Argus object, VFTS 1022, has been observed at 13 epochs and falls outside the plotted range. Middle panel: number of lines available for absolute RV measurements. Bottom panel: time elapsed between the first and last suitable epochs for RV measurement.  The solid line corresponds to objects observed with Medusa while the dashed lines indicate objects observed with the  Argus IFU.}
  \label{fig: sample}
\end{figure}

\begin{figure}
  \centering
  \includegraphics[width=\columnwidth]{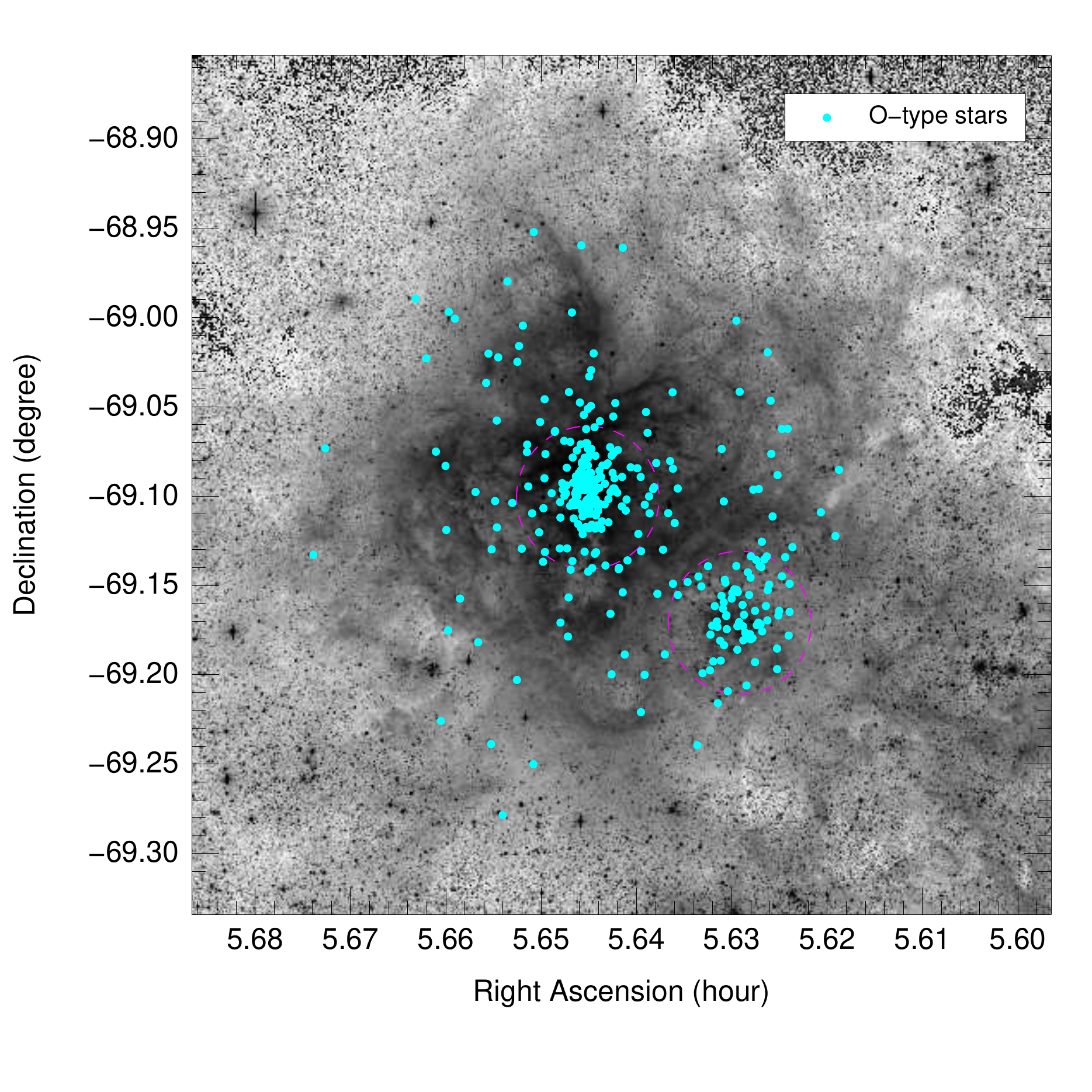}
  \caption{Distribution of the O-type stars in the field of view. The population is dominated by the central cluster, NGC~2070, that contains 57\%\ of our sample. NGC~2060, at 6.7\arcmin\ to the South-West, is slightly older and accounts for 22\%\ of the sample, while the remaining 21\%\ is spread throughout the field of view. The dashed circles, with a 2.4\arcmin-radius each, show the adopted regions for NGC~2070  and NGC~2060 (see Table~\ref{tab: bf}). }
  \label{fig: left}
\end{figure}
 In this paper, we use radial velocities (RVs, see Sect.~\ref{sect: rv}) and a variability analysis to investigate the multiplicity properties of the 360 O-type stars in the VFTS sample. We first establish the observed spectroscopic binary fraction in our sample (Sect.~\ref{sect: mult}). Using a Monte-Carlo method, we correct for the observational biases and we constrain the intrinsic binary fraction and period and mass-ratio distributions (Sect.~\ref{sect: intr_mult}). Possible correlations of these quantities with spatial distribution,  magnitude and spectral type are investigated in Sect.~\ref{sect: correl}. Our findings are discussed in Sect.~\ref{sect: discuss} and summarised in Sect.~\ref{sect: ccl}.


\section{Observations, sample and RV measurements} \label{sect: rv}

   \subsection{Data overview}
The VFTS data and the data reduction process are extensively described in \citetalias{ETHB11}. Here, we provide a brief description of the VFTS campaign and observational settings related to the present study. Using nine different plate configurations, VFTS has collected multi-epoch fibre spectroscopy of about 800 early-type stars in the Tarantula region, covering a 25\arcmin\ diameter field of view centered on R136. Each observation consisted of two 30-min exposures taken back to back. In this paper, we analyze the O-type star population observed with the Medusa fibres and the low resolution settings LR02 and LR03 of the Giraffe multi-object spectrograph. These data provide  continuous coverage of the 3960-5070~\AA\ region at a spectral resolution of about 0.6~\AA. The higher resolution data obtained in the \halph\ region are not used in the present analysis. 

Most of the time sampling is provided by the LR02 observations (3960-4565~\AA, $ \lambda/\lambda \Delta \approx 7000$), with five pointings covering time scales of days to months and one additional pointing taken about 300 days after the first observations. The LR03 setup (4500-5070~\AA, $\lambda / \Delta \lambda=8500$) was observed three times with no specific time constraint. As a result some of the configurations were observed consecutively to LR02 data while others are spread out and provide additional time sampling. For operational reasons, one LR02 plate configuration -- field C -- was reobserved a seventh time in Oct 2010, extending the temporal baseline to almost 700 days for 48 O stars in our sample. The signal-to-noise ratio (\snr) is however lower than average and this last pointing provides little additional information. We have performed the complete analysis presented in Sect.~\ref{sect: mult} and \ref{sect: intr_mult} with and without these supplementary data and found no significant differences in the results. In the rest of the paper, we ignore that epoch in order to preserve the homogeneity of the temporal sampling for all stars in our data set.

The combined effect of chromatic atmospheric diffraction and of seeing may strongly impact the fraction of the starlight that is injected into the Medusa fibres. As a consequence, the total response of the observing system may present significantly different shapes from one exposure to another. Each 30-min spectrum is therefore normalised individually using the procedure described in Appendix~\ref{app: norm} that provides a high degree of automatisation and guarantees an homogeneous handling of the data.

 Contiguous observations were then combined to improve the \snr, yielding a typical number of six epochs per targets  and providing a minimum timebase of about 300 days for most of the stars. Statistics of the number of epochs, diagnostic lines for RV measurements and timebase are provided in Fig.~\ref{fig: sample}.

   \subsection{O-type stars in VFTS}

The sample analyzed in this work includes all O-type stars observed during the VFTS campaign and ignore the B-type stars, the Wolf-Rayet stars and the transitional objects to the Wolf-Rayet domain (the so-called ``slash-stars''). The multiplicity properties of the B stars are investigated in a separate paper within the VFTS series (Dunstall et al., in preparation). 

The O-type stars were classified by comparison, at a resolving power of 4000, with spectral standard stars. The detailed spectral classification is presented in Walborn et al. (in preparation). In total, 332 O-type stars observed with Medusa form the bulk of our sample.

The densest regions of 30~Dor, in and around the core cluster R136, were further observed with the Argus integral-field unit (IFU) and the LR02 Giraffe setup. The RVs of these objects were measured using a similar approach to the one described in Appendix \ref{app: rv} and the same set of lines and rest wavelengths \citep[see Table~\ref{tab: rw} and ][]{HBES12}. The Argus RV measurements of 28 additional O stars observed with the Argus IFU have  been incorporated in the present analysis. In total, the sample that we analyze thus contains  360 O-type stars and is the largest homogeneous sample of O stars investigated for multiplicity to date.

Fig.~\ref{fig: left} shows the spatial distribution of the O stars in the 30~Dor field of view.  About half of our sample (184 objects) is concentrated in a 2.4\arcmin-radius around R136, the young and massive cluster at the core of NGC~2070. Seventy-three targets are spatially associated to NGC~2060, a slightly older cluster at 6.7\arcmin\ South-West of R136 and the remaining 103 objects are spread across the field of view.

\begin{table*}
\centering
\caption{List of lines (marked with an `x')  used to obtain the RV measurements listed in Table~\ref{tab: rv}. The full version of the table is available electronically.} \label{tab: line_used}
\begin{tabular}{ccccccccc}
\hline \hline
VFTS    &\hea+{\sc ii}\l4026 &\heb\l4200      & \hea\l4387    & \hea\l4471     & \heb\l4541      &  \heb\l4686      &  \hea\l4713      &  \hea\l4922      \\
\hline
\vspace*{-2mm}\\
014&-&x&x&-&x&-&x&x \\
016&-&x&-&-&x&-&-&- \\
021&-&x&x&-&x&-&x&x \\
042&-&x&x&-&x&-&x&x \\
045&-&x&x&-&x&-&x&x \\
046&-&x&x&-&x&-&x&x \\
047&x&-&x&-&x&x&x&- \\
049&x&-&x&-&-&x&x&x \\
051&x&x&-&-&x&x&-&- \\
055&x&x&x&x&x&x&x&x \\
056&x&x&-&x&x&x&-&- \\
058&x&-&x&x&x&x&x&x \\
059&-&-&x&-&x&-&x&x \\
061&x&x&-&-&-&-&x&x \\
063&x&x&-&-&x&x&-&- \\
064&-&x&x&-&x&-&x&x \\
065&-&x&-&-&x&-&-&- \\
066&-&-&x&-&x&-&x&x \\
067&-&x&x&-&x&-&x&x \\
070&-&-&x&-&-&-&x&x \\
072&-&x&-&-&x&-&-&- \\
073&-&x&x&-&x&-&x&x \\
074&-&x&x&-&x&-&x&x \\
076&-&x&x&-&x&-&x&x \\
077&-&-&x&-&-&-&-&- \\
\dots &\dots     & \dots &  \dots & \dots &\dots     & \dots &  \dots & \dots \\
\hline              
\end{tabular}       
\end{table*}

\begin{table} 
\centering
\caption{Journal of the observations, RV measurements and their 1\s\ error-bars for the stars in our sample. The full version of the table is available electronically.}
\label{tab: rv}
\begin{tabular}{cccc}
\hline \hline
VFTS    & HJD & RV & $\sigma_\mathrm{RV}$ \\
        & $-$2\,400\,000 & (\kms) & (\kms)\\
 \hline 
\vspace*{-2mm}\\
014    & 54815.791  &     276.4  &     1.6 \\ 
014    & 54822.694  &     289.9  &     2.2 \\ 
014    & 54859.720  &     275.6  &     2.4 \\ 
014    & 54891.573  &     274.7  &     1.9 \\ 
014    & 55113.818  &     280.7  &     2.4 \\ 
014    & 54813.776  &     279.8  &     1.5 \\ 
016    & 54817.734  &     191.9  &     0.7 \\ 
016    & 54822.547  &     187.6  &     0.9 \\ 
016    & 54860.594  &     189.1  &     0.9 \\ 
016    & 54890.530  &     188.7  &     2.0 \\ 
016    & 55112.850  &     189.4  &     1.2 \\ 
016    & 54818.776  &     189.8  &     0.7 \\ 
017    & 54767.824  &     285.5  &     4.1 \\ 
017    & 54827.772  &     229.2  &     6.5 \\ 
017    & 54828.751  &     240.5  &     4.8 \\ 
017    & 54860.643  &     283.4  &     2.8 \\ 
017    & 54886.616  &     318.2  &     6.8 \\ 
017    & 55114.859  &     252.0  &     15.5\\ 
017    & 54828.821  &     235.9  &     3.4 \\
017    & 54836.563  &     260.2  &     1.5 \\
\dots     & \dots & \dots & \dots \\
\hline
\end{tabular}
\end{table}

\subsection{Completeness} \label{sect: completness}
The list of potential targets to be observed by the VFTS was selected based on a simple magnitude cut-off at $V=17$ (see Paper I). The allocation of the MEDUSA fibres to VFTS sources was subject to further constraints related to the physical size of the magnetic buttons supporting the fibres, precluding observation of all sources in the most crowded parts of the field-of-view. The VFTS completeness is about 70\%\ of the total O-star population in the field of view and, importantly, does not show a bias towards the brighter end of the magnitude distribution. Beside the physical constraint on the fibre allocation, the VFTS sample is complete for the O-type stars, except for the late O-type objects that suffer from a significantly larger extinction -- at least one magnitude -- than typical in the field of view.

   \subsection{RV measurements} 

The radial velocities of our targets are obtained by measuring the Doppler displacement of spectral lines.
 Because the RVs are a key ingredient of this study, details of the method are provided in Appendix \ref{app: rv}, including alternative approaches that have been considered. Our method of choice, Gaussian fitting of selected sets of \he\ lines, is prompted by four criteria: (i) robustness, (ii) internal consistency across the full range of O spectral sub-types, and the ability to provide (iii) absolute measurements and (iv) error bars. An extension to the standard line-by-line Gaussian fitting is obtained by simultaneously fitting all the data available for a given object (i.e., all the epochs and spectral lines). The different lines fitted at a given epoch are required to provide the same Doppler shift while the shape of the individual lines is kept constant through all the epochs. This approach provides two significant improvements compared to line-by-line fitting: a strong reduction of the number of parameters of the fit and  a significant gain in robustness in the RVs measured from data with poor \snr. 

Special attention has been given to identify a set of lines that provide consistent RV measurements across the range of O spectral sub-types. Lines displaying systematic deviations as a result of, e.g.,  line blending or wind effects have been discarded (see Appendix \ref{app: rv_line} for a detailed discussion). The best consistency is observed between RVs obtained from the \heb\ll4200, 4541, \hea\ll4387, 4713 and 4922 lines. Whenever possible, we have thus obtained absolute RV measurements based on these five lines or a subset of these lines depending on the spectral type, wavelength coverage and degree of nebular contamination. The spectral lines used for each object are reported in Table~\ref{tab: line_used} while Table~\ref{tab: rv} provides the journal of the observations and, for each epoch, the measured RV and 1\s\ uncertainty.

The typical number of spectral lines used in measuring the RVs is between two and five (Fig.~\ref{fig: sample}). For VFTS 051, 400, 444, 453, 531, 565 and 587, none of these lines were of sufficient quality. RVs for these targets were measured using the \hea+{\sc ii}\l4026 and/or \heb\l4686 lines. Because these are late O-type stars the measured RVs are probably only weakly affected by temperature and/or wind effects but caution should be applied in using the measured RVs in absolute terms. A number of objects displayed double-lined profiles (see Sect.~\ref{sect: sb2}). Measuring RVs for both components sometimes required the use of lines that are not suited to provide absolute RVs. Because SB2s are not suited for absolute RVs unless a full orbit determination can be achieved, we relax our constraints for these objects and include other \he\ lines as indicated in Table \ref{tab: line_used}. For the SB2 cases, Table  \ref{tab: rv} lists the RV of the earliest component of the pair. 

\begin{table}
\centering
\caption{Overview of the results of the multiplicity analysis. Cols.~2 and 3 provide the maximum significance of the RV variations and their maximum amplitudes, respectively. Col.~4 indicates whether LPV are detected. Col.~5 reports on the morphology of the lines and the last column indicates whether the targets are revealed as multiple sources by the HST. The full version of the table is available electronically.} \label{tab: vartest}
\begin{tabular}{cccccc}
\hline \hline
VFTS & \multicolumn{2}{c}{RV variability}  &  LPV  & Morphology & HST  \\
     & $\sigma_\mathrm{detect}$ & $\Delta RV$ \\
     &                          & (\kms) \\
\hline
\vspace*{1mm}
 014  &    5.27 &    15.28 &   y    &     $\Delta$Ampl &  --   \\
 016  &    3.69 &    --    &   --   &     --    &  --   \\
 021  &    2.92 &    --    &   --   &     --    &  --   \\
 042  &    9.47 &    30.13 &   y    &     --    &  --   \\
 045  &   21.96 &    92.09 &   y    &     --    &  --   \\
 046  &    0.59 &    --    &   --   &     --    &  --   \\
 047  &   35.28 &   325.20 &   y    &     SB2  &  --   \\
 049  &   24.21 &   144.04 &   y    &     SB2  &  --   \\
 051  &    2.09 &    --    &   --   &     --    &  --   \\
 055  &   52.41 &   295.27 &   y    &     SB2  &  --   \\
 056  &    4.08 &   143.21 &   --   &     SB2  &  --   \\
 058  &   27.57 &   173.52 &   y    &     SB2  &  --   \\
 059  &   22.37 &   148.58 &   y    &     --    &  --   \\
 061  &   75.04 &   560.25 &   y    &     SB2  &  --   \\
 063  &   12.55 &    62.15 &   y    &     SB2  &  --   \\
 064  &    4.65 &    14.84 &   y    &     --    &  --   \\
 065  &    1.23 &    --    &   --   &     --    &  --   \\
 066  &    4.31 &    46.84 &   --   &     Asym:&  --   \\
 067  &    0.90 &    --    &   --   &     --    &  --   \\
 070  &    3.05 &    --    &   --   &     --    &  --   \\
 072  &    3.28 &    --    &   --   &     --    & \na  \\
 073  &    5.97 &    31.72 &   --   &     --    &  --   \\
 074  &    2.65 &    --    &   --   &     --    &  --   \\
 076  &    1.83 &    --    &   --   &     --    &  --   \\
 077  &    2.58 &    --    &   --   &     --    &  --   \\
\dots &   \dots &    \dots & \dots  & \dots     & \dots \\
\hline
\end{tabular}
\end{table}

\section{Multiplicity analysis}\label{sect: mult}
In this section we investigate the multiplicity of the stars in our sample using various approaches that combine variability analysis (Sect.~\ref{sect: var}) and the search for composite sources (Sect.~\ref{sect: comp}) revealed either by multiple signatures in their spectrum or by HST imaging. Sect.~\ref{sect: obs_bf} establishes the observed spectroscopic binary fraction and Sect.~\ref{sect: btt09} compares our results with earlier studies. The results of the multiplicity analysis are summarised in Table~\ref{tab: vartest}, the full version of which is available online.

\subsection{Variable sources}\label{sect: var}

   \subsubsection{RV variability}\label{sect: RVvar}
To identify objects displaying significant RV variations, we use a statistical test. The null hypothesis of constant RV is rejected if, for a given star, any two RV measurements deviate significantly from one another, i.e.:
\begin{equation}
\sigma_\mathrm{detect}=\max \left( \frac{|v_i - v_j|}{\sqrt{\sigma_i^2+\sigma_j^2}} \right) >  4.0,\label{eq: c2}\\
\end{equation}
where $v_i$ and $\sigma_i$ are the RVs and their 1\s\ errors for a given object at epoch $i$. The confidence threshold of 4.0 provides a probability of about one false variability detection in every 1000 objects, given the sampling and accuracy of our measurements. This corresponds to an expected impact on the measured fraction of RV variable objects of $10^{-3}$. We find that 165 sources (46\%) display significant RV variability.

Keeping all other things equal, Eq.~\ref{eq: c2} provides a more sensitive criterion  to detect variability than a comparison of the $v_i$'s with the average RV of the object. The results of Eq.~\ref{eq: c2} are also equivalent to those of a variability test based on the goodness of fit of a constant RV model \citep[see e.g.][]{HBES12}  for over 96\%\ of our sample.

\begin{table}
\centering
\caption{Overview of the results of various variability and multiplicity criteria used in this paper}
\label{tab: mult_overview}
\begin{tabular}{cccc}
\hline
\hline
Variability            & Constant           & Weak variability           & Variable         \\
\hline 
\vspace*{1mm}
RV variability         & 195               & 39                         & 126               \\
TVS analysis           & 253               & 15                         & 96                \\
\hline 
Multiple signature     & Single/Isolated   & Candidate                  & Multiple          \\
\hline 
Profile morphology     & 261               & 24                         & 63                \\
HST images$^{a}$      & 303               & 5                          & 24                \\
\hline 
\end{tabular}\\
NOTES: $a$. Nineteen stars have no HST images.
\end{table}

   \subsubsection{Line profile variability} \label{sect: lpv}
We also search for line profile variability (LPV) using a Time Variance Spectrum (TVS) analysis \citep{FGB96}. 
The TVS analysis is sensitive to variations in the amplitude, shape and position of the line profile, as well as to inconsistency in the  nebular correction between the various epochs and to technical defects such as cosmic rays and continuum localisation. The method also allows us to verify the consistency of the normalisation procedure described in Appendix~\ref{app: norm}. We adapt the original TVS approach of \citet{FGB96} to take into account the known error bars at each pixel, rather than to rely on the overall \snr\ of the spectrum. This is needed because the \snr\ is variable as a function of wavelength and depends on the chosen setup. We thus define :
\begin{equation}
TVS(\lambda)=\frac{N}{N-1}\sum_i \frac{(F_i(\lambda)-<F(\lambda)>)^2}{\sigma_i^2(\lambda)} / \sum_i \frac{1}{\sigma_i(\lambda)},\label{eq: tvs}
\end{equation}
where $F_i$ and $\sigma_i$ are the observed normalised flux and its 1\s\ error as a function of wavelength $\lambda$ at epoch $i$. $<F>$ is the weighted average flux computed over all available epochs.  The statistical significance threshold is adopted to be $\alpha=0.01$, i.e.:
\begin{equation}
TVS(\lambda) > \frac{N}{N-1} P_{\chi^2}(\alpha,N-1) / \sum_i \frac{1}{\sigma_i(\lambda)},
\end{equation}
where $P_{\chi^2}(\alpha,N-1)$ gives the cutoff value in a $\chi^2$ distribution with $N-1$ degrees of freedom such that the probability that a random variable exceeds $P_{\chi^2}(\alpha,N-1)$ is equal to $\alpha$.
In our data, most of the  variability detected through the TVS analysis is produced by non-optimum nebular correction. It results from the combined effects of the limited spectral resolution, the spatially variable level of nebular emission and the fact that FLAMES does not provide a nebular spectrum close to the target.

In addition to residuals of the nebular correction, the TVS detects significant variability  in 96 cases. By far, most cases (92) are also associated with significant RV variations so that the TVS analysis does not significantly add to the number of detected variable stars.

\subsection{Composite sources} \label{sect: comp}
   \subsubsection{Double-lined binaries} \label{sect: sb2}
The spectra obtained for each target were visually inspected to search for the signature of double-lined and asymmetric profiles. Despite the limited \snr\ of the individual epoch spectra, double-lined profiles were clearly identified in 50 cases. Higher multiplicity profiles were not observed. Asymmetric or variable profiles are observed for 13 objects. Candidate SB2s and objects for which asymmetric profiles are suspected represent another 21 cases. For pronounced SB2s, RV measurements were attempted using two Gaussians per profiles. Reliable fits could be achieved in 43 of the 50 cases. In two cases the stronger component  -- likely the primary, more massive star -- shows no RV variation. These objects are thus added to the list of binary candidates although one cannot confirm the physical link between the two component of the pairs. 

   \subsubsection{HST observations} \label{sect: hst}
We investigate the targets for visual companions using HST data taken with WFC3/UVIS as the primary camera with ACS/WFC in parallel, both using the F775W filter. These data constitute the first epoch of a proper motion program in the region \citep[GO 12499, PI: Lennon, see][]{dMLS12,SLG12}. It provides a $\sim 16' \times 12'$ mosaic centered on R136 covering over 90\% of the O stars in the Tarantula Survey. We find that 19 O stars observed with Medusa and six with ARGUS have one or several companions that are close and bright enough, for the FLAMES spectra to be a composite of multiple sources. These and an additional five multiple candidates are listed in Table~\ref{tab: vartest}.  

The HST point spread function, of $\sim$0.1', corresponds to a physical separation of about $5\times10^3$ AU at the distance of 30~Dor.  HST thus  probes  a very different, non-overlapping, separation range compared to the spectroscopic binaries discussed in Sect.~\ref{sect: var}. Five of the visual pairs identified by HST also show significant RV variations, indicating that at least one of the components of the visual system is also a spectroscopic binary

\begin{figure}
  \centering
  \includegraphics[width=\columnwidth]{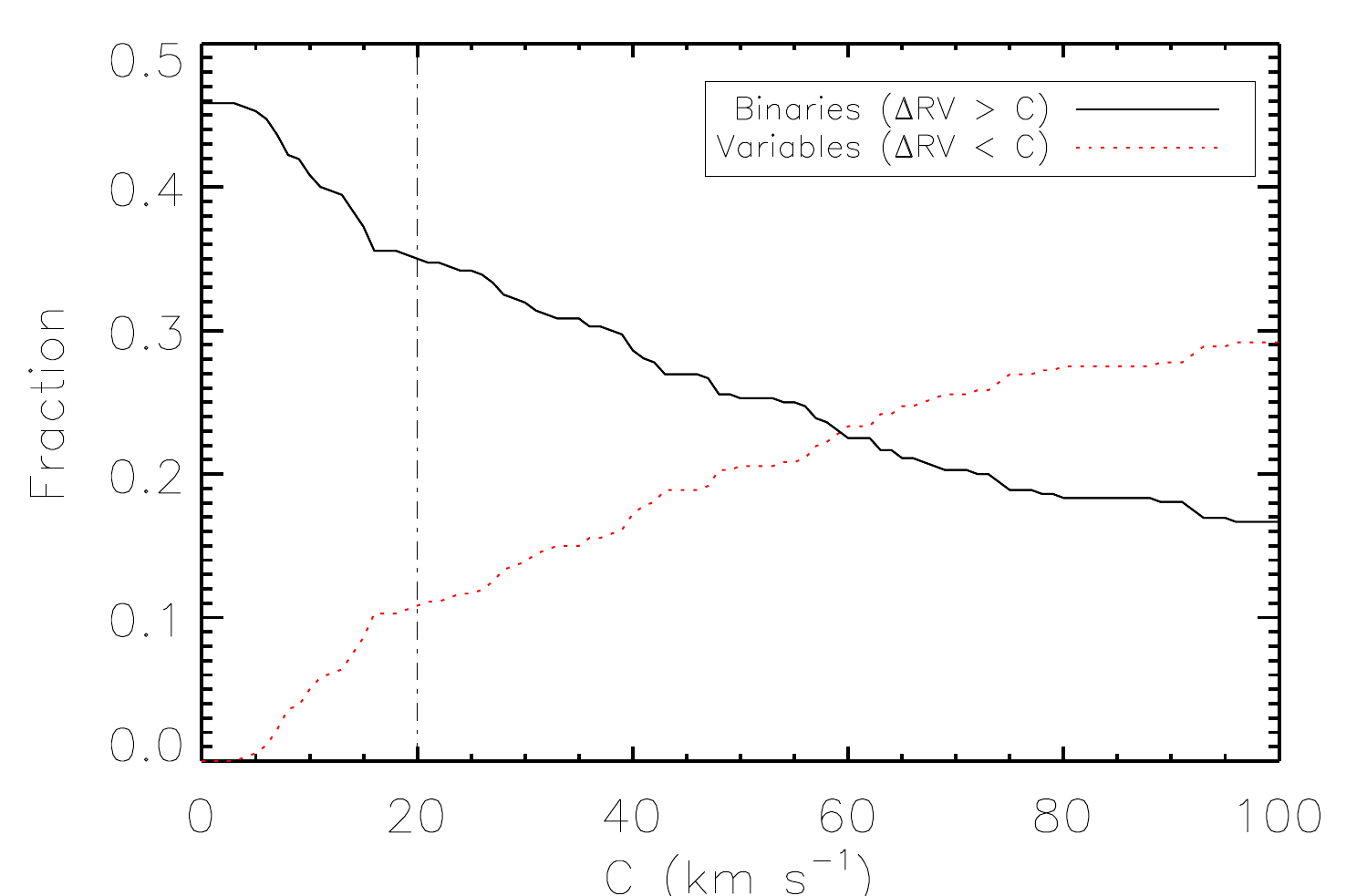}
  \caption{Variations of the fraction of systems below and above the critical RV variation amplitude $C$ separating the spectroscopic binaries (solid line) from low-amplitude RV variables (dotted line). The vertical dashed-dotted line indicates the adopted threshold of $C=20$~\kms.}
  \label{fig: bf_c4}
\end{figure}

   \subsection{Observed binary fraction} \label{sect: obs_bf}

Of all the diagnostics discussed above, RV variability is by far the most robust test to identify variable sources in our data. Complementary tests (Sects.~\ref{sect: lpv} to \ref{sect: hst}) do not significantly change the number of detected systems. Furthermore the RV variability tests defined by Eq.~\ref{eq: c2} do not rely on subjective appreciations such as e.g., the visual inspection of the line profile morphology. This is a necessary condition to model the observational biases by means of Monte-Carlo simulations (Sect.~\ref{sect: MC}). In the following we will thus focus on spectroscopic systems that reveal themselves through their RV variability.

The test provided by Eq.~\ref{eq: c2}  only identifies significant RV variations. To distinguish cases where the observed variability is genuinely caused by orbital motion in a binary system from potential intrinsic variability due to e.g. photospheric or wind effects, we impose a minimum amplitude threshold  $C$ on any significant deviations computed from any individual pair of epochs. I.e., an object is considered a spectroscopic binary if at least one pair of RV measurements satisfies simultaneously 
\begin{equation}
 \frac{|v_i - v_j|}{\sqrt{\sigma_i^2+\sigma_j^2}}  >  4.0  \hspace*{3mm}\mathrm{and}\hspace*{3mm}|v_i - v_j|> C. \label{eq: bin2}
\end{equation}

Fig.~\ref{fig: bf_c4} shows the influence of $C$ on the identified binary fraction. It reveals a kink at about 15~\kms\ that may mark the transition between a regime were photospheric variability has a significant contribution ($\Delta RV < 15$~\kms) and the regime of genuine spectroscopic binaries. A crude extrapolation of the contribution of the binaries into the low-amplitude $\Delta RV$ regime suggests the presence of about 5\%\ of genuine binaries with $\Delta RV < 15-20$~\kms. Similarly, we estimate that  photospheric variabilities affect as little as 6 to 7\%\ of our sample.

In Section.~\ref{sect: intr_mult} we will correct for the observational biases and consider the fraction of long period systems that potentially fall below the $\Delta RV$ cutoff. As a consequence, the exact value of the adopted cutoff is of little importance but has to be large enough to avoid a significant number of false detections due to measurement errors and/or photospheric variability.

Because photospheric variations in supergiants can mimic variations with amplitudes of up to 20~\kms\ \citep[e.g.][]{RCN09},  we conservatively adopt $C=20$~\kms.  Eleven per cent of the sample (39 stars) are thus identified as variable but feature an amplitude of the significant RV variations that remains below 20~\kms. These objects might be long period binaries or stars with photospheric variability and, as discussed above, we estimate an equal contribution of both categories.

Given Eq.~\ref{eq: bin2} with $C=20$~\kms, the observed spectroscopic binary fraction identified through RV variation of the component with the strongest line -- likely the primary -- is thus $35.0\pm2.5$\%, where the 1\s\ error bar gives the statistical error that results from the size of the sample \citep{SGE09}.

\begin{figure}
\includegraphics[width=\columnwidth]{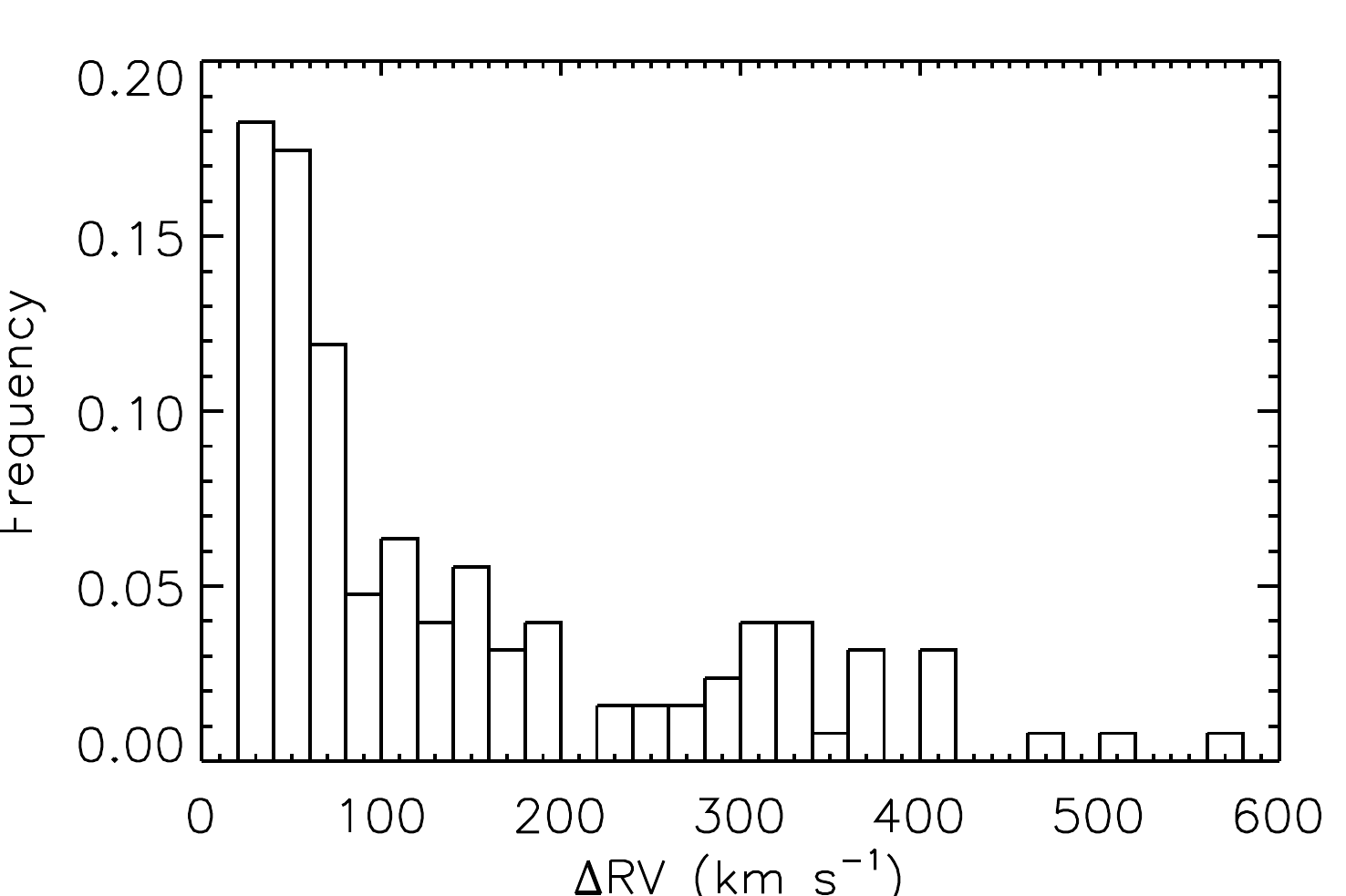}
\caption{Histogram of the peak-to-peak amplitude RV variations of each object. Only significant variations according to Eq.~\ref{eq: bin2} are considered. Therefore the first bin is empty.} \label{fig: dRV}
\end{figure}
\begin{figure}
\includegraphics[width=\columnwidth]{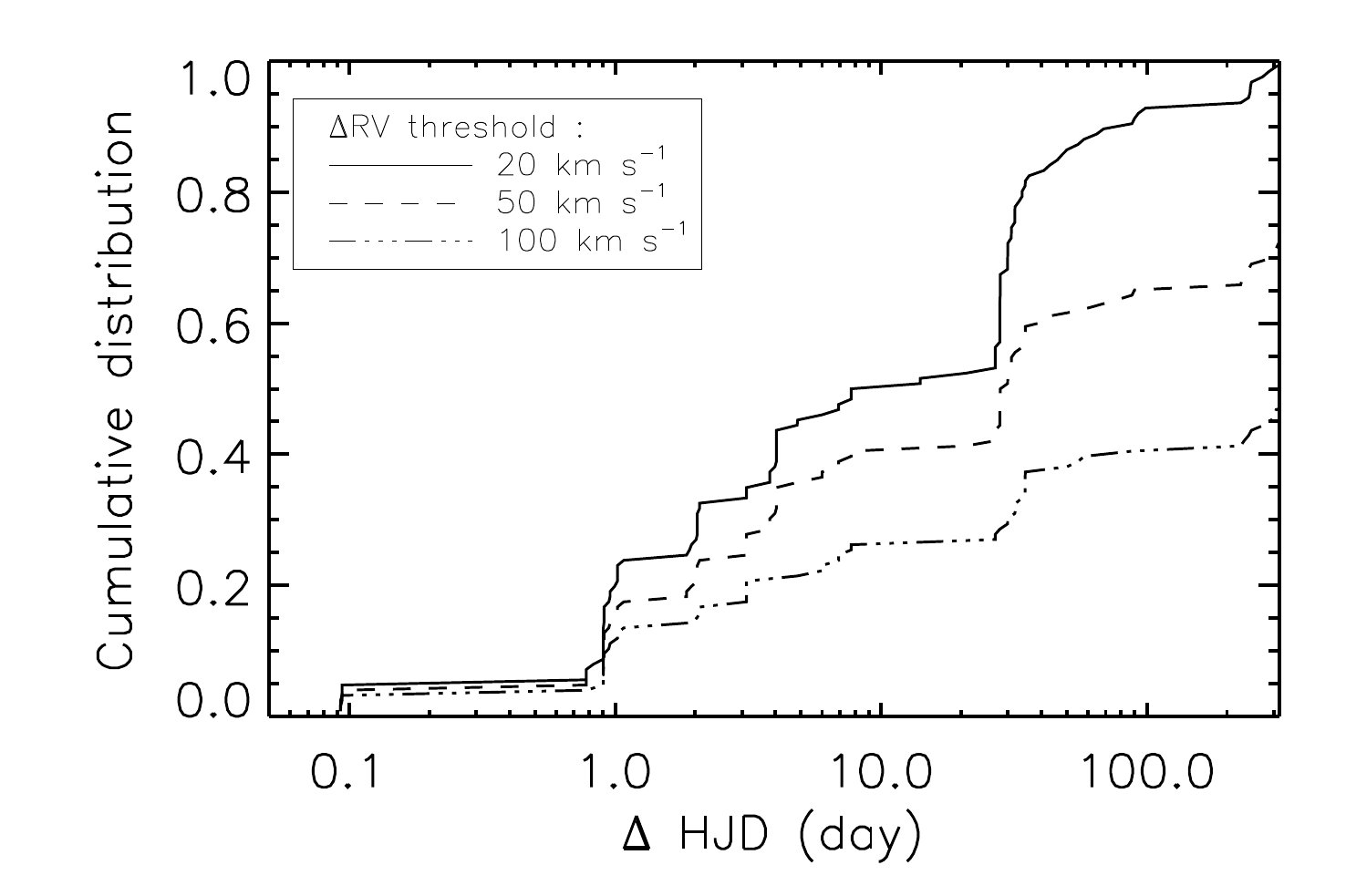}
\caption{Cumulative distribution of the minimum time difference $\Delta$HJD for a binary to display significant RV variations according to Eq.~\ref{eq: bin2} and for different RV amplitude thresholds ($C$) in the RV variation amplitudes. The cumulative distributions are normalised to the total number of binaries.}\label{fig: dHJD}
\end{figure}

\subsection{Comparison with earlier results} \label{sect: btt09}

\citet[][\citetalias{BTT09}]{BTT09} reported on a similar study of 54 O and early B stars in NGC~2070. They obtained $R\sim3700$ seven-epoch spectroscopy covering a wavelength range equivalent to the coverage of our LR02+LR03 settings. Using the external-to-internal velocity dispersion ratio $\sigma_\mathrm{E}/\sigma_\mathrm{I}$ and a threshold of 3\, they identified 17 RV variables and nine additional SB2 binaries and binary candidates for a maximum detected binary fraction of 0.48. The results of \citetalias{BTT09} thus  suggest a larger binary fraction than the one we observed in VFTS. To understand this difference, we perform a detailed comparison of the VFTS and \citetalias{BTT09} approaches.

We first apply our variability and binary criteria on \citetalias{BTT09}'s RV measurements. We detect 24 RV variable objects among the 48 objects listed in their Table~2.  The difference comes from the fact that the ratio  $\sigma_\mathrm{E}/\sigma_\mathrm{I}$ is artificially reduced if one measure has a significantly worse accuracy that the others. All of the stars identified as RV variables have peak-to-peak amplitudes in excess of 20~\kms\ and thus qualify as spectroscopic binaries according to the  criteria of Eq.~\ref{eq: bin2}. Applying our method on the \citetalias{BTT09} RV data set, we measure a spectroscopic binary fraction of 50.0\%\ while \citetalias{BTT09} obtained 35.4\% by using the  $\sigma_\mathrm{E}/\sigma_\mathrm{I}$ ratio. Two of the seven binaries not identified as RV variable by \citetalias{BTT09} are still accounted for in their total binary count because of asymmetry in their line profiles. Even though we require a stricter (4\s) threshold, we conclude that our binary criteria are slightly more sensitive than the one used by \citetalias{BTT09} based on $\sigma_\mathrm{E}/\sigma_\mathrm{I}>3$.

As a second step in comparing our results with those of \citetalias{BTT09} we compute the binary fraction from the overlapping sample between \citetalias{BTT09} and the present work, which are 34 of the 54 objects studied by \citetalias{BTT09}. Of these 34 objects, \citetalias{BTT09} reported RVs for 30 of them among which they identified 12 binaries (40\%) through RV variations. Among the same set of 30 stars, VFTS identify 13 binaries (43\%) and 6 (20\%) low-amplitude variable objects, with $\Delta$RV ranging from 7 to 19~\kms.

On the overlapping sample, the two studies agree well, with VFTS being slightly more sensitive due to the improved RV accuracy and statistical criteria. The significantly lower binary fraction observed  in the whole VFTS sample relative to the work of \citetalias{BTT09} is attributed to the focus of \citetalias{BTT09} on the brightest stars. We will show in Sect.~\ref{sect: Kmag} that fainter O-type objects have an intrinsically lower binary fraction. 

\begin{table}
\centering
\caption{Overview of properties of the Monte-Carlo grid. From left to right we list the physical parameter, its corresponding probability density function (pdf) and domain of applicability, the name of the pdf variable and its allowed range  and step size. The last row gives the allowed range and step size of the binary fraction.}
\label{tab: MC}
\begin{tabular}{lllccc}
\hline 
\hline 
Parameter                   &  pdf               & Domain    & Var.           &   Range     & Step \\
\hline 
\vspace*{1mm}
$P$ & $(\log_{10} P/\mathrm{d})^{\pi}$ & 0.15 - 3.5      & $\pi$    & $-$2.50 - $+$2.50 & 0.05 \\
$q$ &               $q^{\kappa}$       & 0.1 - 1.0       & $\kappa$ & $-$2.50 - $+$2.50 & 0.05 \\   
$e$ &               $e^{\eta}$         & 10$^{-5}$ - 0.9 & $\eta$   & $-$0.5 (fixed)    & $n/a$ \\
\fbin                                 & $n/a$           & $n/a$    & \fbin     & $+$0.20 - $+$1.00 & 0.01 \\  
\hline 
\end{tabular}  
\end{table}

\begin{figure*}
  \centering
  \includegraphics[width=\textwidth]{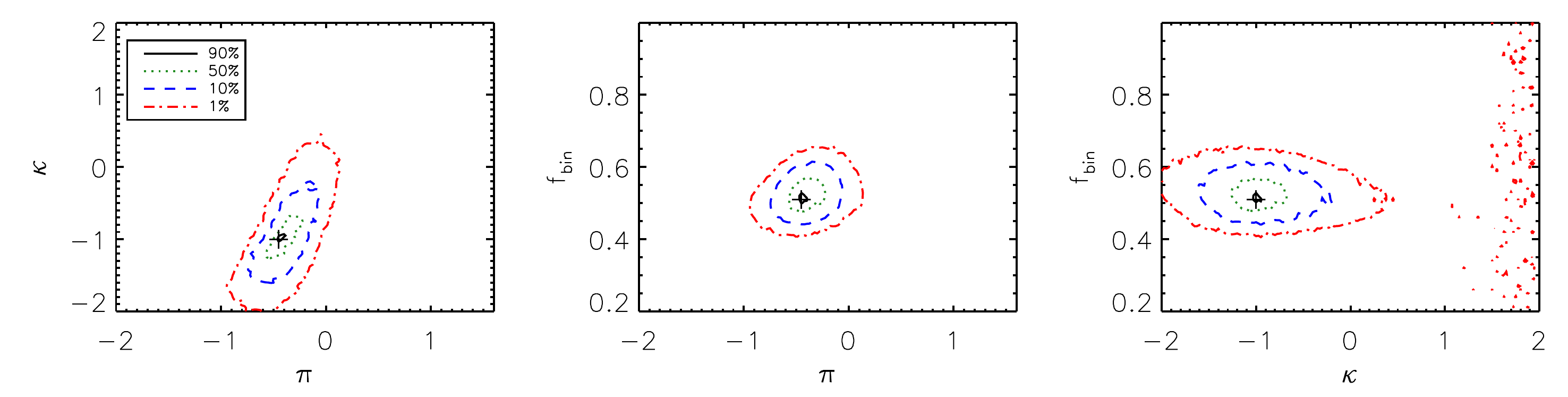}
  \caption{Projections of the global merit function $\Xi'$  on the various pairs of planes defined by \p, \k\ and \fbin. The absolute maximum is indicated by a cross ($+$). Contours indicate loci of equal-values of the merit function, expressed as a fraction of its absolute maximum (see legend). }
  \label{fig: jan_2d}
\end{figure*}

\begin{figure*}
  \centering
  \includegraphics[width=.8\textwidth]{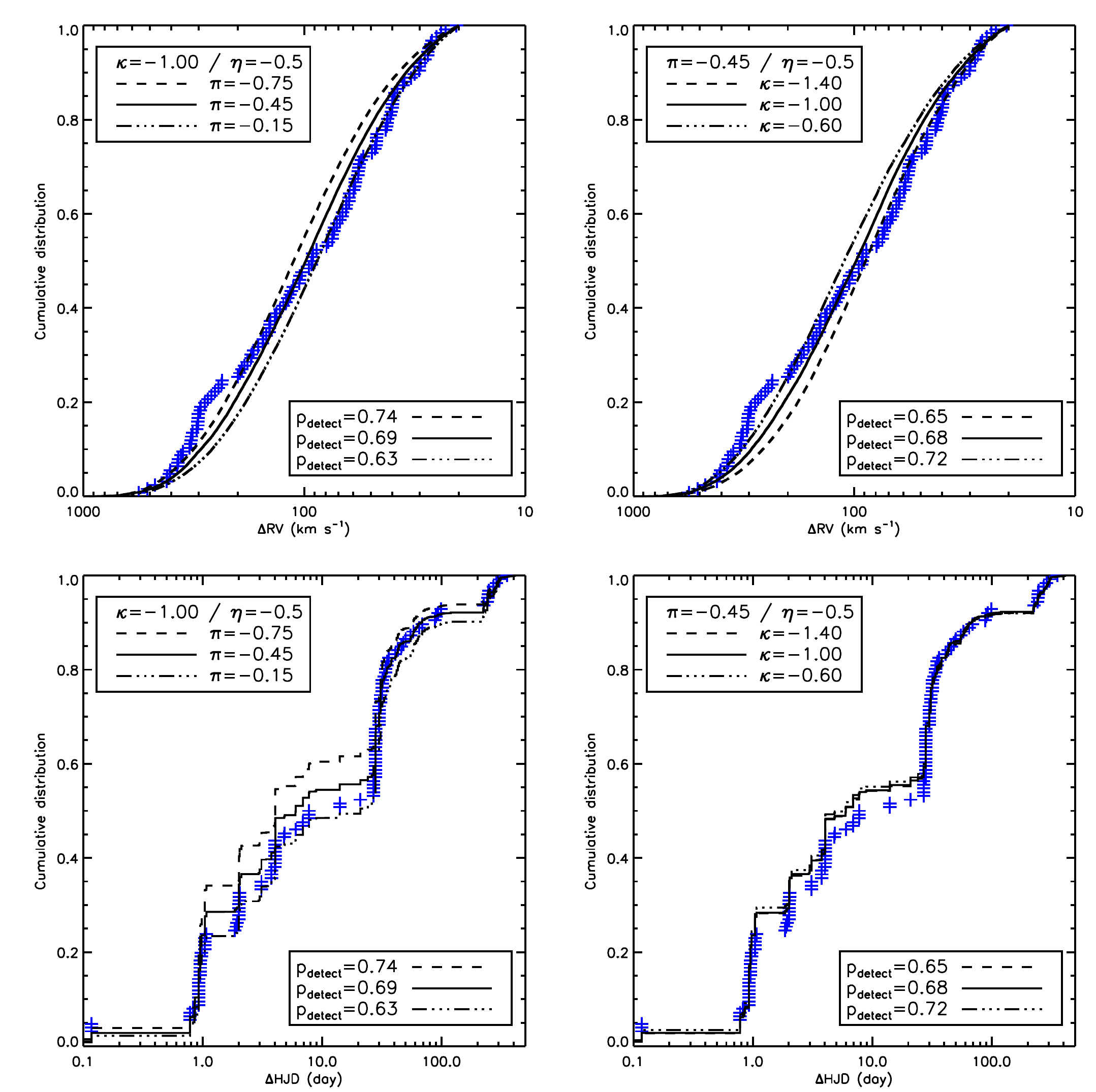}
  \caption{Comparison between the observed (crosses) and simulated (lines) cumulative distributions of the peak-to-peak RV amplitudes (top row) and of the variability time scales (bottom row). In the left- (resp.\ right-) hand panels, we vary the exponent $\pi$ of the period distribution (resp.\ $\kappa$ of the mass-ratio distribution) by $\pm$1\s. The upper-left legends indicate the values of $\pi$, $\kappa$\ and $\eta$ considered in each panel. The bottom-right legends give the overall VFTS detection probability for the adopted parent distributions.}
  \label{fig: fig1}
\end{figure*}

\section{Intrinsic multiplicity properties} \label{sect: intr_mult}
The detected binary population provides a minimum estimate of the true binary population in the Tarantula region. The number of undetected binaries depends on the sensitivity of our campaign in terms of RV accuracy and time sampling and on the intrinsic distribution of the orbital parameters. The problem of correcting for the observational biases is thus ill-posed and we employ here a Monte-Carlo approach to estimate the impact of the observational biases and to provide constraints on the intrinsic binary fraction and distributions of orbital parameters. 

\subsection{Monte-Carlo method} \label{sect: MC}
In our Monte-Carlo approach, we simulate massive star populations with intrinsic multiplicity properties randomly drawn from adopted parent distributions. Accounting for observational biases due to sampling and measurement uncertainties we search for sets of distributions that reproduce the properties of the observations in three aspects: the observed binary fraction, the peak-to-peak amplitude $\Delta RV$ of the RV variations (Fig.~\ref{fig: dRV}) and the minimum time scale $\Delta$HJD for significant RV variation to be observed (Fig. \ref{fig: dHJD}).

The method is detailed in Appendix~\ref{app: MC}. In short, simulated and observed distributions are compared by means of Kolmogorov-Smirnov (KS) tests. The detected binary fraction predicted by the simulations is compared to the observed fraction using a Binomial distribution \citep{SdMdK12}. A global merit function ($\Xi'$) is constructed by multiplying the two KS probabilities obtained for the $\Delta$RV and $\Delta$HJD distributions and the Binomial probability:
\begin{equation}
\Xi' = P_\mathrm{KS}(\Delta RV) \times P_\mathrm{KS}(\Delta HJD) \times B(N_\mathrm{bin},N,f_\mathrm{bin}^\mathrm{simul}), \label{eq: merit}
\end{equation}
where  $f_\mathrm{bin}^\mathrm{simul}$ is the simulated fraction of detected binaries, $N_\mathrm{bin}$ is the detected number of binaries in the observations and $N$ is the population size.

Following \citet{KoF07} and \citet{SdMdK12}, we use power-laws to describe the intrinsic distribution functions of orbital parameters: $f(\log_{10}P/\mathrm{d}) \sim (\log_{10} P)^{\pi}$, $f(q) \sim q^{\kappa}$ and $f(e) \sim e^{\eta}$. The exponents $\pi$, $\kappa$ and $\eta$ are then varied to explore different shapes for the parameter distributions. Because the global merit function is insensitive to the eccentricity distribution (see Fig.~\ref{fig: param}), we cannot constrain $\eta$ and we adopt the eccentricity distribution measured using the O star populations of Galactic young open clusters \citep{SdMdK12}, i.e.\ $f(e)\sim e^{-0.5}$. We thus explore the behaviour of $\Xi'$ in the three-dimensional parameter space defined by the intrinsic binary fraction and distributions of the orbital periods and mass-ratios. The exploration domains and mesh sizes are given in Table~\ref{tab: MC}.

\subsection{Results} \label{sect: res}

Fig.~\ref{fig: jan_2d} displays the merit function ($\Xi'$) projected on planes defined by the various pairs of parameters. It shows that the optimum is well defined within the explored ranges. The best agreement between the observed and simulated distributions is obtained for $\pi=-0.45\pm0.30$, $\kappa=-1.00\pm0.40$ and a intrinsic binary fraction of $f_\mathrm{bin}=0.51\pm0.04$. The quoted errors have been estimated as the 1$\sigma$ confidence interval on the retrieved parameters of 50 synthetic data sets randomly drawn from parent distributions close to our best fit distributions (see Appendix~\ref{app: merit} and the lower part of Table~\ref{tab: mctest}).  The comparison between the simulated and observed cumulative distribution functions (CDFs) of the peak-to-peak RV amplitudes and of the variability time scales is presented in Fig.~\ref{fig: fig1}. 

While the overall shapes of the observed and simulated functions are well matched, some features of the observed distributions are not fully reproduced by the simulations. The CDF of the peak-to-peak RV amplitudes presents an overabundance of systems with $\Delta RV$ in the range 300 to 400~\kms. This could be reproduced by increasing the fraction of short period systems, i.e.\ a more negative $\pi$ exponent, but the CDF of the variability time scale would then not be properly reproduced. Alternatively, a flatter distribution of mass-ratios would help to improve the CDF($\Delta RV$) fit in the 300 to 400~\kms\ range but would overestimate it in the lower amplitude range ($\Delta RV < 100$~\kms). 

We also attempt to reproduce the overabundance of systems with $\Delta RV$ in the range 300 to 400~\kms\ by including specific populations of short period systems, of equal mass systems, of high-mass primaries but none of these, nor any combinations of them, were able to reproduce the observed localised bump.  

 The bump in the CDF($\Delta RV$) curve comprises less than 15 systems and can also result from statistical fluctuations. Indeed, the individual $P_\mathrm{KS}$ probabilities are already large: $P_\mathrm{KS}(\Delta RV) \approx 0.4$ and $P_\mathrm{KS}(\Delta HJD) \approx 0.8$. We thus consider the present fit to provide an adequate representation of the data.

The value of the period distribution index is remarkably close to the value $\pi_\mathrm{GOC}=-0.55\pm0.22$ found in the O star population of Galactic open clusters \citep[GOCs, ][]{SdMdK12}. The intrinsic binary fraction is 51\%\ in VFTS compared to $69\pm9$\%\ in the GOCs. Finally, the mass-ratio distribution obtained in the present study favours lower mass companions while the mass ratios in the \citeauthor{SdMdK12} study were equally distributed ($\kappa_\mathrm{GOC}=-0.1\pm0.6$). Both studies do however agree within 2\s. The implications of these results are discussed in Sect.~\ref{sect: discuss}. 

\subsection{Sensitivity domain of the VFTS campaign} \label{sect: bias}

We use the method of \citet{SGE09} to estimate, given the distributions of the orbital parameters that we find, the sensitivity of our campaign with respect to the orbital period, the mass ratio, the eccentricity and the primary mass. 

Compared to the \citet{SGE09} approach, we now include the measurement uncertainties obtained at each epoch in the simulations and we implement the binary detection criteria given by Eqs.~\ref{eq: c2} and \ref{eq: bin2} to consistently follow the binary detection strategy laid out in this paper. 

\begin{figure}
\includegraphics[width=\columnwidth]{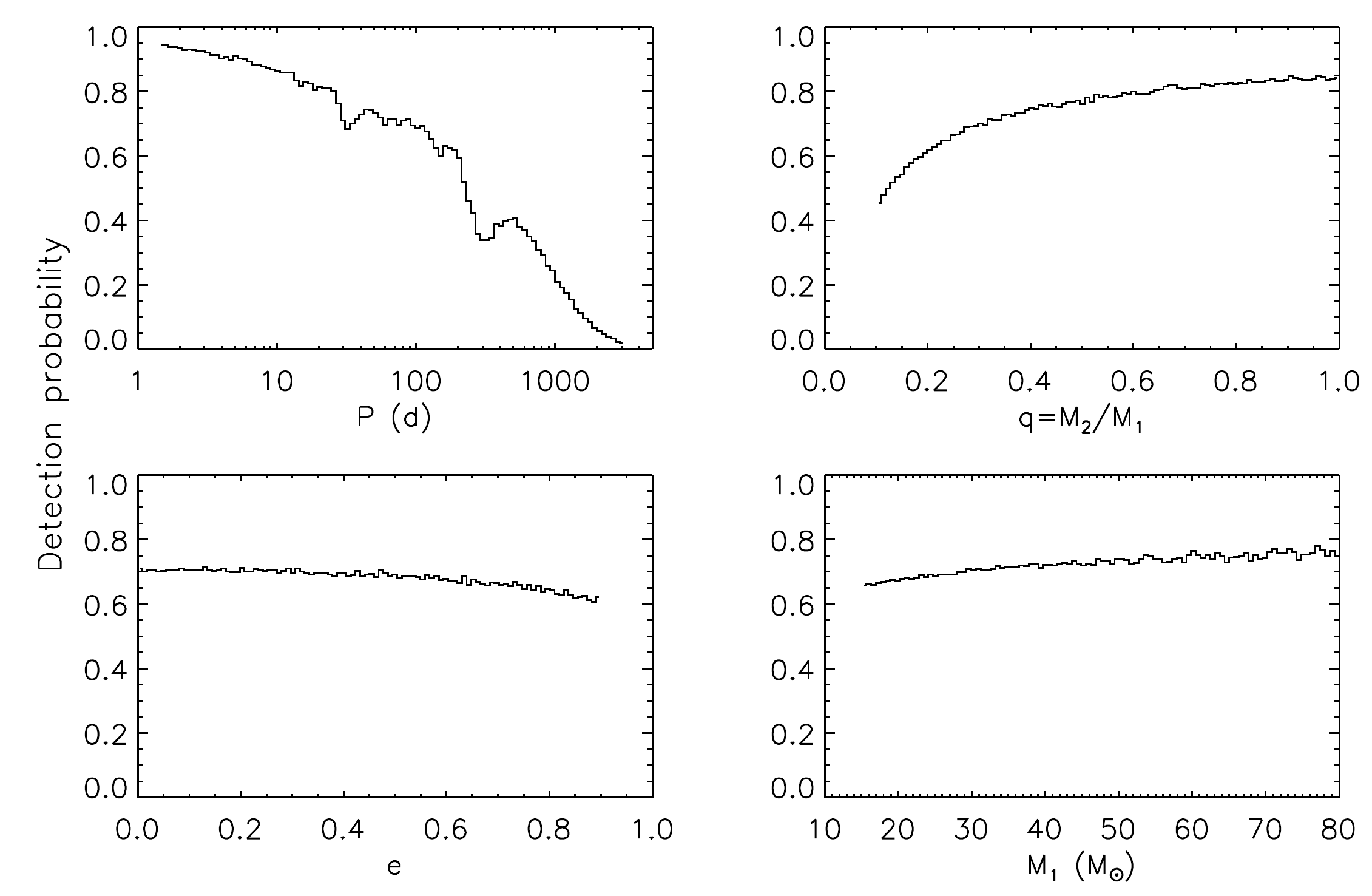}
\caption{Overview of the binary detection probability achieved by the VFTS campaign as a function of the orbital period, the mass ratio, the eccentricity and the primary mass. [See Sect.~\ref{sect: bias} for more details.]}
\label{fig: detect}
\end{figure}

The detection probability curves in  Fig.~\ref{fig: detect} are projections from the multi-dimensional space onto each of the four dimensions investigated. The binary detection probability is a strong function of orbital period. It moderately depends on the mass ratio except for systems with a large mass difference. The detection probability is also weakly affected by the primary mass as well as by the eccentricity. Because we have not taken into consideration the fact that high eccentricity systems may  preferentially have a long period, we underestimate the detection rate at very high eccentricities ($e>0.7$).

\section{Binary fraction} \label{sect: correl}
In this section, we discuss how the binary fraction correlates with the spatial distribution,  brightness, spectral-type and luminosity class of the stars in our sample. Supplementary figures supporting this discussion are provided in Appendix~\ref{app: LC}.

\begin{figure}
\includegraphics[width=\columnwidth]{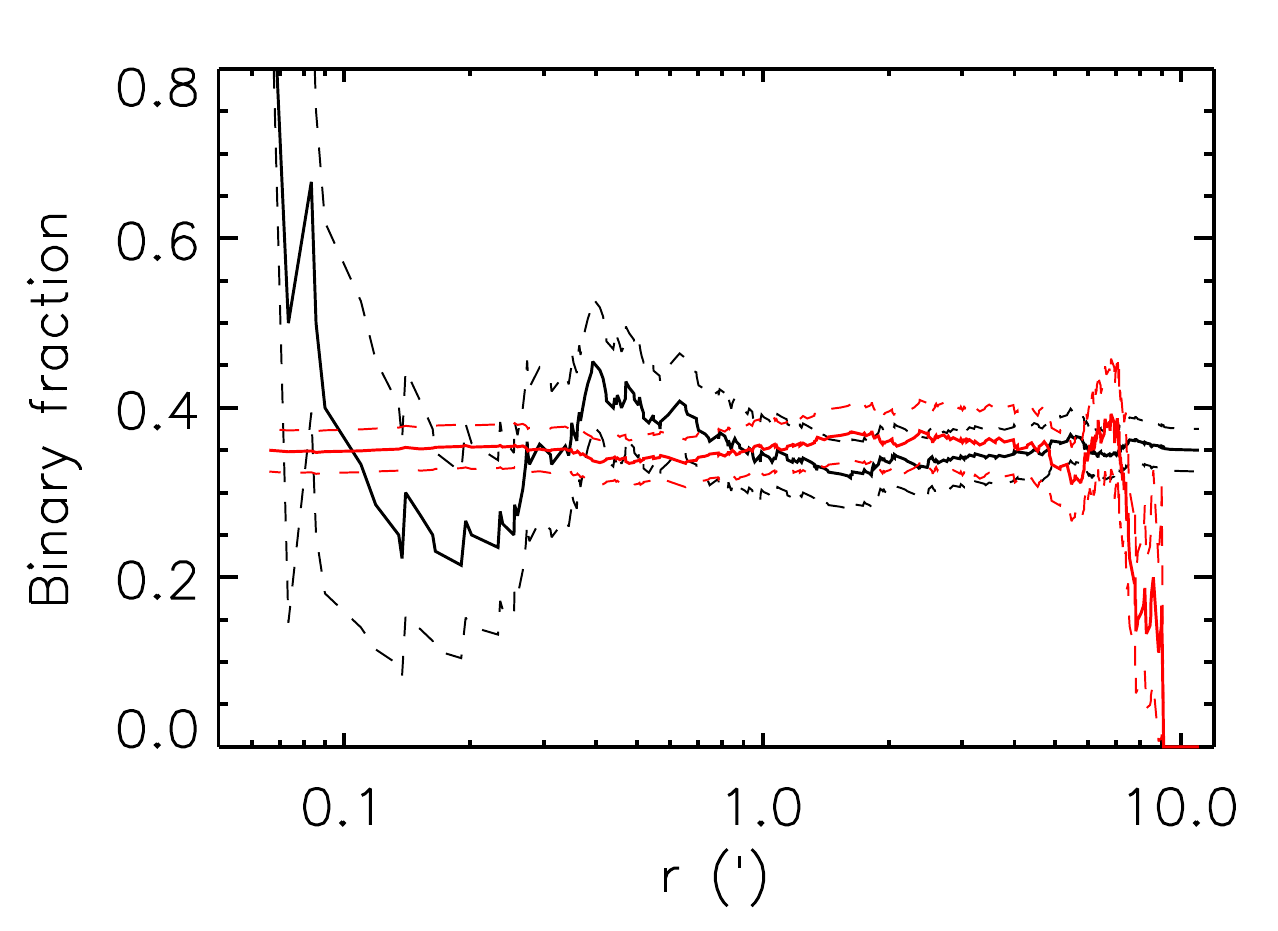}
\caption{Cumulative radial distribution of the observed binary fraction as a function of the distance to the centre of the field. The black solid line shows the distribution computed inside-out, and the grey/red line shows the distribution computed outside-in. The dashed curves indicate the $\pm 1\sigma$\ confidence interval on both curves.}  \label{fig: bf_cumul}
\end{figure}

\begin{table}
\centering
\caption{Observed fraction of RV variables and binaries among different O star populations in 30~Dor.} 
\label{tab: bf}
\begin{tabular}{lcc}
\hline
Population (size) & RV variables (Eq.~\ref{eq: c2}) & Binaries (Eq. ~\ref{eq: bin2}) \\
$C$ (\kms)      & 0.0         &  20.0 \\ 
\hline 
\vspace*{1mm}
Overall       (360)  & $0.458 \pm 0.026$  & $0.350 \pm 0.025$ \\ 
NGC~2070$^a$  (184)  & $0.429 \pm 0.036$  & $0.332 \pm 0.035$ \\
NGC~2060$^b$   (73)  & $0.562 \pm 0.058$  & $0.384 \pm 0.057$ \\
Remaining$^c$ (103)  & $0.437 \pm 0.049$  & $0.359 \pm 0.047$ \\
\hline
\end{tabular}\\
\flushleft {\sc notes:} a. Defined as a 2.4\arcmin\ radius around R136 ($\alpha=05:38:42.0$, $\delta=-69:06:02$); b. Defined as a 2.4\arcmin\ radius around $\alpha=05:37:45.0$, $\delta=-69:10:15$; c. All stars not included in NGC~2060 nor NGC~2070.
\end{table}

\begin{figure*}
  \centering
  \includegraphics[width=.32\textwidth]{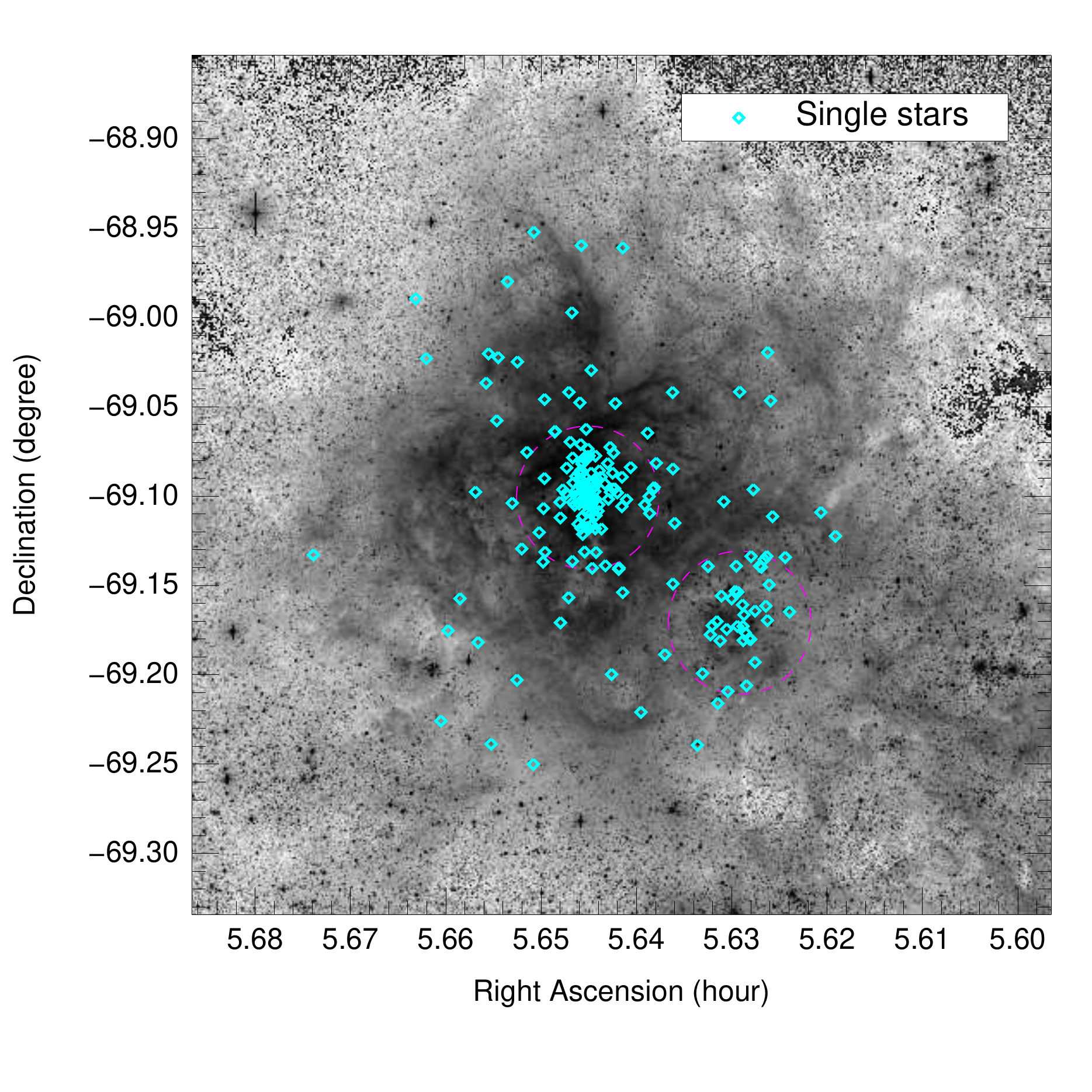}
  \includegraphics[width=.32\textwidth]{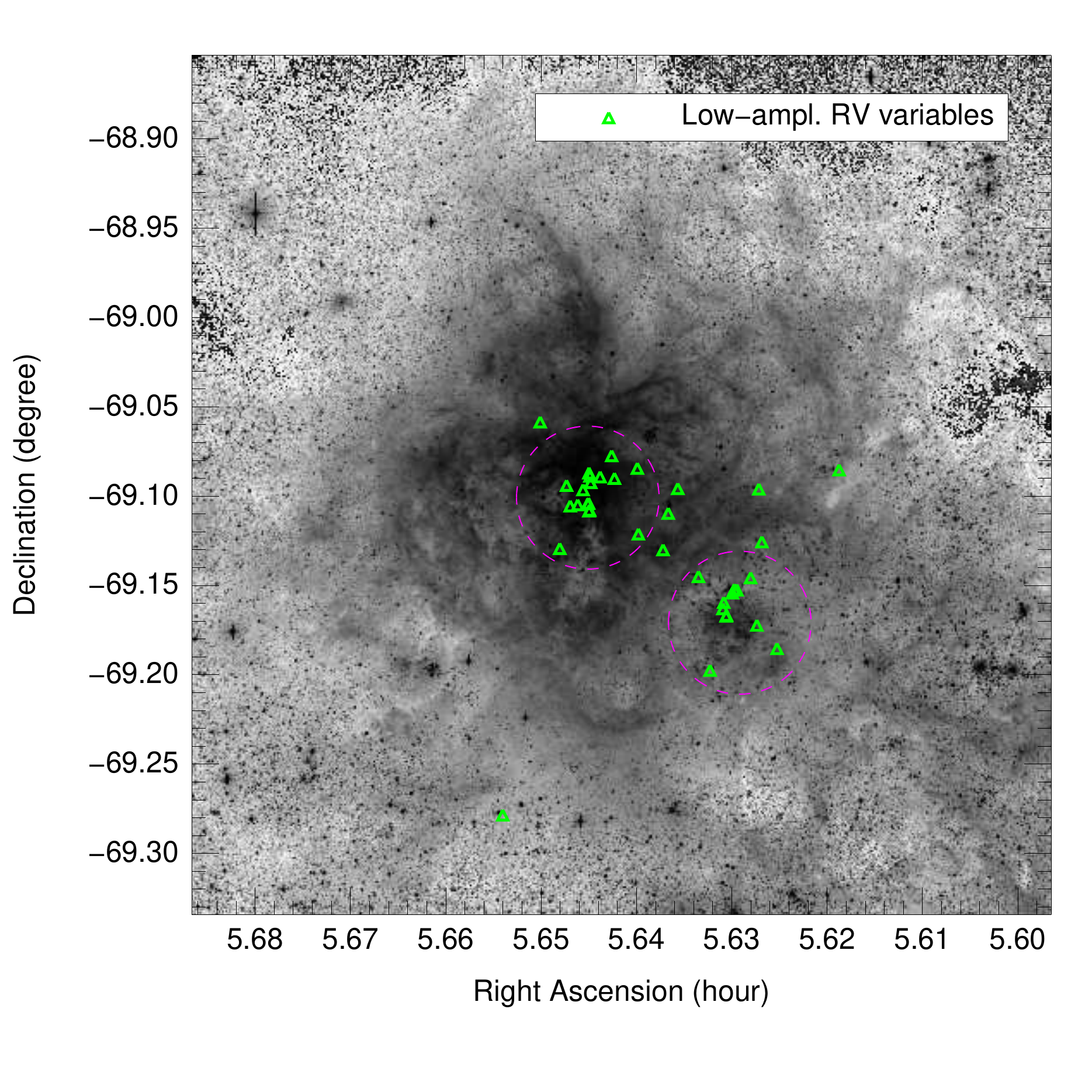}
  \includegraphics[width=.32\textwidth]{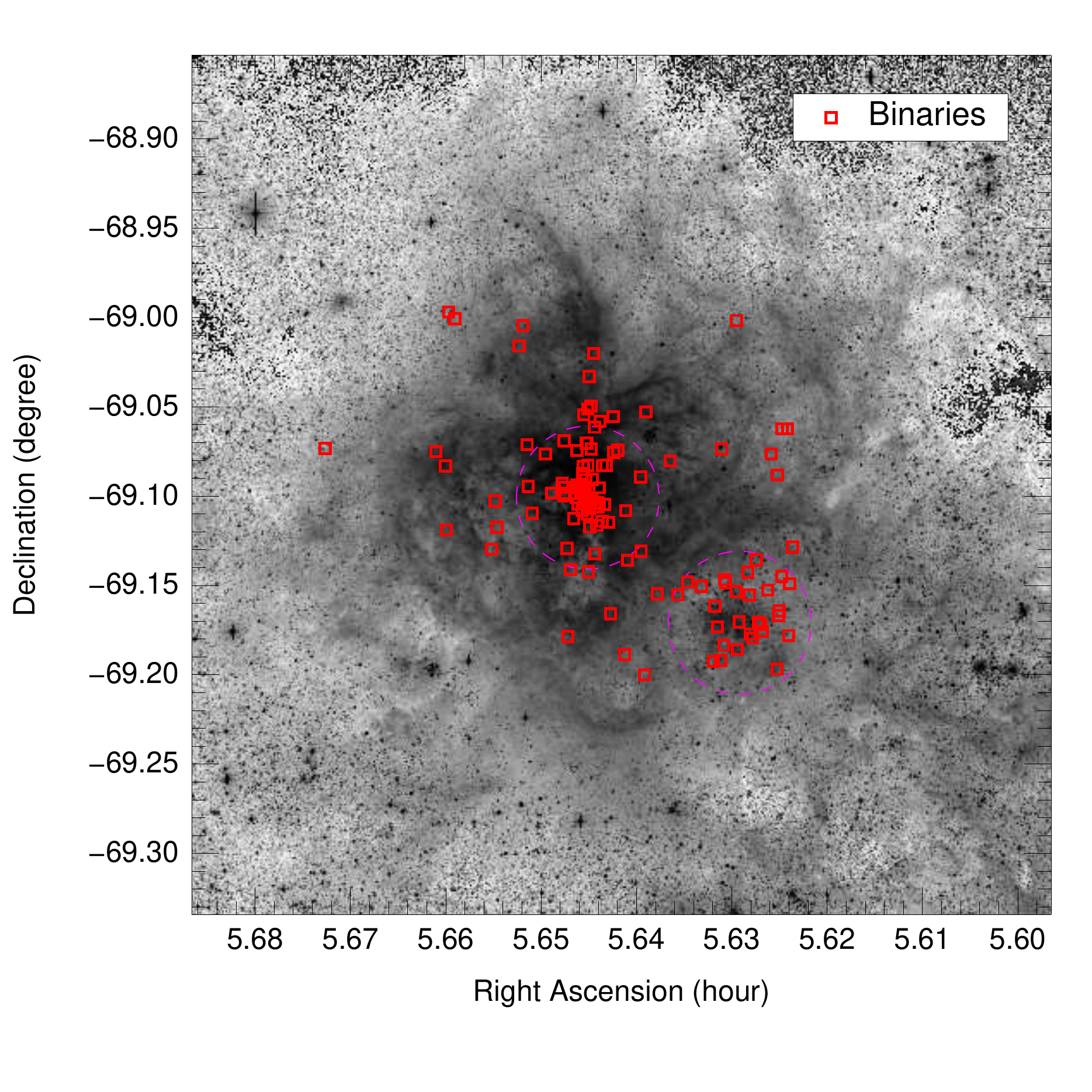}
  \caption{From left to right, distributions of the RV constant, likely single stars, the binary candidates (low-amplitude RV variables) and the binary stars in the 30~Dor field of view. }
  \label{fig: left_2d}
\end{figure*}

   \subsection{Spatial variations}

We search for variations in the observed binary fraction throughout the field of view using various techniques. First we divide the stars in three samples based on their spatial location: stars within a 2.4\arcmin\ radius from the centre of NGC~2060 and NGC~2070, and stars outside these two clusters. NGC~2060 tends to display a slightly larger fraction of low-amplitude RV variable stars (Table~\ref{tab: bf}). This suggests a larger number of objects affected by photospheric variability, likely supergiants, reflecting the older age of the cluster, hence the larger fraction of supergiants (see also Fig.~\ref{fig: bf_sg}). When the sample is limited to stars with large amplitude RV variations, the genuine binaries, there are no obvious spatial differences between the various populations.

We also look for spatial variations in the field of view using a 2D smoothing. We do not identify significant variations save for two trends. There are slightly fewer binaries towards the Southern and, to a lesser extent,  Northern parts of the field of view, as well as some enhancement in between NGC~2070 and NGC~2060. These trends are significant at the 95\%\ level ($\approx1.7\sigma$), but do not reach the 2$\sigma$ confidence level. 

Finally, we also build the radial cumulative distribution of the binary fraction, starting both from the inside and from the outside (Fig.~\ref{fig: bf_cumul}). The centre of the field of view is taken at the core of R136, as defined in Table~\ref{tab: bf}. Overall, there is no significant variability but for the edges of the field. The binary fraction for the 22 outermost objects is $0.14\pm0.07$ which is just at 3\s\ from the average binary fraction in the whole sample. A visual impression of the lack of binaries in the outer regions can  be obtained by comparing the locations of the single and binary stars as shown in Fig.~\ref{fig: left_2d}.

\begin{figure}
\includegraphics[width=\columnwidth]{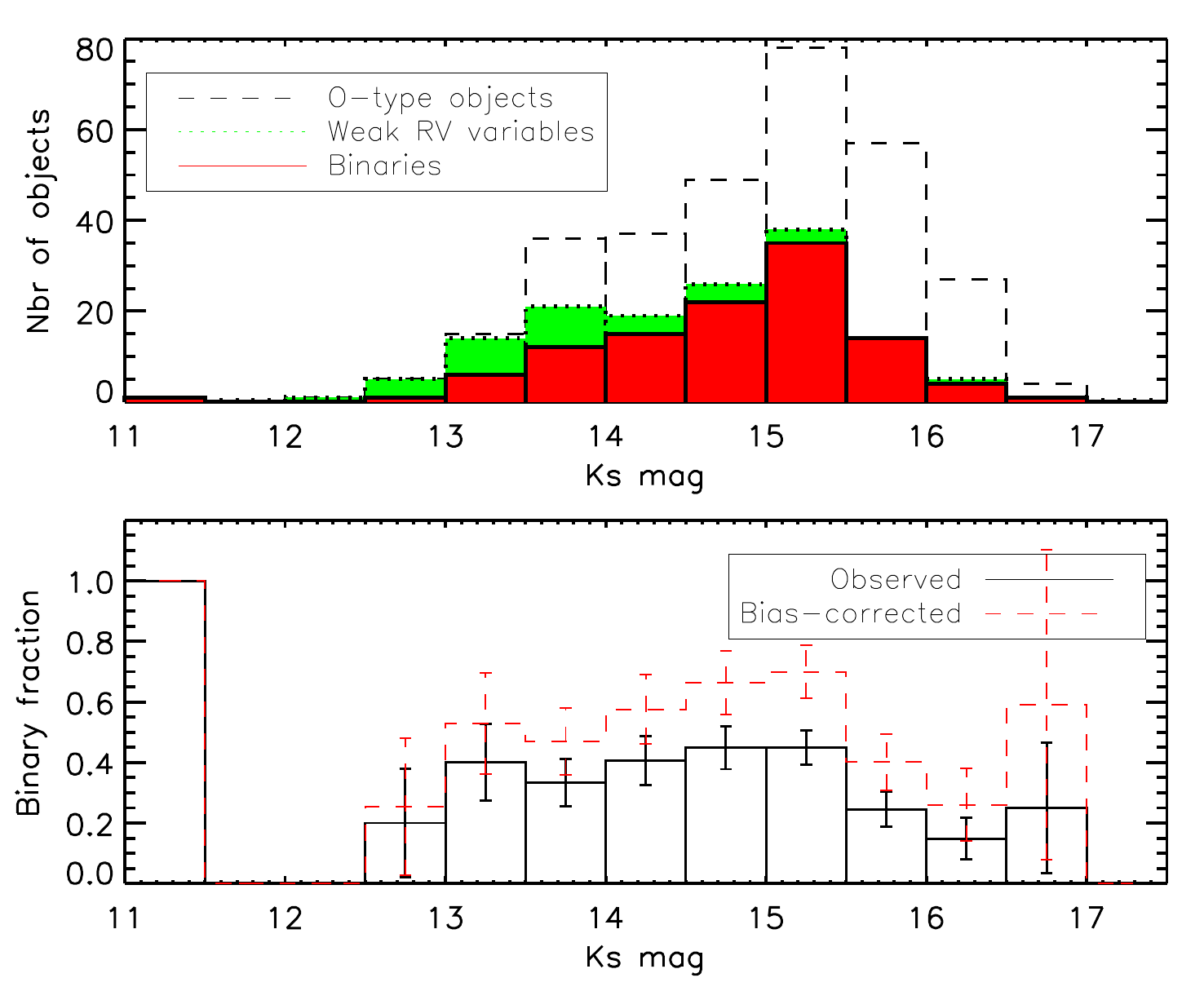}
\caption{Top panel: Number of objects (dashed line) and of binaries (solid line) per $K_\mathrm{s}$-band magnitude bin. Bottom panel: Observed binary fraction in corresponding magnitude bins (solid line) and intrinsic binary fraction (dashed line). Error bars are plotted whenever a bin contains more than one star. Fig.~\ref{fig: bf_Kmag_lc} provides similar figures showing the dependence with luminosity class. } \label{fig: bf_Kmag}
\end{figure} 

   \subsection{Brightness dependence} \label{sect: Kmag}

Objects in our sample span five magnitudes in the $V$ band. We use the $V$-band photometry published in \citetalias{ETHB11} to investigate any variations of the binary fraction with magnitude.  We also employ $K_{\rm s}$-band photometry from the VISTA Magellanic Clouds (VMC) Survey \citep{CCG11} to the same end.  Specifically, we use PSF-fitting photometry of the 30~Dor VMC Survey tile from \citet{RKG12}, with cross-matches to the VFTS sources listed by Zaggia et al. (in preparation).  $K_{\rm s}$-band magnitudes from the PSF-fitting analysis are unavailable for 15 VFTS stars, either due to crowding, blending or saturation issues.  VMC photometry is also missing for the Argus targets and these are henceforth excluded from the present analysis. Where possible, the VMC Survey results are supplemented with 2MASS photometry using the cross-matches given in \citetalias{ETHB11}. 

The analysis of both bands provides similar results. Because of the variable extinction in the region (Ma\'{i}z Apell\'aniz et al., in preparation) we focus here on the $K_\mathrm{s}$ data only. Fig.~\ref{fig: bf_Kmag}  displays the magnitude distributions of the VFTS O stars and the observed binary fraction in various magnitude bins. The latter seems relatively uniform for objects brighter than $K_\mathrm{s}=15.5$~mag. 
For dimmer objects, the observed binary fraction starts to decrease and reaches 16\%\ at the magnitude cut-off of our survey: $K_\mathrm{s}\approx16.5$~mag, corresponding to $V=17$~mag. 

The apparent lower binary fraction for fainter stars can be explained by the combination of two observational effects: \begin{enumerate}
\item[(i)] spectra from fainter stars have a lower \snr, thus  RV measurements are on average less accurate and small RV variations are more difficult to detect;
\item[(ii)] binaries are typically brighter than single stars of the same spectral type by up to 0.75~mag, depending on the brightness of the companion. Binary systems thus tend to avoid the faintest bins, which then appear to present a lower binary fraction.
\end{enumerate}

To investigate the effect of (i) and estimate the intrinsic binary fraction in each bin, we compute the average detection rate achieved in different magnitude bins using the Monte-Carlo method of \citet{SGE09}. We adopt the orbital parameter distributions obtained previously (Sect.~\ref{sect: MC}). However, we do not take into account the likely dependence of the primary mass with the magnitude, i.e.\ we keep the primary mass range fixed to 15-80~\msun\ for all considered intervals. This impacts by a few per cent the detection probabilities listed in Table~\ref{tab: Kmag}. When the results are computed for all of the O stars, a slightly different binary fraction is found than obtained in Sect.~\ref{sect: MC}.  Still, Table~\ref{tab: Kmag} indicates that the lack of binaries at $K_\mathrm{s}>15.5$ is an intrinsic effect, which can not be explained by the lower probability to detect binaries among the fainter objects in our sample.

\begin{table}
\centering
\caption{Number of objects, observed binary fraction detection probability and intrinsic binary fraction as a function of the $K_\mathrm{s}$ magnitude.}
\label{tab: Kmag}
\begin{tabular}{cccccc}
\hline
$K_\mathrm{s}$ mag range & Size & $f_\mathrm{bin}^\mathrm{obs}$ & $P_\mathrm{detect}$ & \fbin \\
\hline
\vspace*{1mm}
$<14.0$    &  50 & $0.345\pm0.106$ & 0.731 & 0.472\\
14.0--14.5 &  34 & $0.405\pm0.129$ & 0.705 & 0.575\\
14.5--15.0 &  42 & $0.449\pm0.106$ & 0.676 & 0.664\\
15.0--15.5 &  68 & $0.449\pm0.084$ & 0.635 & 0.706\\
15.5--16.0 &  53 & $0.245\pm0.115$ & 0.609 & 0.403\\
16.0--17.0 &  24 & $0.148\pm0.177$ & 0.572 & 0.259\\
\hline
\end{tabular}
\end{table}

To investigate the effect of (ii) we simulate the magnitude distribution of an O-type population made-up of 50\%\ single stars and 50\%\ binaries. We find that the intrinsic binary fraction in the brightest bins was typically close to 60\%\ but drops to 20-25\%\ in the last bin, in excellent agreement with the intrinsic binary fraction measured in Table 7. Adding a slightly variable extinction easily allows us to extend the effect over the last two bins. 
If early B-type stars with a similar binary fraction are included in the simulations the  magnitude bins at $K_\mathrm{s}\sim$15.5-16.5 do not show such a drop.

 We thus conclude that the variation of the intrinsic binary fraction with magnitude is simply a consequence of the fact that our sample is defined by a cut-off in spectral type, i.e.\ the O-type sub-sample of the VFTS is not magnitude limited, save for the few stars with severe extinction. We also conclude that O star binary fractions obtained from magnitude-limited samples can be significantly overestimated, as  such a selection criterion picks up extra binaries near the magnitude limit. 

This depletion of binaries for fainter objects is actually responsible for the apparent difference between our overall results and the observed binary fraction obtained by \citeauthor{BTT09}. Our observed binary fraction  among objects  with $K_\mathrm{s}<15.5$ is 41.4\%. Based on the results of Table~\ref{tab: Kmag}, we estimate an intrinsic binary fraction of 61.4\%\ in that brightness range.

\begin{table}
\centering
\caption{Observed fractions of low-amplitude RV variables and binaries as a function of the luminosity class.}
\label{tab: bf_lc}
\begin{tabular}{lcc}
\hline
Luminosity class (size) & low-ampl.\ RV variables & Binaries \\
\hline 
\vspace*{-2mm}\\
V to I (293)   & $0.11$  & $0.38$ \\ 
V/IV   (186)   & $0.06$  & $0.38$ \\
III    (77)    & $0.13$  & $0.42$ \\
II/I   (30)    & $0.40$  & $0.23$ \\
\hline
\end{tabular}\\
\end{table}

\begin{figure}
\includegraphics[width=\columnwidth]{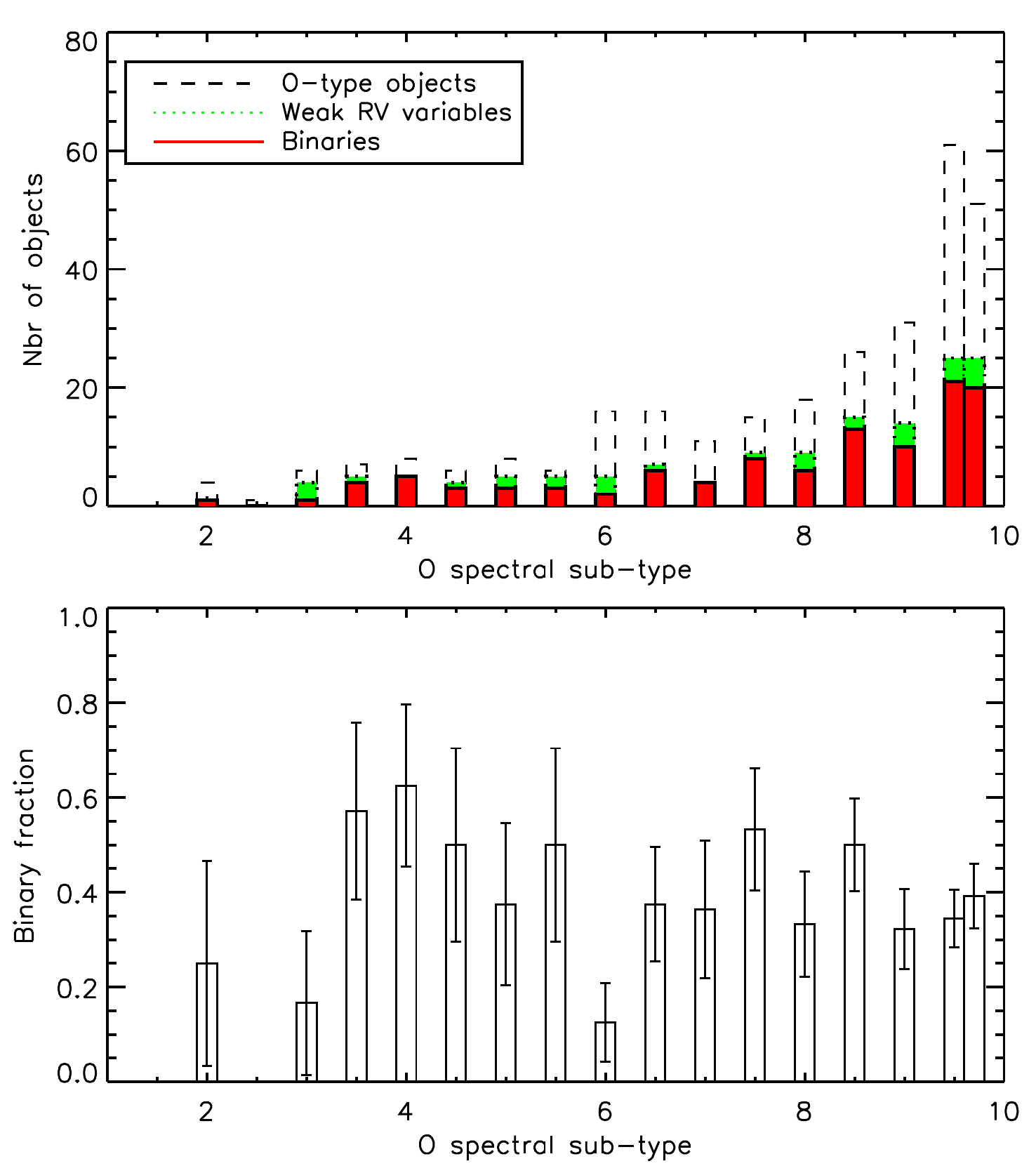}
\caption{Top panel: Number of objects (dashed line) and of binaries (solid line) per O spectral sub-type. Bottom panel: Observed binary fraction as a function of the spectral sub-types. Error bars are plotted whenever a bin contains more than one star. Fig.~\ref{fig: bf_spt_lc} provides similar figures for the different luminosity classes. } \label{fig: bf_spt}
\end{figure}

\begin{figure}
\includegraphics[width=\columnwidth]{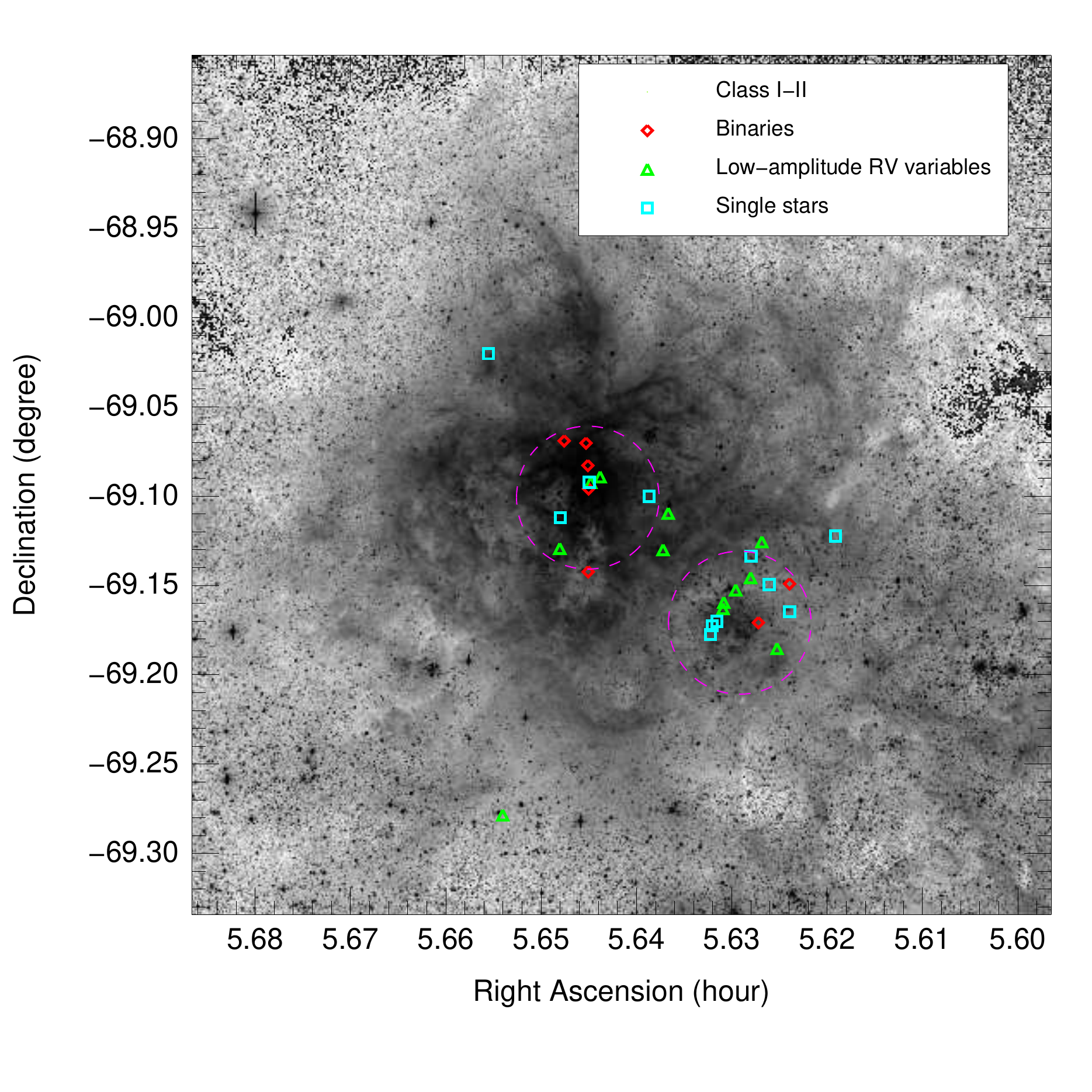}
\includegraphics[width=\columnwidth]{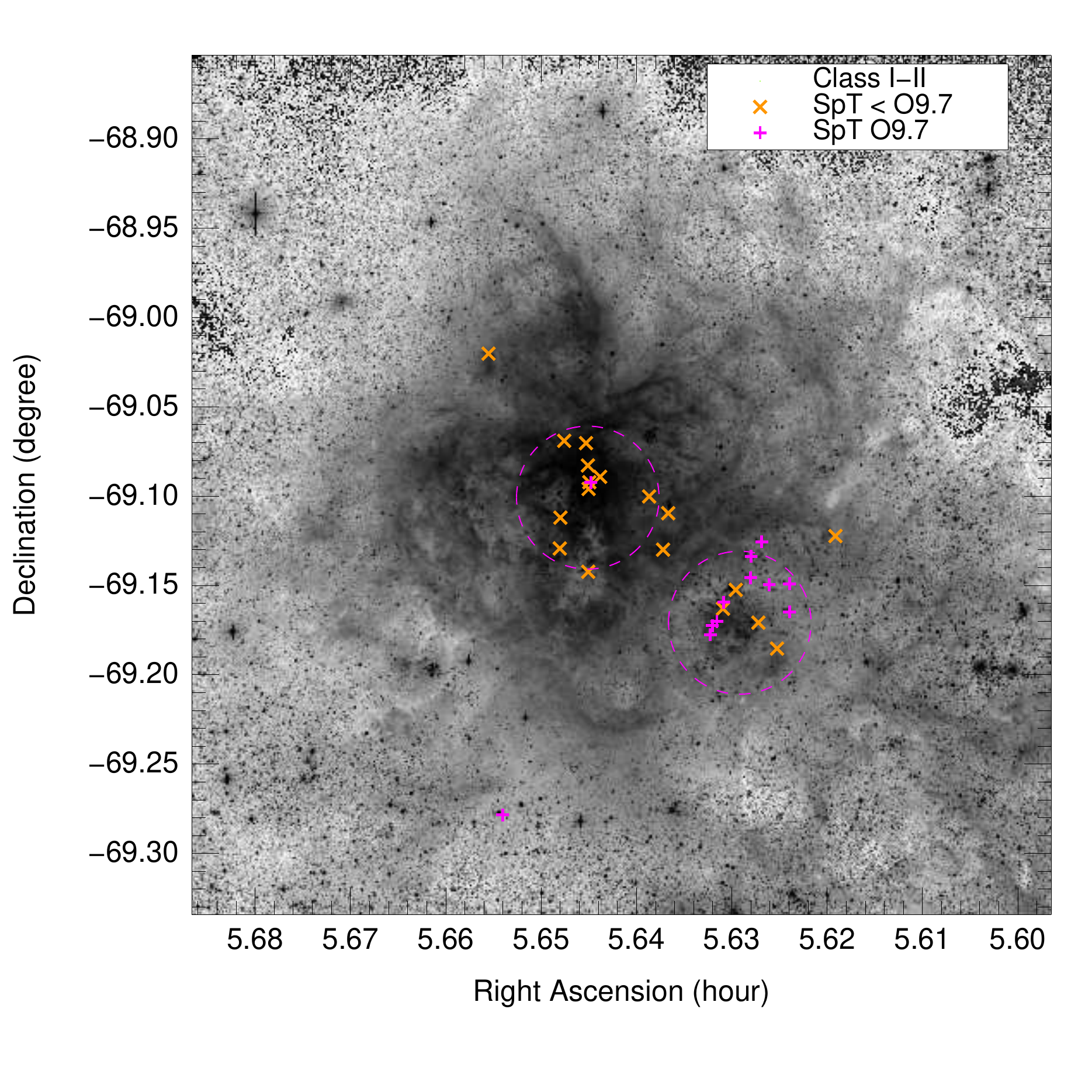}
\caption{Spatial distribution and multiplicity properties of the supergiants in the field of view.} \label{fig: bf_sg}
\end{figure}

   \subsection{Spectral sub-type and luminosity class dependence}\label{sect: bf_spt}
   
   To investigate possible variations of the detected binary fraction with the spectral sub-type and luminosity class, we ignore all objects which spectra have a too poor \snr\ or are too heavily contaminated by nebular lines to yield decent spectral classification. We also ignore the Argus sources that have a more limited spectral coverage and thus a less accurate spectral classification, and the seven Onfp stars, that present peculiar spectra as noted by the qualifier `p' appended to their spectral type. We are left with 295 out of 328 Medusa objects (i.e., 90\%\ of the Medusa sample). These objects are distributed among the luminosity classes as follows:  30 I-II objects, 77 III objects and 186 IV-V objects. Fig.~\ref{fig: bf_spt} shows the observed binary fraction as a function of  spectral sub-type and Fig.~\ref{fig: bf_spt_lc} splits Fig.~\ref{fig: bf_spt} according to luminosity class. Table~\ref{tab: bf_lc} indicates the overall low-amplitude RV variable and binary fraction in each category. Most of the fluctuations are compatible with the sample sizes so that the binary fraction seems homogeneous across the spectral sub-types in the various samples. With two binaries and one low-amplitude RV variable object in a sample of seven, the binary fraction of the Onfp stars is also compatible with that of the main sample.

The O6 stars display a significantly lower binary fraction than the average of the population. Inspection of Fig.~\ref{fig: bf_spt_lc} reveals that the drop in the binary fraction at O6 is due to the dwarfs, with only one out of 13 O6~V stars being detected as a binary, thus a binary fraction of $0.08\pm0.07$.

 The binary fraction among supergiants (23\%) is smaller than among dwarfs and giants ($\sim$40\%, Table~\ref{tab: bf_lc}). Interestingly, these numbers are roughly in line with expectations from binary evolution. The larger radii of I and II stars may imply that the shortest period binaries ($P<2-3$~d) have already interacted. Because post-interaction systems are most likely to appear as single stars in RV studies (de Mink et al., in preparation), the fraction of detected binaries is thus expected to be lower.  

With only one binary out of 12 objects (corresponding to an observed binary fraction of $0.08\pm0.08$) but a large fraction (42\%) of objects showing  low-amplitude RV variability, the O9.7~I and II stars are actually responsible for most of the differences between the binary fraction among supergiants and among dwarfs and giants. Fig.~\ref{fig: bf_sg} compares the spatial location of the O9.7 I-II objects with the other class I-II objects in the field. It reveals that the O9.7 supergiants preferentially belong to NGC~2060 while the hotter supergiants are distributed according to the ratio of the populations of NGC~2060 and NGC~2070. The data associated with the O9.7 supergiants are of a high quality, therefore exclude the possibility that the sample suffers from a higher detection threshold than the other supergiants.  It may thus be interesting to speculate that the O9.7~I-II population in NGC~2060 contains a large fraction of apparently single post-interaction stars.

\subsection{Summary}

The binary fraction displays a high degree of homogeneity across the different populations. Only three sub-categories deviate significantly from the mean:
\begin{enumerate} 
\item[i.] The overall binary fraction is lower in the outer region ($r>7.8$\arcmin) of the field of view ;
\item[ii.] The binary fraction is lower for the fainter stars in the sample ($K_\mathrm{s}>15.5$), an apparent effect that results from observational biases and the definition of the sample in terms of spectral type; 
\item[iii.] The O9.7 I-II stars, mostly located in NGC~2060, are predominantly single, though a relatively large fraction show low-amplitude RV variability.

\end{enumerate}

\section{Discussion}\label{sect: discuss}

\begin{figure}
  \centering
  \includegraphics[width=\columnwidth]{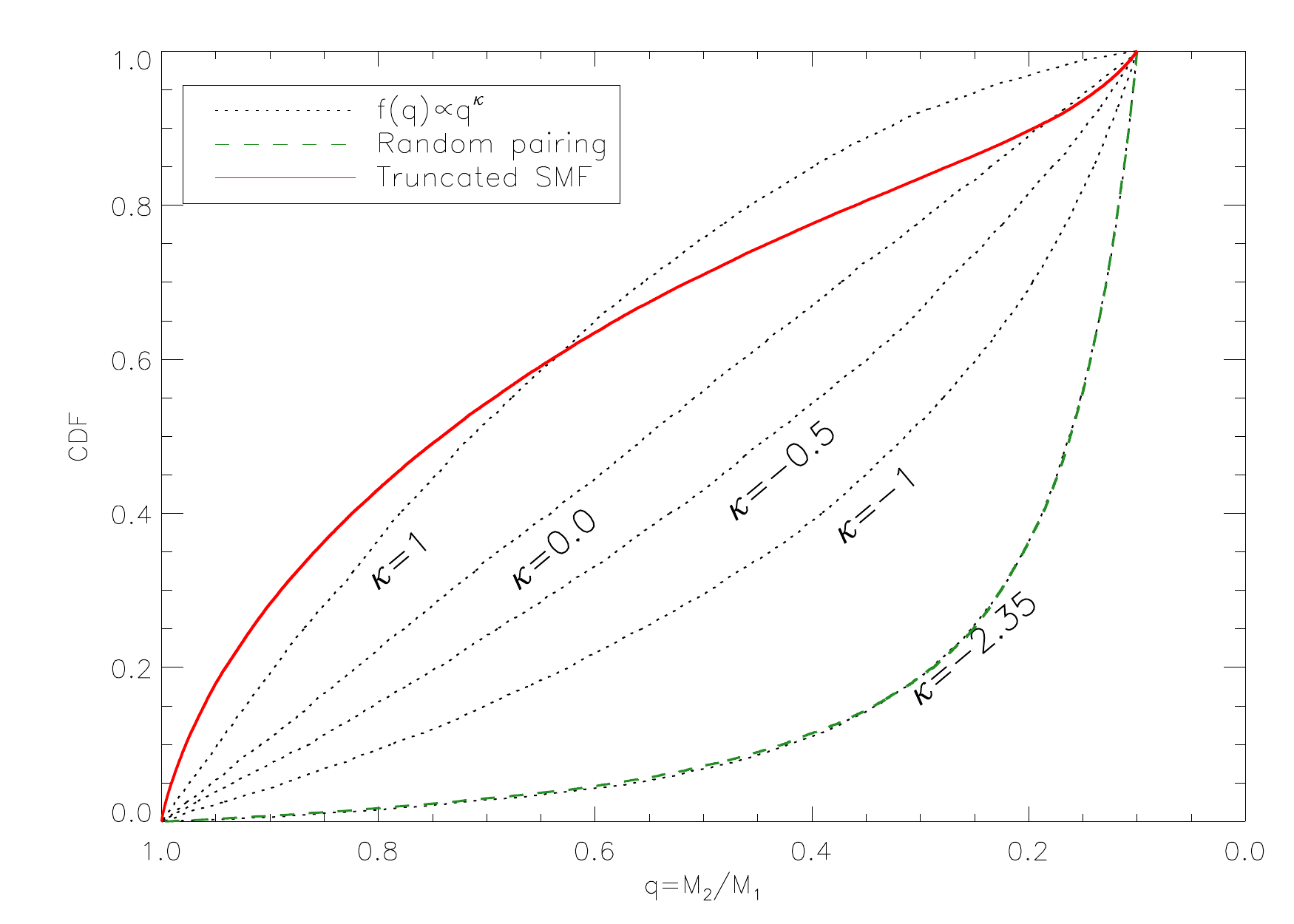}
  \caption{Mass-ratio distributions obtained using different pairing mechanisms (solid and dashed lines) compared to those obtained with a power-law density distribution with different \k\ values (dotted lines).}
  \label{fig: pairing}
\end{figure}

\subsection{Nature of the pairing process}
We investigate two different pairing processes to associate the components of the binary systems and we compare the results of the mass-ratio distributions of simulated populations with the one derived from the observations. \begin{enumerate} 
\item[i.]{\it Random pairing:} Primary and secondary masses are independently drawn, between 0.05 and 80~\msun, from a mass function  with a slope of $-$2.35. Stars are randomly  associated in pairs;
\item[ii.]{\it Truncated secondary mass function:} The primaries masses are drawn as above. The secondary masses are drawn from a mass function truncated at the primary mass. 
\end{enumerate}
To mimic an O star selection criterion, only pairs for which at least one of the components has a mass larger than 15~\msun\ are kept. In case (i), the positions of the primary and secondary are exchanged if the latter is the most massive component of the pair, so that $q \leq 1$ is respected. The $q$ distribution is then truncated at $q=0.1$  to confine to the same parameter range as discussed in this paper.

Fig.~\ref{fig: pairing} illustrates the  mass-ratio distributions obtained. None of the tested pairing processes produces a distribution function that resembles $f(q)\propto q^\kappa$ with $-1<\kappa<0$. It also shows that, as expected, random pairing results in $\kappa=-2.35$ and that the truncated secondary mass function comes close to  $\kappa=1.0$.

\subsection{On the universality of the multiplicity properties} \label{sect: MW}

Our results can be discussed in the broader context laid out by recent studies of two different O star samples in the Milky Way. First, \citet{KoF07} attempted to model the multiplicity properties in Cygnus OB2 using the results of a multi epoch RV campaign targeting 120 OB stars (900 epochs in total). Because they only adjusted the $\chi^2$ distribution, the authors obtained a degenerate solution in the exponent of the period and mass-ratio distributions and they chose to fix $\pi$ at 0.0. The separation regime considered in their modelling extends at least up to 100~A.U. and, for some of their simulations, to 10\,000 AU, i.e, quite beyond the sensitivity range of their spectroscopic observations. In the considered parameter ranges, \citeauthor{KoF07} obtained a large binary fraction, at least 80\%, and $\kappa$ values ranging from $-0.6$ to 1.0. Because of the different approaches and the different parameter ranges considered, direct comparison with the current results is hazardous. 

Second, \citet{SdMdK12} studied the multiplicity properties of the O star populations in nearby young open clusters. They performed a similar Monte-Carlo analysis as done here. Most of the binaries identified in their sample have constraints on their orbital parameters. This allowed them to test directly the period and mass-ratio distributions. They obtained an intrinsic binary fraction of $0.69\pm0.09$ and values of $\pi=-0.55\pm0.22$ and $\kappa=-0.1\pm0.6$ for the power-law exponents of the period and mass-ratio distributions.

The  period distributions obtained in both studies are in excellent agreement. The mass-ratio distribution in \citeauthor{SdMdK12} is almost flat while that of the VFTS sample is more abundant in lower mass-ratios. Both results are not inconsistent within 2\s\ and, as discussed in Sect.~\ref{app: diagn}, the present approach is only weakly sensitive to \k. The main difference resides in the intrinsic binary fraction which seems significantly lower in the Tarantula region. This aspect will be discussed in Sect.~\ref{sect: evol}.

\subsection{Evolutionary implications} \label{sect: evol}
Assuming that today's distributions are representative of the distributions at birth and using a series of assumptions on critical period and mass-ratio values that delimit various binary interaction scenarios,
the integration of the distribution functions of Sect.~\ref{sect: MC} provides an  estimate the evolutionary fate of the 30~Dor O star population. For comparison purposes, we adopt the same set of assumptions as in \citet{SdMdK12}, namely: the maximum period at which significant binary interaction takes place is 1500~d. Binaries with periods less than 6 days will initiate mass-transfer while the primary is still on the main sequence (Case A mass transfer). Among these systems,  binaries with periods less than two days will merge. Between 2 and 6 days, binaries with mass-ratios less than 0.65 will also merge. Similarly a small fraction (5\%) of case BC mass transfer (i.e., mass transfer occurring after  the main sequence) will lead to coalescence. In the remaining interaction cases, the primary will be stripped from most/all of its envelope while the secondary will gain mass and angular momentum and will be spun up to critical rotational velocities. 

Under these assumptions and given an intrinsic binary fraction of 0.51 in the range $0.15<\log_{10} P/\mathrm{d}<3.5$, we estimate that 53\%\ of all the stars born as O-type star belong to a binary system with a period less than 1500~d. The evolution of these stars is strongly affected by binary interaction: 18\%\ of the O stars will merge with a companion, 27\%\ will be stripped from their envelope and 8\%\ are expected to be spun up. About one third of the binary interaction thus results in coalescence of the two companions. 

Compared to the frations of stars in the various binary evolution channels derived in the Milky Way \citep{SdMdK12}, we obtain from the VFTS population an equivalent merger rate but lower rates of stripping and spin-up. In the above, we have implicitly assumed that  today's distributions are representative of the distributions at birth. In an environment such as 30~Dor, this assumption may not hold. The presence of different populations in the field, some already quite old, and the fact that at least 5\%\ of the O star population is running away, either because of dynamical interaction or supernova kick, suggest that a fraction of the current O star population has already undergone significant dynamical and/or evolutionary interaction.

In particular, today's binary fraction might be dragged towards lower values by the presence of runaways and post-interaction objects. As mentioned already the latter are indeed expected to be predominantly single (de Mink et al., in preparation). Runaways and post-interaction objects would need to represent about 20\%\ of the current 30~Dor population to reconcile the  binary fraction in 30~Dor (51\%) with that of the Galactic open cluster population \citep[69\%][]{SdMdK12}. Establishing the star formation and dynamical history of the Tarantula region would constitute a major step in deciding whether such a high fraction is plausible.

It remains that binary interaction effects have a critical influence on the evolutionary path of more than half the stars born as O-type stars. These effects need to be considered to better interpret massive star populations seen in integrated light and to accurately explore the frequency of progenitors of compact objects, high-mass X-ray binaries, hydrogen-poor supernovae and long-duration gamma-ray bursts.

\section{Conclusions} \label{sect: ccl}
In this paper, we investigate the multiplicity properties of a sample of 360 O-type stars observed by multi-epoch spectroscopy in the framework of the VLT-Flames Tarantula Survey. Forty-six per cent of our sample present significant RV variations and, for 35\%, the detected variations have an amplitude larger than 20~\kms\ and are very likely spectroscopic binaries. 

The observed binary fraction present a high degree of homogeneity across the field of view and among the various spectral-types and luminosity classes. Three sub-groups however show a significantly lower binary fraction:
\begin{enumerate} 
\item[i.] the binary fraction is lower in the outer region ($r>7.8$\arcmin; $\approx$117~pc) of the field of view. This outer population may be dominated by single stars ejected from the core of the region;
\item[ii.] the binary fraction is lower for the fainter stars in the sample ($K_\mathrm{s}>15.5$), which results from the sample definition and traces the brightness separation between the O and B stars ; 
\item[iii.] the O9.7 I-II stars are predominantly single and belong to NGC\,2060, the older O-star cluster in the field.  Though speculative at the moment, a large fraction of the O9.7 I-II stars may be post-interaction stars.
\end{enumerate}

We use a Monte-Carlo method to correct for the observational biases and to constrain the intrinsic multiplicity properties of the O-type star population. We obtain an intrinsic binary fraction of $f_\mathrm{bin}=0.51\pm0.04$. The most likely period distribution, $f(\log_{10} P/\mathrm{d})\sim (\log_{10} P/\mathrm{d})^{-0.45}$ in the interval $\log_{10} P/\mathrm{d} \in [0.15,3.5]$, favours shorter period systems compared to a flat distribution in $\log_{10} P/\mathrm{d}$. Similarly the mass-ratio distribution, $f(q) \sim q^{-1.0}$ in the interval $q \in [0.1,1.0]$, favours lower-mass companions but we note that our method  only provides weak constraints on the $q$ distribution.

The period distribution obtained here is strikingly similar to the one obtained for Galactic O-type binaries, possibly hinting a universal property among massive binaries. The intrinsic fraction of binaries with periods less that $10^{3.5}$~d ($0.51 \pm 0.04$) seems lower than obtained for the Galactic O-type binaries ($0.69 \pm 0.09$) albeit the two results agree within 2\s. Alternatively it could reflect the fact that the binary properties in the Tarantula region have already been significantly affected by dynamical and/or stellar evolution that would predominantly lead to either a merger event or disrupt the binary. Quantitatively understanding if and how binary evolution and dynamical interaction have affected the multiplicity properties in the Tarantula region would help to decide whether the observed differences are nature or nurture. A key ingredient in this process will be a better understanding of the star formation history of the region.

Our results emphasize again that multiplicity is one of the main characteristics of massive stars and that it needs to be taken into account to properly predict the evolution of entire populations of massive stars. Dynamical interactions and evolutionary effects that could have
  possibly affected our sample would be expected to, predominantly,
  decrease the observed number of binaries; our derived binary fraction
  can therefore be seen as a lower limit to the binary fraction at birth.  This suggests that the most frequent final product of high-mass star formation is an O + OB binary and that massive star formation theories not only have to explain the formation of high mass stars but also have to reproduce their multiplicity properties.

\begin{acknowledgements} SdM acknowledges NASA Hubble Fellowship grant HST-HF-51270.01-A awarded by STScI, operated by AURA for NASA, contract NAS 5-26555.S.E. STScI is operated by AURA, Inc., under NASA contract NAS5-26555. JMA acknowledges support from  the Spanish Government Ministerio de Educaci\'on y Ciencia through grants AYA2010-15081 and AYA2010-17631 and  the Consejer\'{\i}a de Educaci\'on of the Junta de Andaluc\'{\i}a through grant P08-TIC-4075. AH ackowledges support by the Spanish MINECO under grants AYA2010-21697-C05-04 and Consolider-Ingenio 2010 CSD2006-00070, and by the Canary Islands Government under grant PID2010119. MG acknowledges financial support from the Royal Society and VHB acknowledges support from the Scottish Universities Physics Alliance (SUPA) and from the Natural Science and Engineering Research Council of Canada (NSERC). Finally, HS acknowledges support from the SARA Computing and Networking Services. 
\end{acknowledgements}

\bibliographystyle{aa} 
\bibliography{/home/hsana/Desktop/Dropbox/Dropbox/literature} 

\Online
\begin{appendix}

\section{Normalisation}\label{app: norm}
In this section, we describe the semi-automatic procedure used to normalise the spectra discussed in this paper. The adopted algorithm uses all continuum points after  automatically rejecting stellar and interstellar lines (both in emission and absorption) and cosmetic defects (bad columns, cosmic rays).

The global trend in the spectrum is first removed using a one-degree polynomial fitted to the continuum, resulting in a spectrum scaled around unity.  A polynomial of degree 4 to 8 is then fitted to the scaled continuum.  The error bars on the flux in each pixel are in principle provided by the ESO pipeline. By a comparison of the error spectra with the empirical \snr\ in continuum regions, we find that the pipeline actually overestimates the error spectra by a factor 1.2, which we correct for before using the pipeline-provided errors to weight the fit. The normalisation procedure proceeds as follows: \begin{enumerate}
\item[i.] the spectrum is smoothed using a 21-pixel wide median filter;
\item[ii.] the spectral lines are removed by comparing the results of median and max/min filters applied on the initial spectrum, accounting for the spectrum \snr;
\item[iii.] a first guess-solution of the continuum position is obtained using the smoothed spectrum (ii); 
\item[iv.] a second-guess solution is obtained using the smoothed spectrum (ii) after \s-clipping around (iii); 
\item[v.] the final solution is obtained using the original spectrum (i.e., no smoothing) after \s-clipping around (iv). Manual exclusion of specific regions is allow at this stage to incorporate a priori knowledge on unsuitable continuum regions (e.g., broad diffuse interstellar bands, broad emission regions);
\item[vi.] the spectrum (v) is scaled by the median value computed over its continuum regions. This allows to correct for slight global under/over estimates of the average continuum level which occurs when there are many absorption/emission lines as the line wings may induce a small but systematic bias. This empirical correction is of the order of a fraction of a per cent;
\item[vii.] the normalisation function is applied to the error spectrum. 
\end{enumerate}

Results are inspected. The degree of the polynomial and/or  the exclusion of  unsuitable continuum regions are modified until a satisfying fit is reached. We estimate that the continuum is, in general, constrained to better than 1\% over the whole wavelength range, save for the edges of the spectra.\\

\section{Radial velocity measurements}\label{app: rv}
   \subsection{Methodology}\label{app: rv_method}

The choice of the RV measurement method is guided by the need to provide consistent, unbiased and homogeneous absolute RVs with representative error bars over the full range of O spectral sub-types and given the specificity of the data set. In our case, only a few lines are suitable for RV measurements. The number of epochs, the number of lines and their quality change from one star to another as a function of the star's spectral type and brightness and the degree of nebular contamination. The \snr\ also varies significantly from one star to another, from one epoch to another and over different spectral regions.

Three RV techniques were considered: line moments, cross-correlation and Gaussian fitting. Tests were performed both on synthetic data and on a small set of objects taken from our survey. Synthetic spectra for a range of temperatures, gravities, projected rotational velocities (\vsini) and \snr\ representative of the VFTS O star data  were generated using \fw\ \citep{PUV05}  and degraded to the resolution of our survey. We briefly discuss the pros and cons of the three approaches below.

 Line moments turn out to be very sensitive to residuals of the nebular correction  and the accuracy of the method is not explored further. Auto cross-correlation (Dunstall et al., in preparation) provides accurate RVs and is robust against residuals of the nebular correction. However, for stars with less than four lines suitable for RV measurement, cross-correlation has intrinsic difficulties in providing accurate error bars. Gaussian fitting provides slightly less accurate RVs than cross-correlation, but with more reliable error bars for stars with only  few lines of sufficient quality. However, the performance of Gaussian fitting degrades strongly with the quality of the lines (i.e., low \snr\ or high \vsini). 

To improve the performance of the Gaussian fitting, we
    developed a tool to allow us to fit all the available lines at all
    epochs (and both wavelength set-ups) simultaneously.
 We assume that the line profiles are constant and well described by Gaussians. Each considered line is required to have the same full-width at half-maximum (FWHMs) and amplitude ($A$) at all epochs. All the lines of a given epoch are further required to display the same RV shift. If $L$ is the number of considered lines and $N$ the number of epochs, the number of parameters in the fit is thus : $N+2 \times L$. Typical values for $N$ are 18 and 6, depending on whether consecutive exposures are taken individually or summed up. Values for $L$ vary between 2 and 5. This yields a maximum number of parameters of the order of 28 in the present approach, to be compared with the $3 \times N \times L=270$ parameters of the standard approach that adjusts each line separately. In essence the proposed method allows the line profile parameters (FWHMs and amplitudes) to be bootstrapped by the best quality spectra, improving the RV measurements of the lowest \snr\ data. 

For simplicity and robustness in the SB2 cases (see below), we have chosen to fit the full line profile using Gaussians. Intrinsic He profiles are however not always well represented by Gaussian shapes. For slow rotators ($v_\mathrm{rot} \sin i \le 80$~\kms), the shape of the line is dominated by the instrumental profiles that is well approximated by a Gaussian. The profile of moderately rotating stars ($v_\mathrm{rot} \sin i \le 300$~\kms) is also well matched by a Gaussian, but with a small difference in the line wings. For fast rotators, rotationally broadened line profiles deviate significantly from the Gaussian shape. However, because we only attempt to measure the position of the centre of the lines and because the fitted lines are symmetric to within first-order, no significant bias is  expected from this approximation.

The fitting method is based on least-square estimates where the merit function accounts for the error spectrum to provide individual error bars for each pixel. The optimisation relies on a modified Levensberq-Marquardt algorithm that requires a first estimate of the parameters. For constant RV stars and single-lined (SB1) binaries, the method is found to be robust against the initial guesses and we have taken $v=270$~\kms, FWHM$=3.0$~\AA\ and $A=0.1$ to initialize the fit. The allowed ranges for the various parameters are as follow: $v \in [-300,+700]$~\kms, $A \in [0.0,0.6]$ and FWHM $\in [0,10]$~\AA. Only absorption lines were considered.  For the  doubled-lined spectroscopic binaries (SB2) in our sample, we follow the same approach with two Gaussian components per line profile. For some of the objects the initial guesses needed to be iterated once or twice to improve the quality of the SB2 fit. 

\begin{figure*}
  \centering
  \includegraphics[width=\columnwidth]{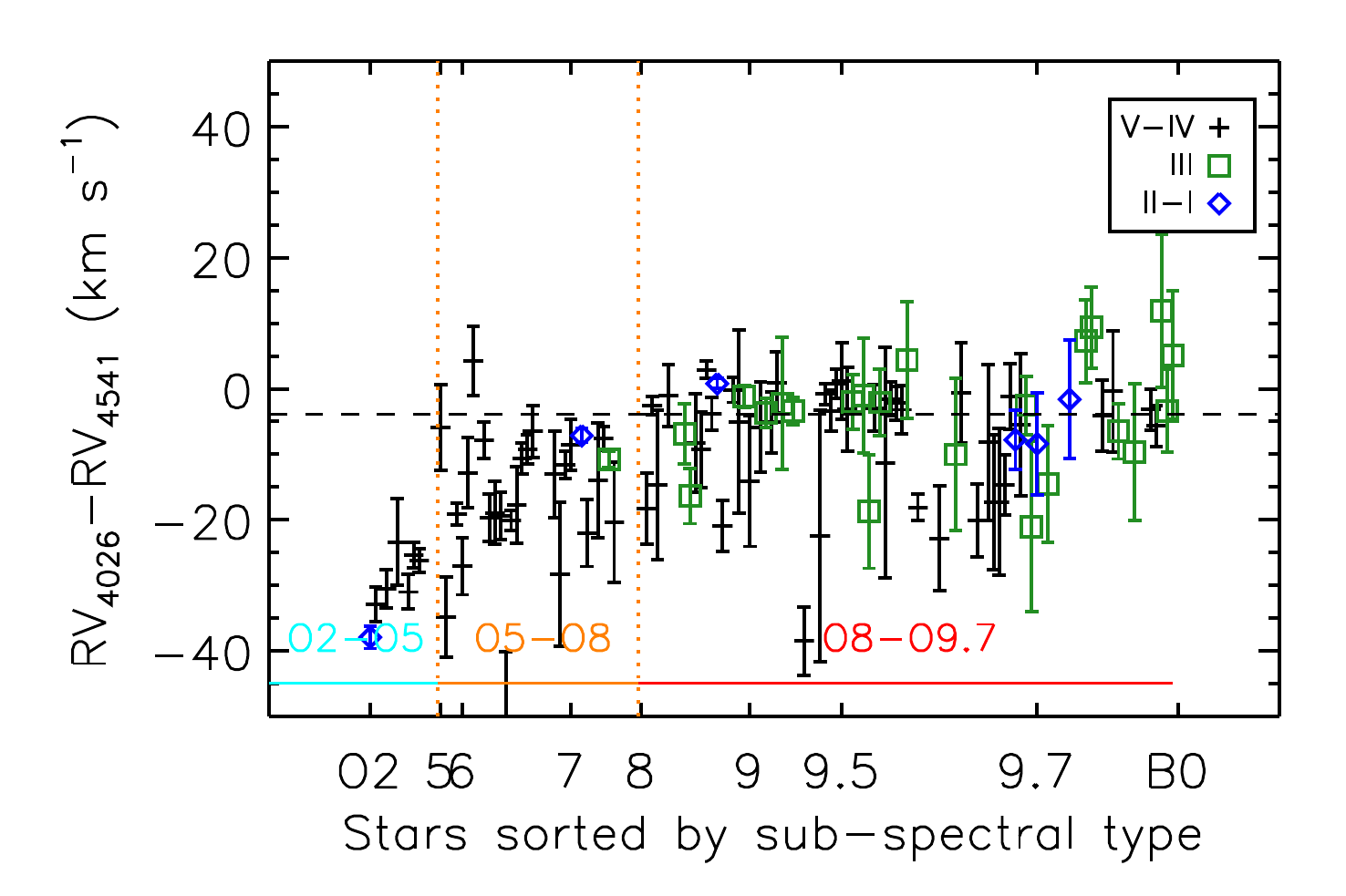}
  \includegraphics[width=\columnwidth]{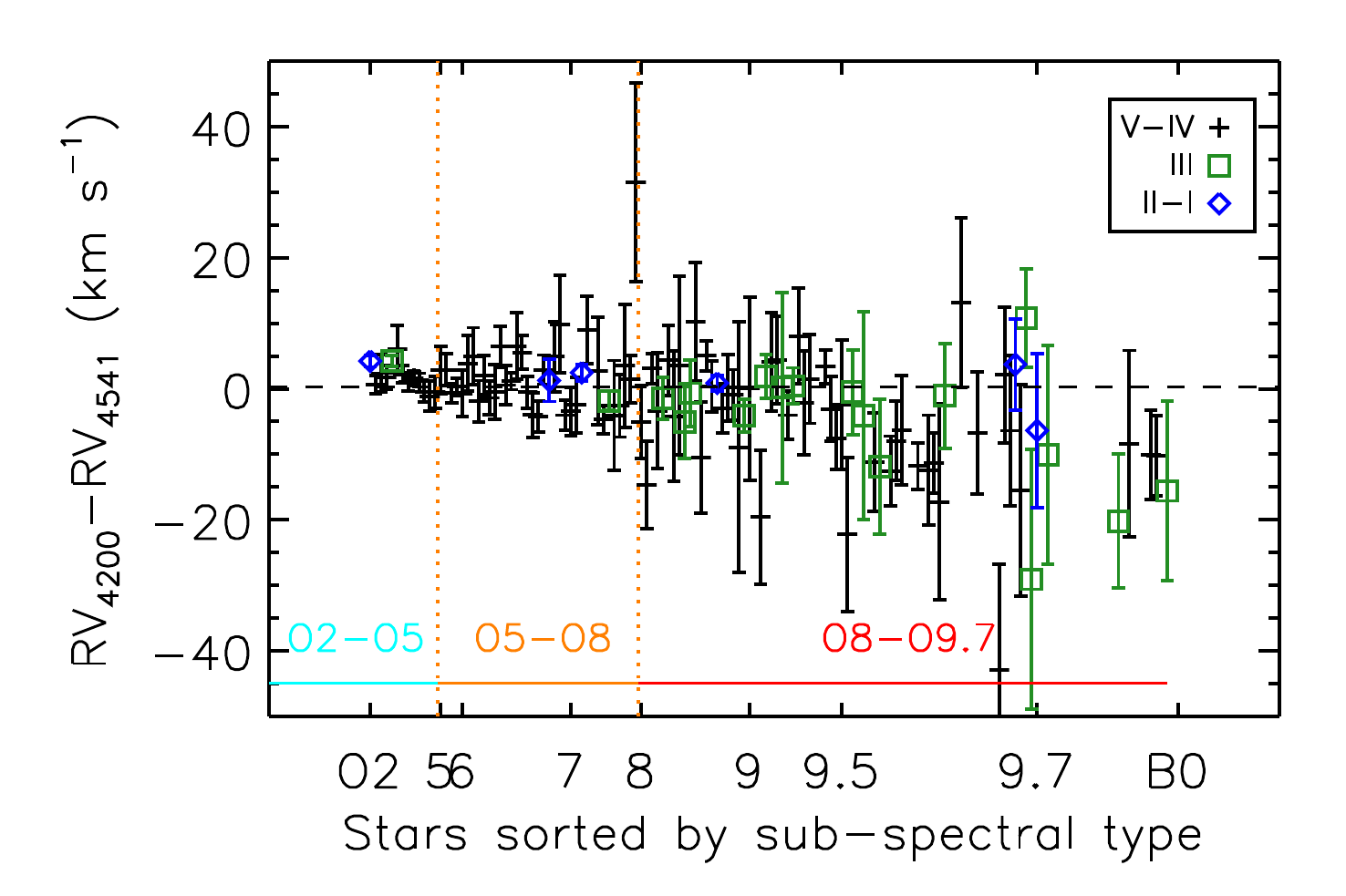}
  \includegraphics[width=\columnwidth]{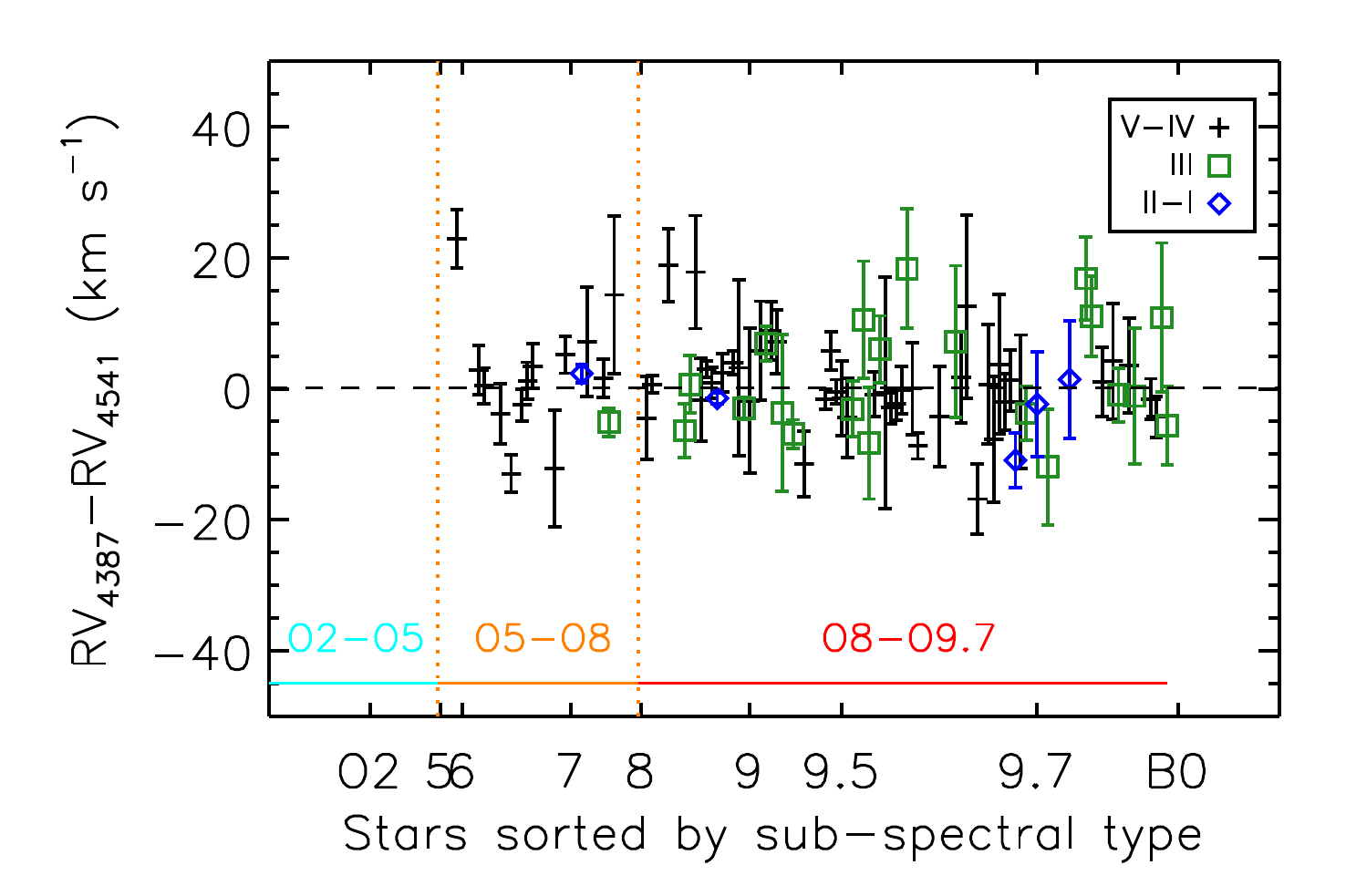}
  \includegraphics[width=\columnwidth]{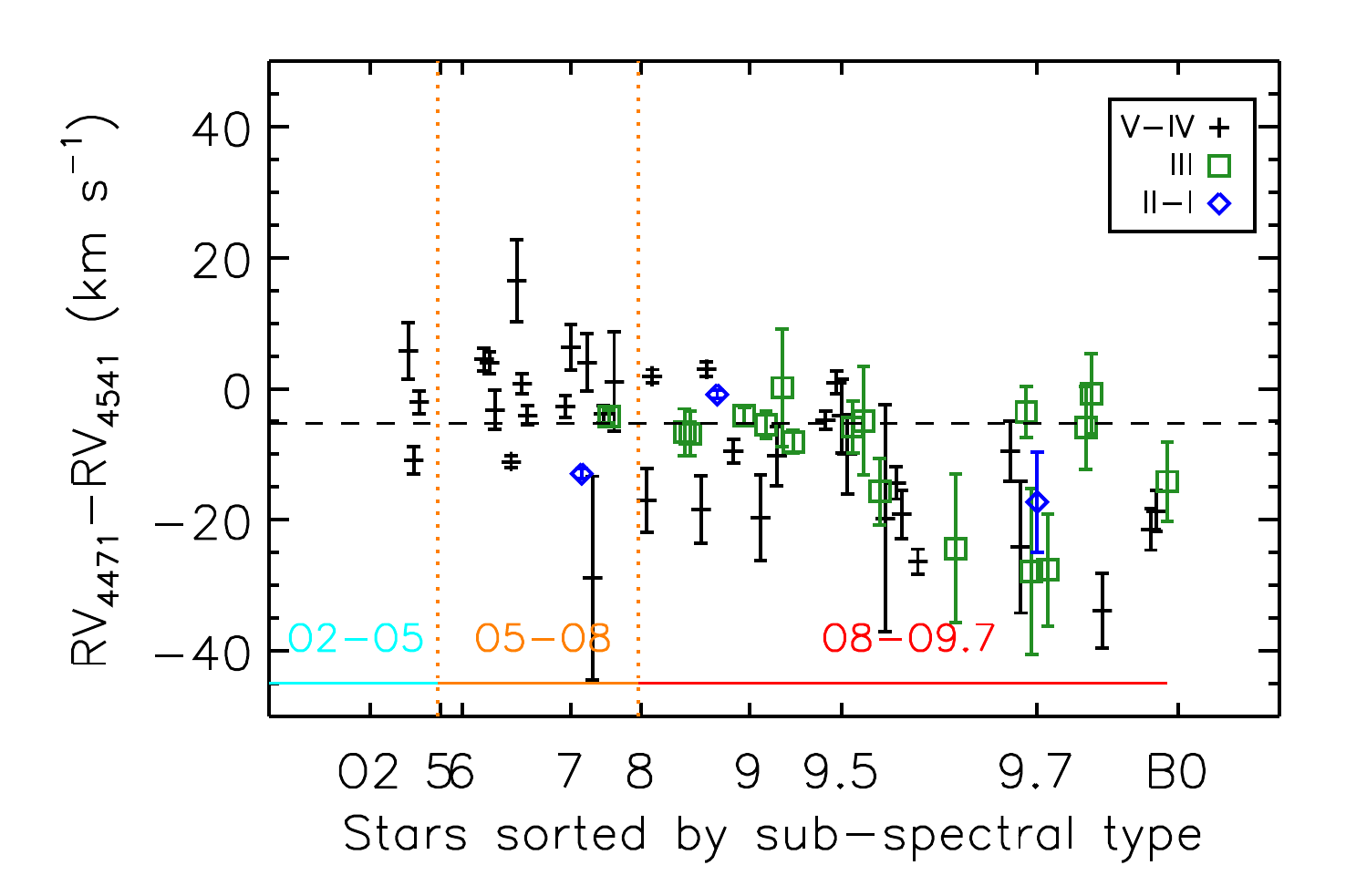}
  \includegraphics[width=\columnwidth]{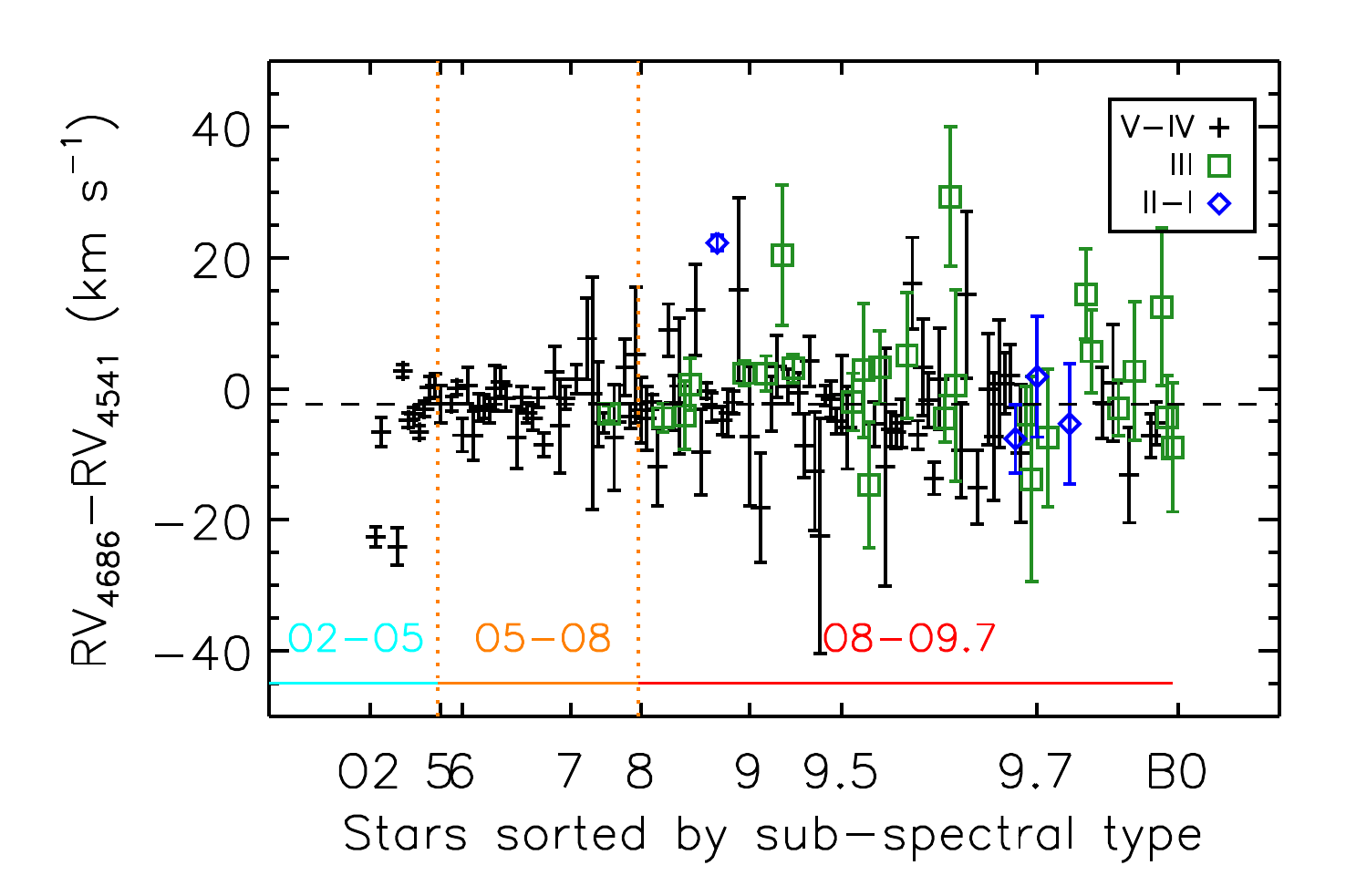}
  \includegraphics[width=\columnwidth]{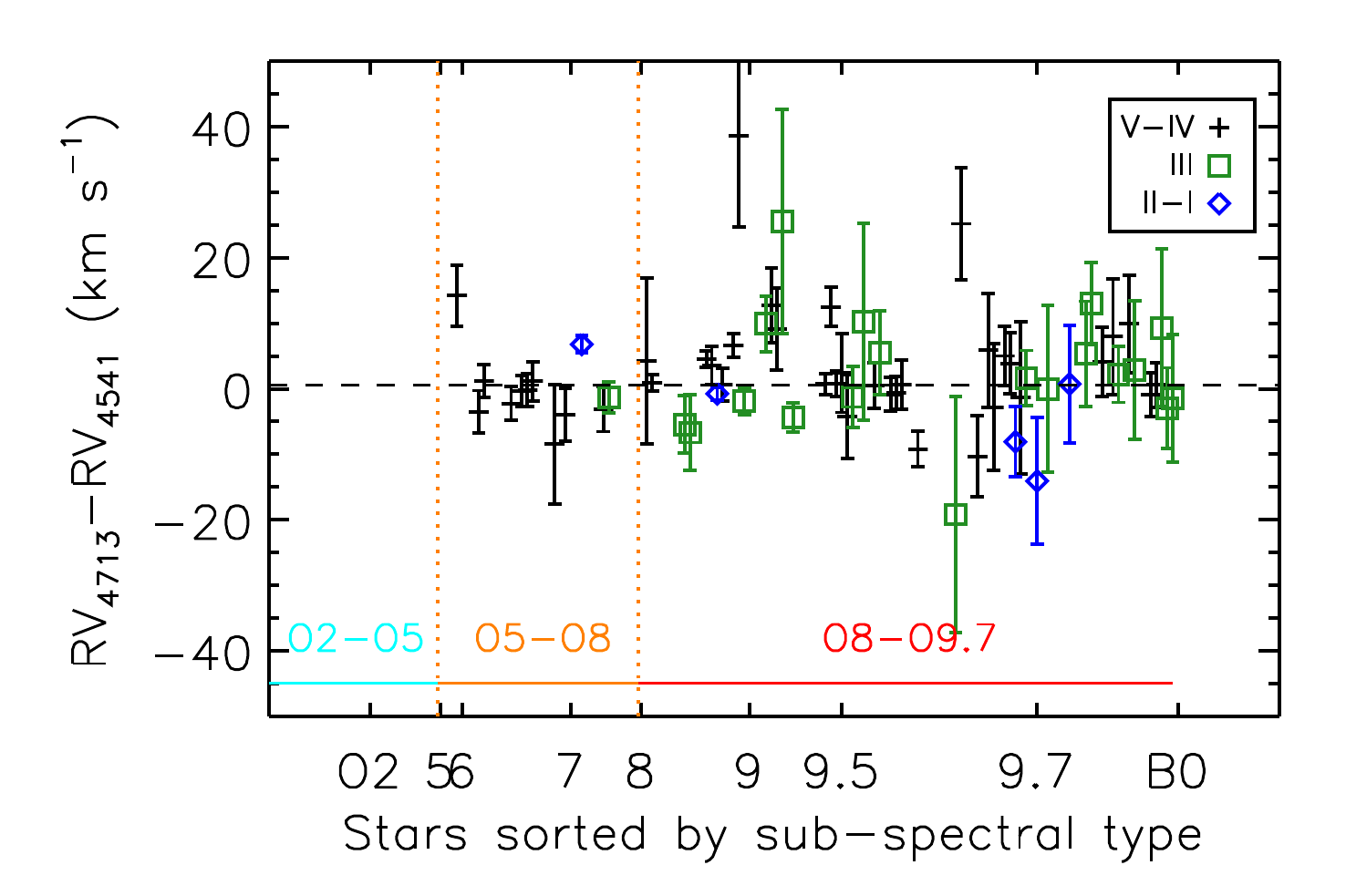}
  \includegraphics[width=\columnwidth]{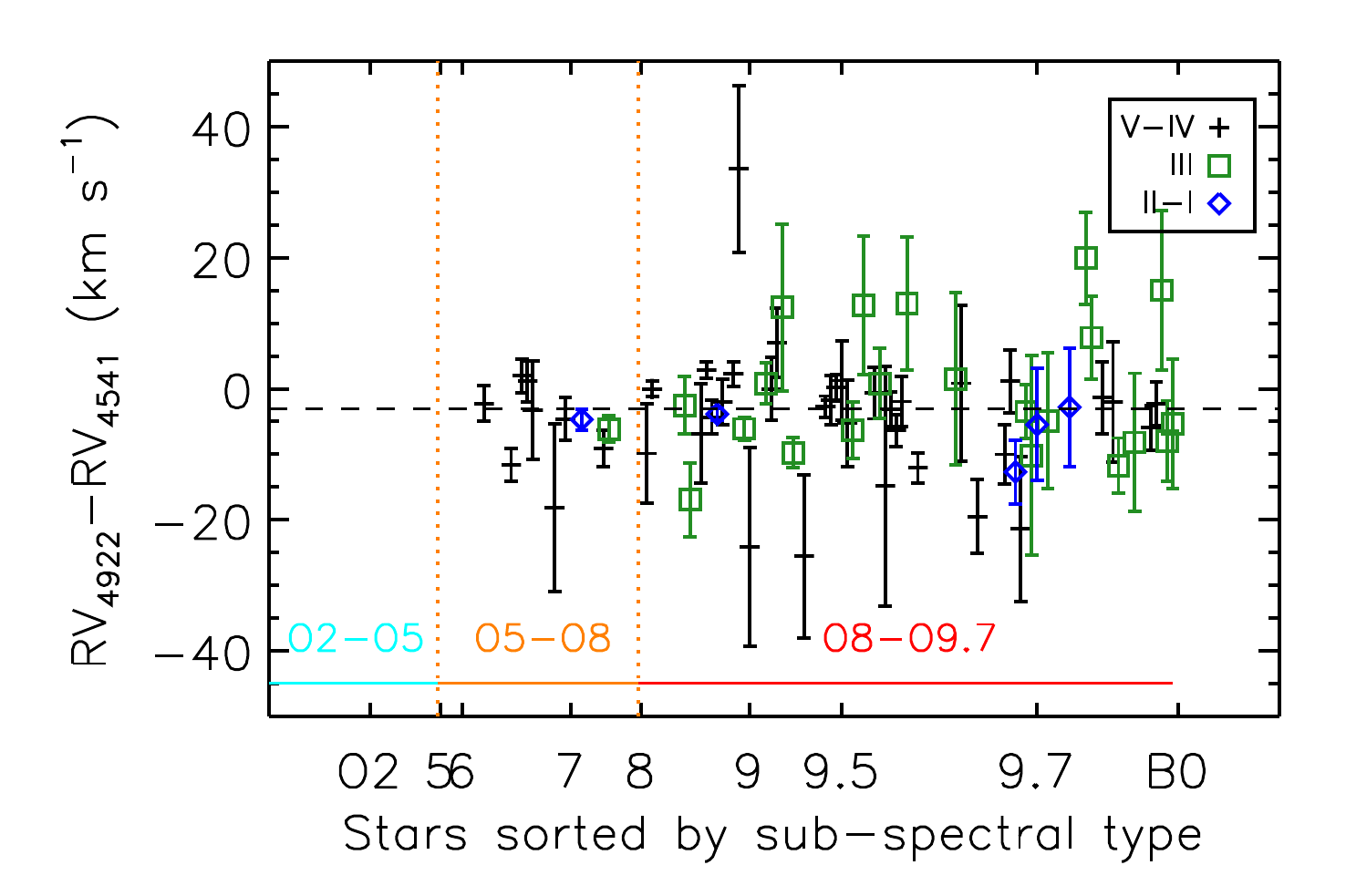}
  \includegraphics[width=\columnwidth]{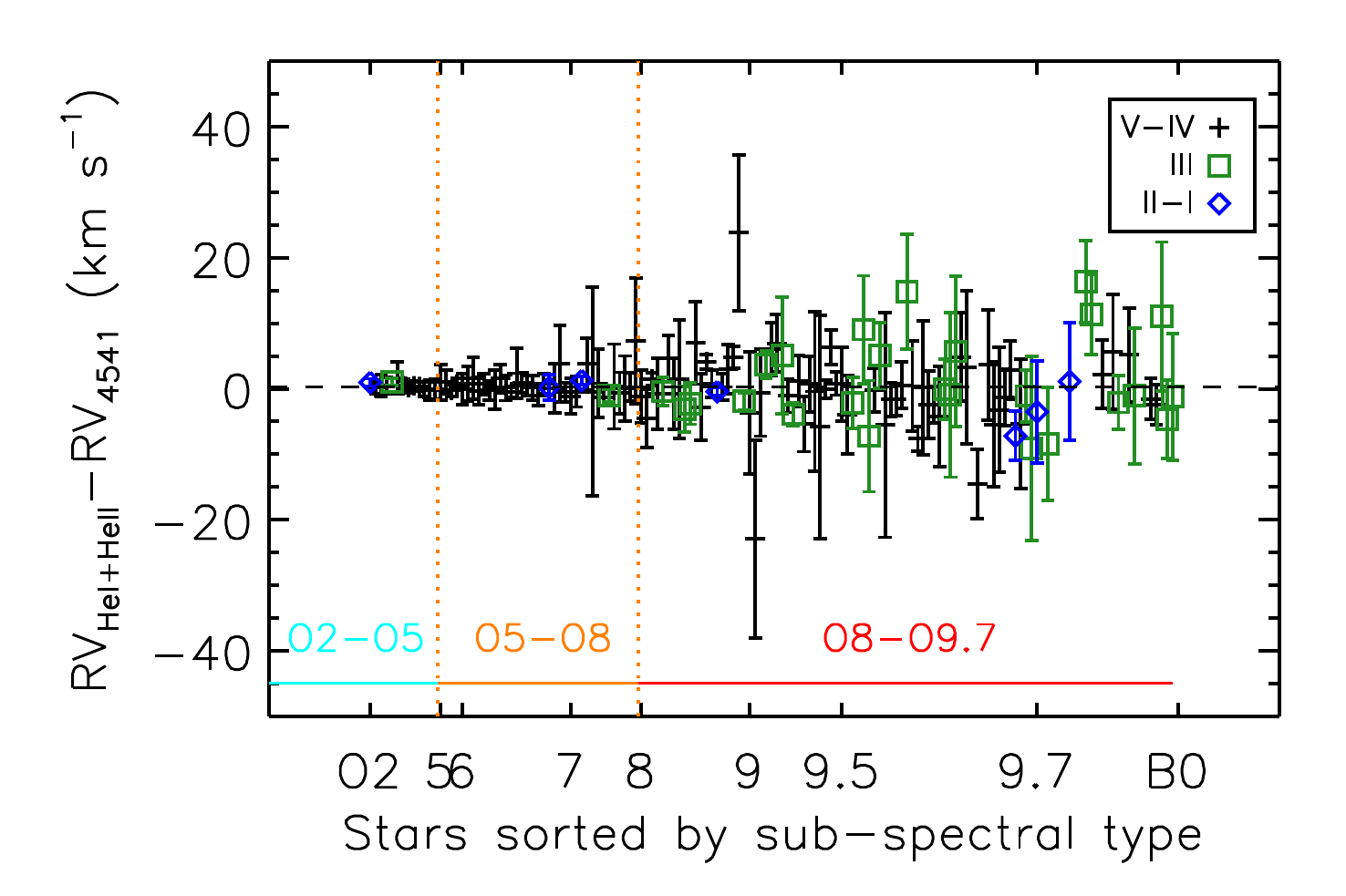}
  \caption{Difference between the RVs measured from individual \he\ lines and those measured from \heb\l4541 as a function of the spectral sub-type. The panel at the bottom right corner compares RVs obtained using all available He lines (except \hea+{\sc ii}\l4026 and \heb\l4686) with the \heb\l4541 RVs. The dashed line indicates the average of all comparisons weighted by their uncertainties. Only stars with constant RVs and data points with $\sigma_{\Delta RV}<20$~\kms\ have been included.}
  \label{fig: RVcomparA}
\end{figure*}

\begin{figure*}
  \centering
  \includegraphics[width=\columnwidth]{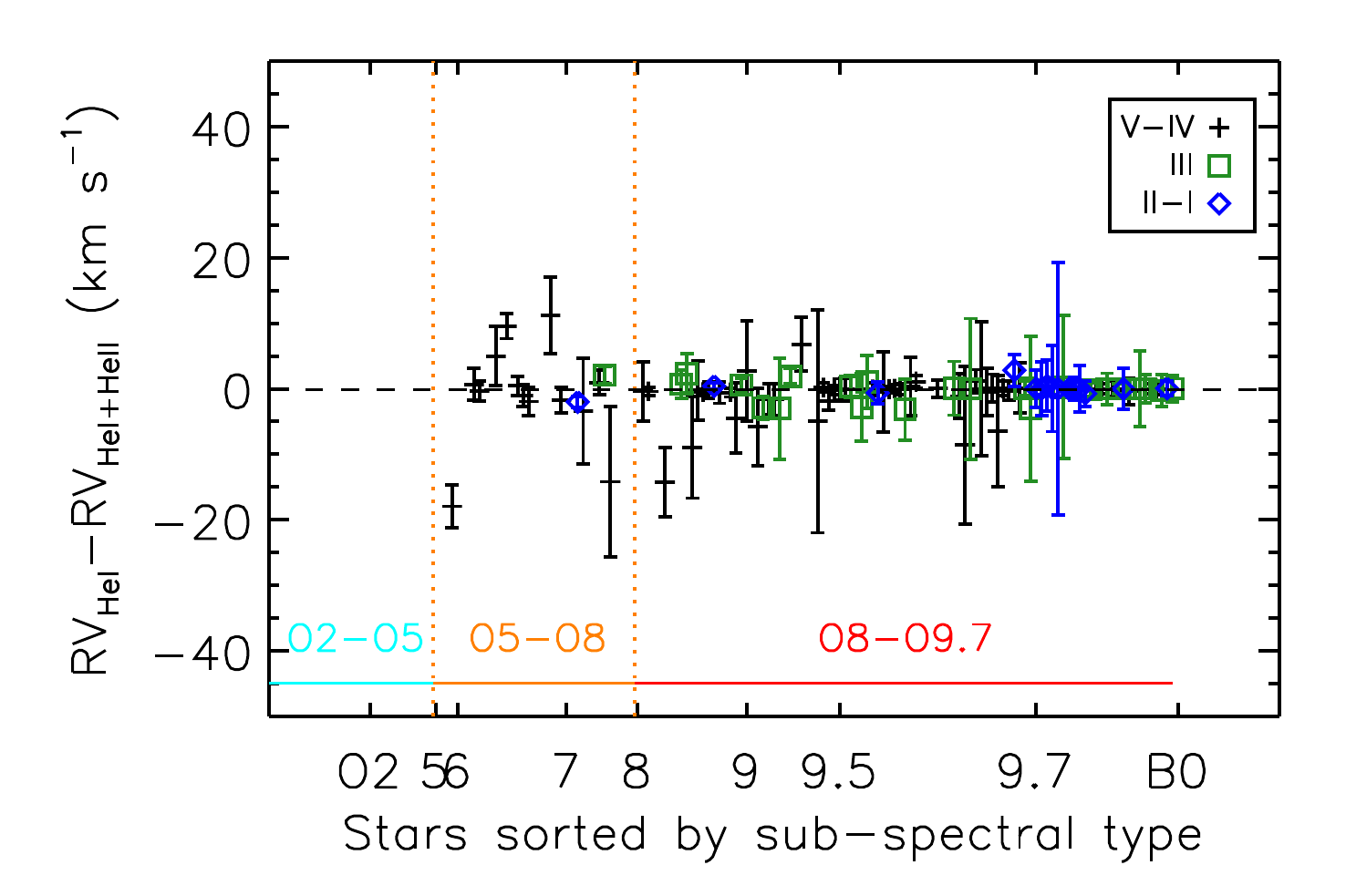}
  \includegraphics[width=\columnwidth]{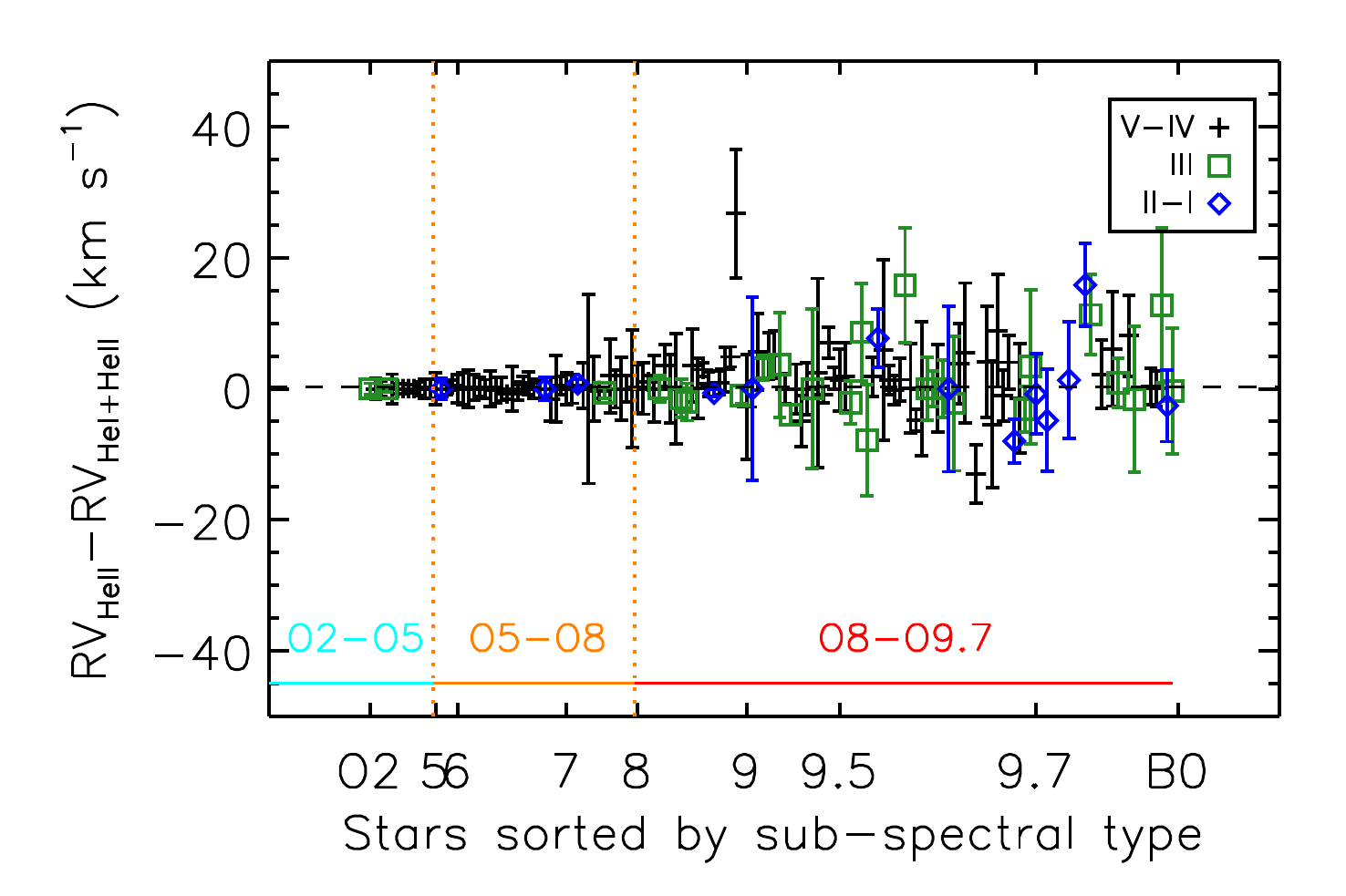}
  \caption{Comparison of the RVs measured using simultaneously all suitable \hea\ and \heb\ lines (left and right panel, respectively) with the  RVs measured using all suitable He lines.  The \hea+{\sc ii}\l4026 and \heb\l4686 lines have not been used in the measurements. }
  \label{fig: RVcomparB}
\end{figure*}

   \subsection{Line selection}\label{app: rv_line}

In this section, we discuss the choice of spectral lines suitable for relative and absolute RV measurements. 
The spectra of O-type stars are dominated by \ha, \hea\ and \heb\ lines. Hydrogen lines are generally poorer RV indicators than (most of) the He lines because they are broader and more sensitive to wind effects. The hydrogen lines of most of the VFTS O stars are also strongly affected by nebular emission. Metal lines (C, N, O, Si, Mg) are also present in O star spectra but are typically weaker than He lines. The presence of a given metal line is often limited to a small range of spectral sub-types and does not allow for an homogeneous approach across the O star domain. Our analysis is thus focused on the strong \hea\ and \heb\ lines from 4000 to 5000~\AA, i.e.\, \hea\ll4387, 4471, 4713, 4922, \heb\ll4200, 4541, 4686 and the \hea+{\sc ii}\ blend at $\lambda$4026. In the rest of this section, we discuss the respective merits of each of these lines. Two aspects are important to consider: \begin{enumerate}
\item[-] the internal consistency, i.e.\ RV measurements from various lines of a given star should, coherently, reflect the systemic velocity of the star;
\item[-] the homogeneity with respect to the full sample, i.e.\ the selected set of lines should provide RV measurements that can be compared to RVs from other stars even if only a sub-sample of the lines are available. This is important as the spectrum of early O stars is dominated by \heb\ lines and those of late O stars by \hea\ lines.
\end{enumerate}

We first use simulated data to investigate the effect of temperature and wind strength on the measured RVs. Towards earlier spectral types, the \hea+{\sc ii}\ blend  at $\lambda$4026 progressively shifts from pure \hea\ at $\lambda$4026.191 in late O stars to pure \heb\ at \l4025.602 in early O stars. The RV measured from this blend can thus vary by over 30~\kms\ unless the effective rest wavelength of the blend is adapted as a function of the effective temperature of the star. Temperature also has an effect on the accuracy of the RV measurements. \hea\ lines provide  more accurate RVs in mid to late O stars while \heb\ lines are favoured for mid and early sub-types. The most accurate RVs are obtained for mid O-type stars because their spectra feature a larger number of good quality lines.

Of all helium lines, stellar winds mostly affect \heb\l4686, though not all spectral sub-types or luminosity class are equaaly affected by the filling in of the photospheric absorption line by wind emission. We find that RVs obtained from \heb\l4686 are reliable for late and mid main-sequence sub-types. They are biased towards negative values by up the 25~\kms\ in early main-sequence stars. RVs from giants and supergiants are blue-shifted even for late-O stars and the line cannot be used as a probe for absolute RVs.

We also compare RV measurements obtained individually from different lines in a representative sub-sample of the O stars, covering the complete O spectral sub-range. Only stars displaying constant RVs are considered for this test. Because the line  is observed both in the LR02 and LR03 setups and remains of a reasonable strength even in the latest O sub-types, we use the \heb\l4541 line as a reference for the comparison.  Fig.~\ref{fig: RVcomparA} shows the difference between RVs obtained for a given line and those obtained from \heb\l4541 plotted against the spectral sub-type. Temperature effects on \hea+{\sc ii}\l4026 are clearly illustrated. Wind effects turn out to have a limited impact on  \heb\l4686 measurements  mostly because our sample is dominated by main-sequence stars and because many single supergiants present low-amplitude RV variations and are thus excluded from the comparison.

 RVs from the \hea\l4471 line show large discrepancies, which we attribute to the combination of the following factors: \hea\l4471 is a triplet transition, with a forbidden component. It is quasi-metastable, similar to but less extreme than \hea\ll3889, 5876, 10830 and is highly susceptible to wind effects.  For late spectral sub-types it is blended with \ob\ in its blue wing and, among the measured \hea\ lines, it suffers the most from the nebular contamination.

For late spectral types, \heb\l4200 is blended with \nc\l4200.07 (i.e., about 17 km/s redwards from \hea\l4200), which can  explain some of the deviations seen in the
corresponding panel in Fig.~\ref{fig: RVcomparA}. For late O stars, the uncertainties associated with \heb\l4200 RVs are large anyway and the line has a very limited weight in the final RV measurements obtained from simultaneously fitting all available lines (Fig.~\ref{fig: RVcomparB}).

The best consistency is observed between the lines \heb\ll4200, 4541, \hea\ll4387 and 4713.  \hea\l4922 displays good consistency over the range of O spectral sub-types but provides RVs that are on average shifted by 3~\kms\ compared to the other lines. We mitigate this by modifying the effective rest wavelength of the transition.

 As a conclusion, up to five lines can be used for absolute RV measurements (\heb\l4200, 4541, \hea\ll4387, 4713, 4922) depending on the spectral types, \snr\ and nebular contamination. \hea+{\sc ii}\l4026, and, for relatively weak wind stars,  \heb\l4686  can be used for relative measurements.

\begin{table}
  \centering
  \caption{Rest wavelengths ($\lambda_0$) used to computed the RVs.}
  \label{tab: rw}
\begin{tabular}{lcc}
\hline \hline
Line & $\lambda_0$ (\AA) & Reference   \\
\hline
\vspace*{1mm}
\hea+{\sc ii}\l4026 &  4026.072$^a$ &   CLL77   \\
\heb\l4200      &  4199.832 &   PvH/NIST \\ 
\hea\l4387      &  4387.929 &   PvH \\
\hea\l4471      &  4471.480 &   NIST \\
\heb\l4541      &  4541.591 &   PvH/NIST \\
\heb\l4686      &  4685.710 &   PvH \\
\hea\l4713      &  4713.146 &   NIST \\
\hea\l4922      &  4921.931 &   PvH/NIST \\
\hline\end{tabular}\\ \flushleft
{\sc References:} CLL77: \citet{CLL77}; NIST atomic spectra data base: {\tt http://www.nist.gov/pml/data/asd.cfm}; PvH : Peter van Hoof atomic line list ({\tt http://www.pa.uky.edu/$\sim$peter/atomic/}).\\
{\sc Notes:} $a$. Because of the strong blend between \hea\l4026.1914 and \heb\l4025.602, the effective rest wavelength is function of the temperature, thus the spectral type. We use here an observed value which is representative of mid- to late- O-type stars.  
\end{table}

\section{A Monte-Carlo method to constrain the intrinsic orbital parameter distributions}\label{app: MC}

In this section, we provide the details of the  Monte-Carlo method used to constrain the intrinsic orbital parameter distributions. The general philosophy resembles that of \citet{KoF07}, in the sense that it relies on a Monte-Carlo modeling of the orbital velocities. On the one hand, we follow \citeauthor{KoF07} in using KS tests to compare simulated and observed distributions and the diagnostic distributions based on $\Delta RV$ and $\chi^2$ (See Sect. \ref{app: merit}) are similar to, respectively, the $V_h$ and $V_{rms}$ used by \citeauthor{KoF07}.  On the other hand, our method fundamentally  differs in various aspects. First we use the specific measurement errors provided by the RV analysis throughout the complete Monte-Carlo analysis. Second, we do not apply external constraints such as fixing the period distribution or using the fraction of Type Ib/c supernovae corresponding to each model. We thus attempt to simultaneously constrain the intrinsic period and mass-ratio distributions and the intrinsic binary fraction. 

The core of the method is identical to that described in \citet{SdMdK12} but differs in the adopted merit function. \citeauthor{SdMdK12} compared the observed orbital parameter distributions while we will use distributions constructed from the RV database. Indeed most of the detected binaries in 30~Dor have too few RV measurements to compute a meaningful orbital solution, precluding the direct approach used by \citet{SdMdK12}.

\subsection{Method overview}
As discussed in the main text, we use power laws to describe the intrinsic distributions of  orbital parameters:  $f(\log_{10} P/\mathrm{d})\sim (\log_{10} P/\mathrm{d})^\pi$ , $f(q)\sim q^\kappa$ and  $f(e)\sim q^\eta$, the exponents  of which are left as free parameters. 
For each combination of $\pi$, $\kappa$ and \fbin\ in our grid, we draw populations of $N=360$ primary stars using a Kroupa mass function \citep{Kro01} between 15 and 80~\msun, covering thus the expected mass range of O-type stars. Each star is assigned an observational sequence (observing epochs and RV accuracy at each epoch) from our VFTS sample.  A fraction of these stars are randomly assigned to be binaries following a binomial with a success probability \fbin. The orbital parameters of the binaries are randomly drawn as follows :
\begin{enumerate}
\item[-] the period and mass-ratio are taken from the power-law distributions given $\pi$ and $\kappa$;
\item[-] the eccentricity is taken from a power-law distribution with an index given by $\eta=-0.5$ (see discussion in Sect.~\ref{sect: MC});
\item[-] the inclination and periastron argument are taken assuming random orientation of the orbit in space;
\item[-] the time of periastron passage is assumed to be uncorrelated with the observational sampling and  is thus taken from a uniform distribution.
\end{enumerate}

For each binary, the orbital velocities associated with the primary stars are computed taking into account the time sequence of our observations. Measurement uncertainties are randomly added using a Gaussian distribution with a standard deviation given by the associated measurement error. For simulated single stars, only the contribution of the error is taken into account. 

The binary detection criteria of Eq.~\ref{eq: bin2} are then applied  and the {\it simulated} detected binaries are flagged. {\it Simulated} observed distributions (see Sect.~\ref{app: merit}) are built based on the simulated RV measurements of the detected systems and the {\it simulated} measured binary fraction $f_\mathrm{obs}^\mathrm{simul}$ is computed.

For a given combination of $\pi$, $\kappa$ and \fbin\ the process is repeated 100 times to build the parent statistics. The simulated parent distributions are then compared to the observed ones by means of a one-sided KS test. We also compute the binomial probability $B(N_\mathrm{bin},N,f_\mathrm{obs}^\mathrm{simul})$ to detect $N_\mathrm{bin}$ binaries among a population of $N$ stars given the success probability  predicted by the simulations, i.e.\ $f_\mathrm{obs}^\mathrm{simul}$. Adopting $N_\mathrm{bin}$ to be the actual number of detected O-type binaries in VFTS allows us to estimate the ability of the assumed intrinsic parameters to reproduce the observed binary fraction while taking into account the finite size of the sample.

\begin{figure*}
  \centering
  \includegraphics[width=\textwidth]{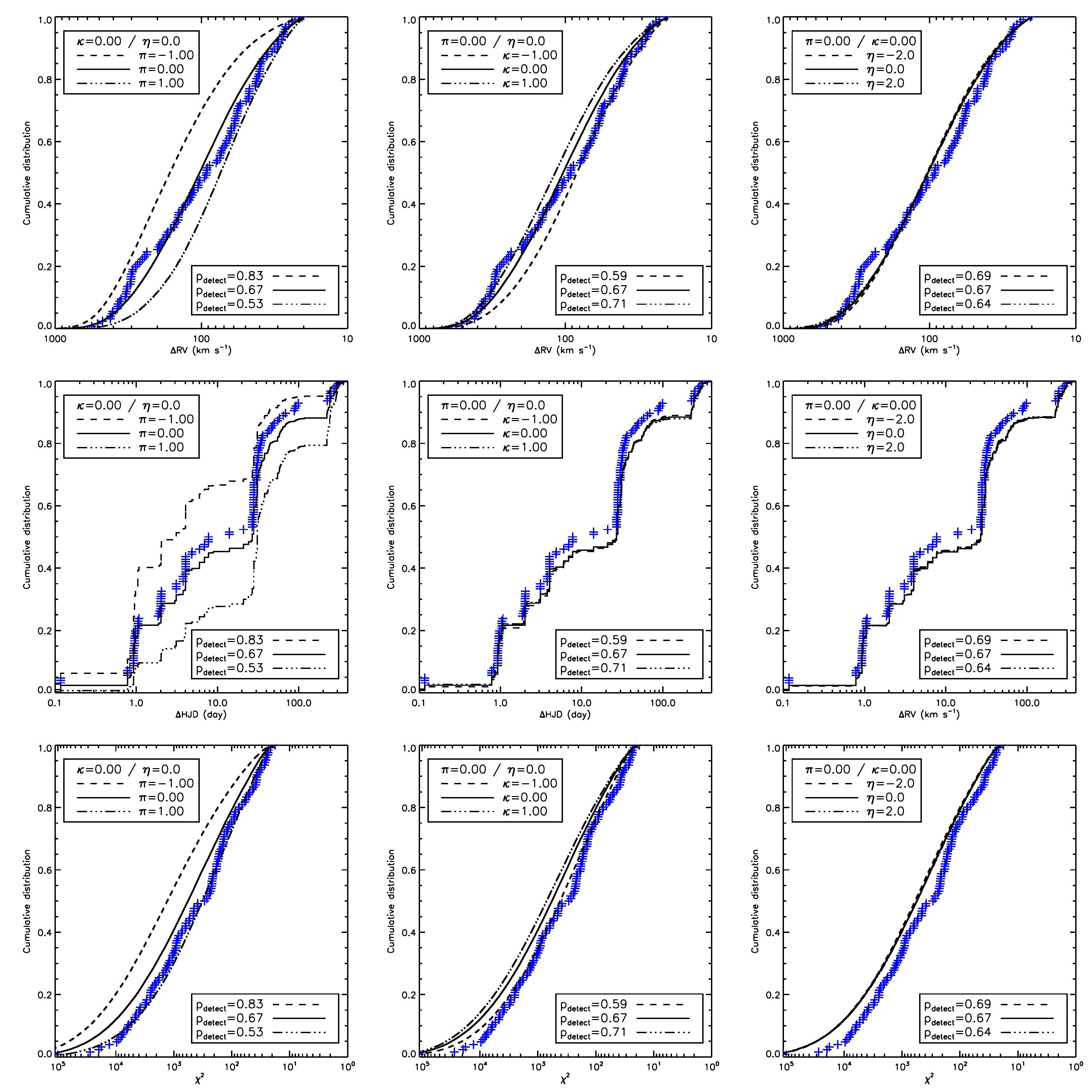}
  \caption{Comparison between the observed (crosses) and simulated (lines) cumulative distributions of the peak-to-peak RV amplitudes (top row), variability time scales (middle row) and \chisq\ (bottom row). Lefthand panels vary the exponent $\pi$ of the period distribution. Panels in the central column vary $\kappa$, the exponent of the mass-ratio distribution and the righthand panels vary $\eta$, the exponent of the eccentricity distribution. $\pi$, $\kappa$ and $\eta$ values are indicated in the upper-left legend of each panel. The bottom-right legends indicate the overall VFTS detection probability for the adopted parent distributions.}
  \label{fig: param}
\end{figure*}

\begin{table*}
\centering
\caption{Overview of the test results using different merit functions (Col.~2). The median, 0.16 and 0.74 percentiles of the retrieved parameters in a set of 50 test runs are given in Cols.~4 to 6. The input values of these parameters are indicated as the second line of the table. The explicit mathematical descriptions are given for the individual approaches (a)-(c) and (l)-(m). Implicit abbreviations are  used otherwise to keep the notation compact.   } 
\label{tab: mctest}
\begin{tabular}{llcccc}
\hline
ID & Merit function & $C$  & \multicolumn{3}{c}{Multiplicity properties}\\
\vspace*{3mm}
&  & (\kms) & \fbin & \p & \k              \\
\hline
\\
&  &         & ($f_\mathrm{bin}^\mathrm{true}=0.5$) & ($\pi_\mathrm{true}=-0.5$) & ($\kappa_\mathrm{true}=0.0$)  \\
\\
(a) &   $P_\mathrm{KS}(\Delta \mathrm{HJD})$                             &  0.0  &  0.22  [0.22:0.22] & $-$2.50  [$-$2.50:$-$2.50] & $-$2.50  [$-$2.50:$-$2.50] \\ 
(b) &   $P_\mathrm{KS}(\Delta \mathrm{RV})$                              &  0.0  &  0.48  [0.46:0.54] & $-$0.50  [$-$1.15:$-$0.10] & $-$0.10  [$-$1.15:$+$1.55] \\ 
(c) &   $P_\mathrm{KS}(\Delta \mathrm{\chi^2})$                          &  0.0  &  0.50  [0.44:0.54] & $-$0.35  [$-$1.80:$-$0.05] & $+$0.15  [$-$1.85:$+$1.50] \\ 
\hline                                                                                                                                                 
(d) &   $(\Delta HJD) \times (\Delta RV)$                               &  0.0  &  0.26  [0.20:0.34] & $-$0.65  [$-$0.85:$-$0.50] & $+$1.55  [$-$0.15:$+$2.40]  \\ 
(e) &   $(\Delta HJD) \times (\chi^2)$                                  &  0.0  &  0.26  [0.20:0.42] & $-$0.65  [$-$1.30:$+$0.00] & $-$0.35  [$-$1.25:$+$1.70]  \\ 
(f) &   $(\Delta HJD) \times (Bin)$                                     &  0.0  &  0.28  [0.20:0.44] & $-$1.10  [$-$2.35:$+$0.00] & $-$1.15  [$-$1.95:$+$2.05]  \\ 
(g) &   $(\Delta RV) \times (Bin)$                                      &  0.0  &  0.40  [0.38:0.46] & $-$2.30  [$-$2.45:$-$1.85] & $-$0.55  [$-$1.80:$+$1.75]  \\
(h) &   $(\chi^2) \times (Bin)$                                         &  0.0  &  0.48  [0.46:0.54] & $-$0.50  [$-$1.15:$-$0.10] & $-$0.10  [$-$1.15:$+$1.55]  \\  
\hline                                                                                                                                                 
(i) &   $(\Delta HJD) \times (\Delta RV) \times (Bin)$                  &  0.0  &  0.50  [0.44:0.54] & $-$0.35  [$-$1.80:$-$0.05] & $+$0.15  [$-$1.85:$+$1.50]  \\  
(j) &   $(\Delta HJD) \times (\chi^2) \times (Bin)$                     &  0.0  &  0.40  [0.38:0.46] & $-$2.30  [$-$2.45:$-$1.85] & $-$0.55  [$-$1.80:$+$1.75]  \\  
(k) &   $(\Delta HJD) \times (\Delta RV) \times (\chi^2) \times (Bin)$  &  0.0  &  0.48  [0.46:0.54] & $-$0.45  [$-$1.20:$+$0.05] & $-$0.10  [$-$1.00:$+$1.80]  \\  
\hline                                                                                                                                                 
(l) &   $P_\mathrm{KS}(\Delta \mathrm{HJD})$                             & 20.0  &  0.50  [0.44:0.52] & $-$0.40  [$-$1.55:$+$0.00] & $+$0.00  [$-$1.65:$+$1.85]  \\  
(m) &   $P_\mathrm{KS}(\Delta \mathrm{RV})$                              & 20.0  &  0.28  [0.20:0.52] & $-$0.75  [$-$1.00:$-$0.50] & $-$0.55  [$-$0.80:$-$0.20]  \\  
(n) &   $P_\mathrm{KS}(\Delta \mathrm{\chi^2})$                          & 20.0  &  0.24  [0.22:0.36] & $-$0.75  [$-$1.00:$-$0.55] & $-$0.55  [$-$0.95:$+$0.05]  \\  
(o) &   $B(N_\mathrm{obs},N,f_\mathrm{sim})$                              & 20.0  &  0.46  [0.42:0.50] & $-$0.65  [$-$0.85:$-$0.50] & $+$1.80  [$+$0.55:$+$2.30]  \\ 
\hline                                                                                                                                                 
(p) &   $(\Delta HJD) \times (\Delta RV)$                               & 20.0  &  0.48  [0.46:0.54] & $-$0.50  [$-$1.30:$+$0.05] & $-$0.10  [$-$1.25:$+$2.05]  \\  
(q) &   $(\Delta HJD) \times (\chi^2)$                                  & 20.0  &  0.48  [0.42:0.52] & $-$0.90  [$-$2.15:$-$0.10] & $-$0.65  [$-$1.90:$+$1.65]  \\  
(r) &   $(\Delta HJD) \times (Bin)$                                     & 20.0  &  0.48  [0.46:0.54] & $-$0.45  [$-$1.20:$+$0.05] & $-$0.10  [$-$1.00:$+$1.80]  \\  
(s) &   $(\Delta RV) \times (Bin)$                                      & 20.0  &  0.50  [0.44:0.52] & $-$0.40  [$-$1.55:$+$0.00] & $+$0.00  [$-$1.65:$+$1.85]  \\  
(t) &   $(\chi^2) \times (Bin)$                                         & 20.0  &  0.50  [0.46:0.54] & $-$0.25  [$-$1.30:$-$0.05] & $+$0.25  [$-$1.20:$+$1.75]  \\  
\hline                                                                                                                                                  
(u) &   $(\Delta HJD) \times (\Delta RV) \times (Bin)$                  & 20.0  &  0.48  [0.44:0.54] & $-$0.75  [$-$1.05:$-$0.55] & $-$0.60  [$-$0.80:$-$0.15]  \\  
(v) &   $(\Delta HJD) \times (\chi^2) \times (Bin)$                     & 20.0  &  0.48  [0.44:0.54] & $-$0.80  [$-$1.00:$-$0.55] & $-$0.60  [$-$1.10:$+$0.05]  \\  
(w) &   $(\Delta HJD) \times (\Delta RV) \times (\chi^2) \times (Bin)$  & 20.0  &  0.48  [0.44:0.54] & $-$0.80  [$-$1.00:$-$0.55] & $-$0.65  [$-$0.80:$-$0.15]  \\  
\\
&  &         & ($f_\mathrm{bin}^\mathrm{true}=0.5$) & ($\pi_\mathrm{true}=-0.5$) & ($\kappa_\mathrm{true}=-0.9$)  \\
\\
(a) &   $P_\mathrm{KS}(\Delta \mathrm{HJD})$       & 0.0 &  0.40  [0.40:0.40] & $-$1.50  [$-$1.50:$-$1.50] & $-$2.00  [$-$2.00:$-$2.00] \\ 
(b) &   $P_\mathrm{KS}(\Delta \mathrm{RV})$        & 0.0 &  0.50  [0.44:0.58] & $-$0.55  [$-$1.15:$+$0.05] & $-$0.70  [$-$1.60:$+$0.10] \\ 
(c) &   $P_\mathrm{KS}(\Delta \mathrm{\chi^2})$    & 0.0 &  0.50  [0.44:0.56] & $-$0.45  [$-$1.10:$-$0.05] & $-$0.80  [$-$1.50:$+$0.10] \\ 
\hline                                                        
(d) &   $(\Delta HJD) \times (\Delta RV)$         &  0.0 &   0.50  [0.44:0.58] & $-$0.55  [$-$1.15:$+$0.05] & $-$0.70  [$-$1.60:$+$0.10]  \\ 
(e) &   $(\Delta HJD) \times (\chi^2)$            &  0.0 &   0.50  [0.44:0.56] & $-$0.45  [$-$1.10:$-$0.05] & $-$0.80  [$-$1.50:$+$0.10]  \\ 
(f) &   $(\Delta HJD) \times (Bin)$               &  0.0 &   0.48  [0.42:0.62] & $-$0.60  [$-$1.15:$+$0.00] & $-$0.70  [$-$1.70:$+$0.10]  \\ 
(g) &   $(\Delta RV) \times (Bin)$                &  0.0 &  0.50  [0.44:0.56] & $-$0.45  [$-$1.15:$+$0.05] & $-$0.70  [$-$1.60:$+$0.20]  \\ 
(h) &   $(\chi^2) \times (Bin)$                   &  0.0 &  0.48  [0.44:0.54] & $-$0.45  [$-$1.20:$-$0.05] & $-$0.70  [$-$1.50:$+$0.10]  \\ 
\hline 
(i) &   $(\Delta HJD) \times (\Delta RV) \times (Bin)$   & 0.0 & 0.50  [0.44:0.56] & $-$0.45  [$-$1.15:$+$0.05] & $-$0.70  [$-$1.60:$+$0.20]  \\ 
(j) &   $(\Delta HJD) \times (\chi^2) \times (Bin)$      & 0.0 & 0.48  [0.44:0.54] & $-$0.45  [$-$1.20:$-$0.05] & $-$0.70  [$-$1.50:$+$0.10]  \\ 
(k) &   $(\Delta HJD) \times (\Delta RV) \times (\chi^2) \times (Bin)$  & 0.0&  0.48  [0.44:0.54] & $-$0.55  [$-$1.25:$-$0.05] & $-$0.80  [$-$1.70:$+$0.20]  \\ 
\hline 
(l) &   $P_\mathrm{KS}(\Delta \mathrm{HJD})$      &  20.0 & 0.48  [0.40:0.64] & $-$0.65  [$-$0.85:$-$0.45] & $+$0.10  [$-$0.90:$+$0.40] \\ 
(m) &   $P_\mathrm{KS}(\Delta \mathrm{RV})$       &  20.0 & 0.56  [0.42:0.80] & $-$0.60  [$-$1.05:$+$0.00] & $-$0.90  [$-$1.50:$+$0.00] \\ 
(n) &   $P_\mathrm{KS}(\Delta \mathrm{\chi^2})$   &  20.0 & 0.50  [0.42:0.84] & $-$0.90  [$-$1.50:$-$0.10] & $-$1.50  [$-$2.00:$-$0.10] \\
(o) &   $B(N_\mathrm{obs},N,f_\mathrm{sim})$       &  20.0 & 0.48  [0.42:0.62] & $-$0.60  [$-$1.15:$+$0.00] & $-$0.70  [$-$1.70:$+$0.10] \\ 
\hline 
(p) &   $(\Delta HJD) \times (\Delta RV)$        & 20.0 & 0.52  [0.42:0.70] & $-$0.80  [$-$1.05:$-$0.55] & $-$1.20  [$-$1.50:$-$0.80]  \\ 
(q) &   $(\Delta HJD) \times (\chi^2)$           & 20.0 & 0.54  [0.40:0.86] & $-$0.80  [$-$1.05:$-$0.50] & $-$1.20  [$-$1.60:$-$0.70]  \\ 
(r) &   $(\Delta HJD) \times (Bin)$              & 20.0 & 0.44  [0.40:0.50] & $-$0.65  [$-$0.90:$-$0.45] & $-$0.10  [$-$1.30:$+$0.30]  \\ 
(s) &   $(\Delta RV) \times (Bin)$               & 20.0 & 0.48  [0.44:0.54] & $-$0.65  [$-$1.05:$-$0.05] & $-$1.00  [$-$1.40:$+$0.00]  \\ 
(t) &   $(\chi^2) \times (Bin)$                  & 20.0 & 0.46  [0.42:0.54] & $-$0.95  [$-$1.40:$+$0.00] & $-$1.20  [$-$1.90:$+$0.20]  \\ 
\hline 
(u) &   $(\Delta HJD) \times (\Delta RV) \times (Bin)$  & 20.0 & 0.48  [0.44:0.52] & $-$0.80  [$-$1.05:$-$0.50] & $-$1.20  [$-$1.50:$-$0.80]  \\ 
(v) &   $(\Delta HJD) \times (\chi^2) \times (Bin)$     & 20.0 & 0.48  [0.42:0.52] & $-$0.75  [$-$1.05:$-$0.50] & $-$1.20  [$-$1.50:$-$0.60]   \\ 
(w) &   $(\Delta HJD) \times (\Delta RV) \times (\chi^2) \times (Bin)$  & 20.0 & 0.48  [0.44:0.52] & $-$0.80  [$-$1.05:$-$0.50] & $-$1.20  [$-$1.50:$-$0.80]  \\ 
\hline
\end{tabular}
\end{table*}

\subsection{Diagnostic distributions}\label{app: diagn}
In this section, we specifically discuss the empirical distributions used as diagnostics. The choice of the merit function will be discussed in the next section. We investigate four diagnostic quantities:\begin{enumerate} 

\item[i.] the distribution of, for each detected binary, the amplitude of the RV signal ($\Delta RV$);
\item[ii.] the distribution of, for each detected binary, the $\chi^2$ of a constant RV model;
\item[iii.] the distribution of, for each detected binary, the minimum time scale for significant RV variations to be seen ($\Delta$HJD);
\item[iv.] the detected binary fraction.
\end{enumerate}

We thus perform  a parameter study to investigate the effect of different exponents on the simulated distributions and on the overall binary detection rate. Figure~\ref{fig: param}  shows that the exponent $\pi$ of the period distribution has the largest impact on the simulated distributions. The exponent $\kappa$ of the mass-ratio distribution is of secondary importance and the exponent $\eta$ has almost no impact on the simulated distribution, albeit affecting the simulated detection rate by a couple of per cent. The distribution of the minimum time scale for the observed RV variations is almost exclusively sensitive to the period distribution but mostly independent of the mass-ratio.

\subsection{Suitable merit functions}\label{app: merit}
At each point of the grid defined by the the intrinsic values $\pi$, $\kappa$ and \fbin, the simulated distributions (i) to (iii) are individually compared to the observed ones using KS-tests while the simulated and observed binary fraction are compared using a binomial statistics. Because we aim at matching all the diagnostic distributions simultaneously, we also combine the individual probabilities in a global merit function. 

We run several test to assess different merit functions (i.e., different ways of combining the individual probabilities) as well as to validate our method. We generate synthetic data from known input distributions, i.e.\ known  $\pi$, $\kappa$ and \fbin, and we apply our code using different merit functions. Input and output distributions are then compared to check the suitability of the various merit functions as well as the general accuracy of the method. The process has been repeated using 50 synthetic data sets in each case. The synthetic data have been generated such that they share the same observational properties (sampling, measurement accuracy) as the VFTS Medusa data. Table ~\ref{tab: mctest} provides an overview of the results of the tests. Because of the computational load of such tests, we use a slightly coarser grid with steps of 0.02 in \fbin. Explored ranges in $f_\mathrm{bin}$, $\pi$ and $\kappa$ and steps in the power law indexes $\pi$ and $\kappa$ are identical to those quoted in Table~\ref{tab: MC}. In addition to appraising the merit function, we also test the need to apply a minimal threshold $C$ for the detected significant RV signal. 

The first striking result is that most merit functions can recover the input binary fraction within a few  percents albeit with various precision. Some of the best results are obtained when including the  $B(N_\mathrm{obs},N,f_\mathrm{sim})$ term in the merit function. 
 
The index $\pi$ of the period distribution is poorly constrained whenever one uses no detection threshold (i.e., $C=0$~\kms). This indicates that the detection threshold is not only useful to reject false detections due to intrinsic profile variability but is also a critical ingredient of the method. Interestingly, $\pi$ is reasonably well constrained by the $\Delta$HJD distribution alone, i.e.\ merit function (l), but a better overall result is obtained when used in conjunction with other diagnostic quantities. 

The $\Delta$RV and $\chi^2$  distributions are useful to refine the $\kappa$ value but the uncertainties remain large. In essence, the   $\Delta$RV and $\chi^2$ distributions carry similar information. Because merit functions using $\Delta$RV tend to behave slightly better than those using $\chi^2$, and because the $\chi^2$ values are by definition much more sensitive to the exact knowledge of the measurement errors, we select our final merit function as given by the product of the $P_\mathrm{KS}$ probabilities computed from the $\Delta$HJD and $\Delta$RV distributions and of the Binomial probability (Eq.~\ref{eq: merit}), hence merit function (u) in Table \ref{tab: mctest}.  

Overall, the retrieved values of $\pi$ and $\kappa$ tend to be smaller than their input values but the correct indexes concur with the 1$\sigma$ confidence intervals quoted in Table \ref{tab: mctest}. These confidence intervals can further be used to estimate the formal uncertainty of the method. Given that the best representation of the data are obtained with $f_\mathrm{bin}=0.51$, $\pi=-0.45$ and $\kappa=-1.00$  (see Sect.~\ref{sect: res}), the bottom part of Table~\ref{tab: mctest} applies and we adopt $\sigma_{f_\mathrm{bin}}=0.04$, $\sigma_\pi=0.3$ and $\sigma_\kappa=0.4$.

\section{Supplementary figures}\label{app: LC}

\begin{figure*}
\includegraphics[width=0.33\textwidth]{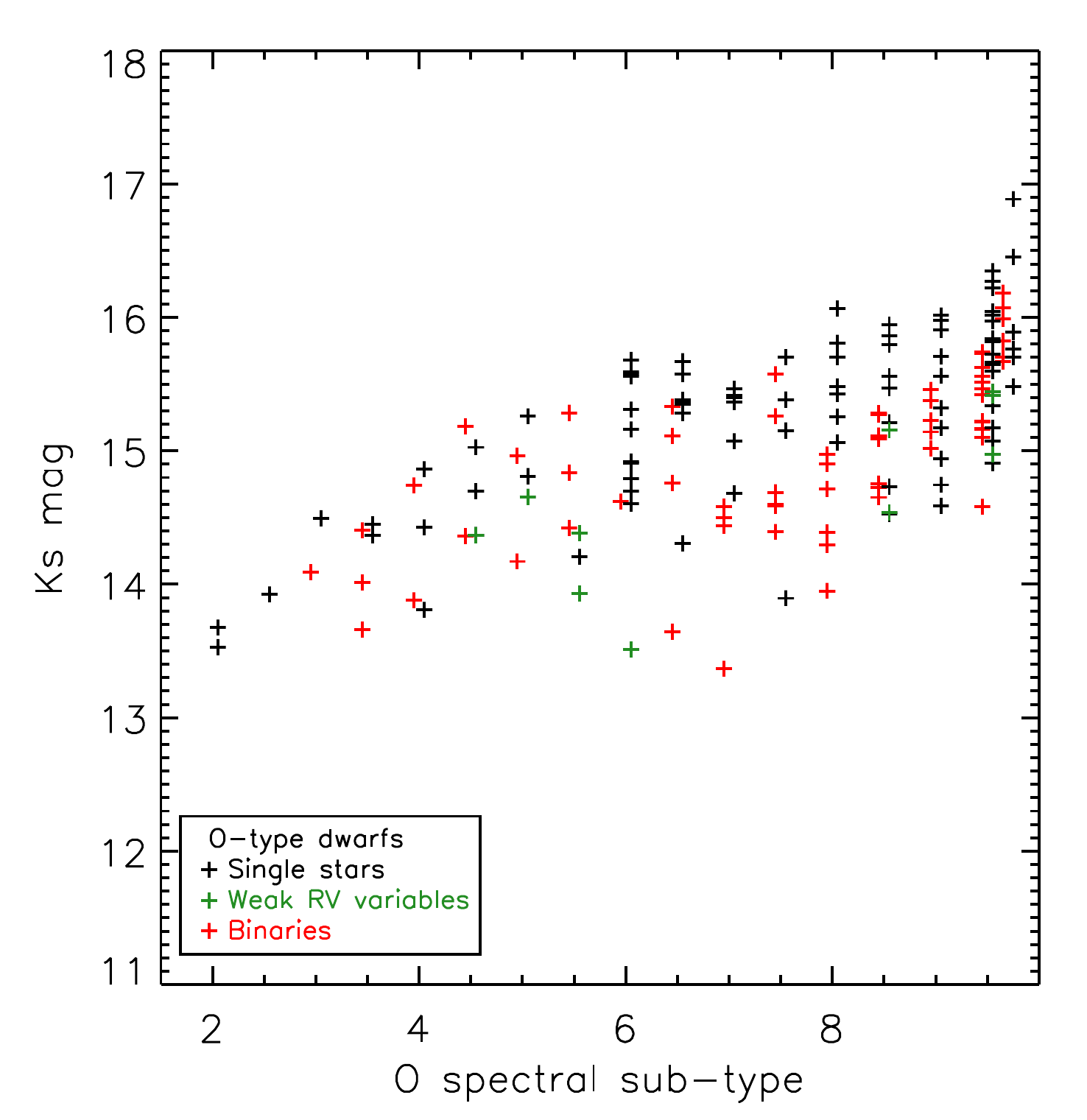}
\includegraphics[width=0.33\textwidth]{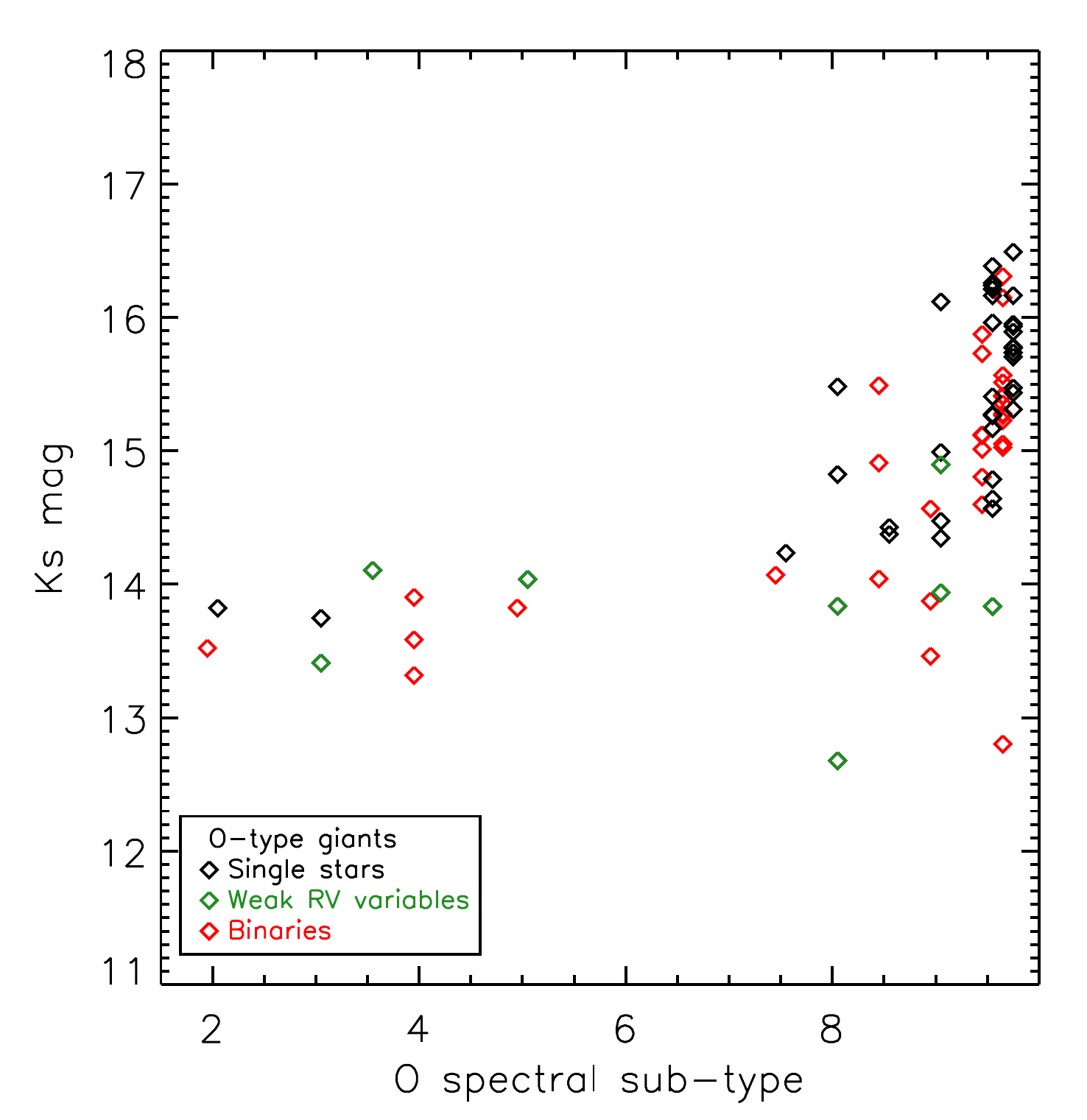}
\includegraphics[width=0.33\textwidth]{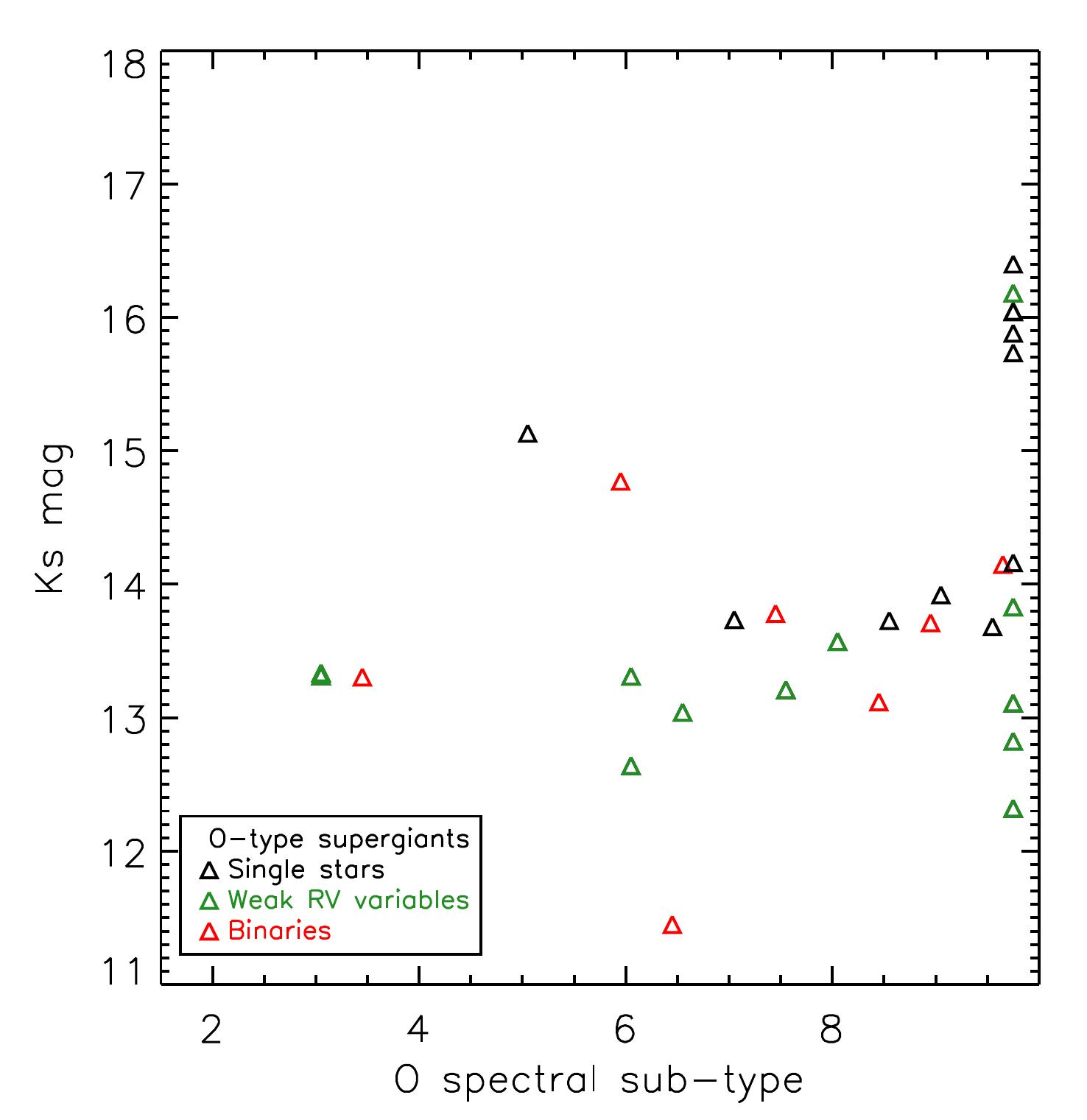}
\caption{$K_\mathrm{s}$-band magnitudes as a function of the spectral sub-types for luminosity classes V-IV (left panel), III (middle panel) and II-I (right panel). Different symbols identify the single stars, the low-amplitude RV variables and the binaries. Symbols for single and binaries of a given spectral-types have been shifted, compared to one another, by a small amount amount along the $x$-axis to improve the clarity of the plots.} \label{fig: spt_Kmag_lc}
\end{figure*}

\begin{figure*}
\includegraphics[width=0.33\textwidth]{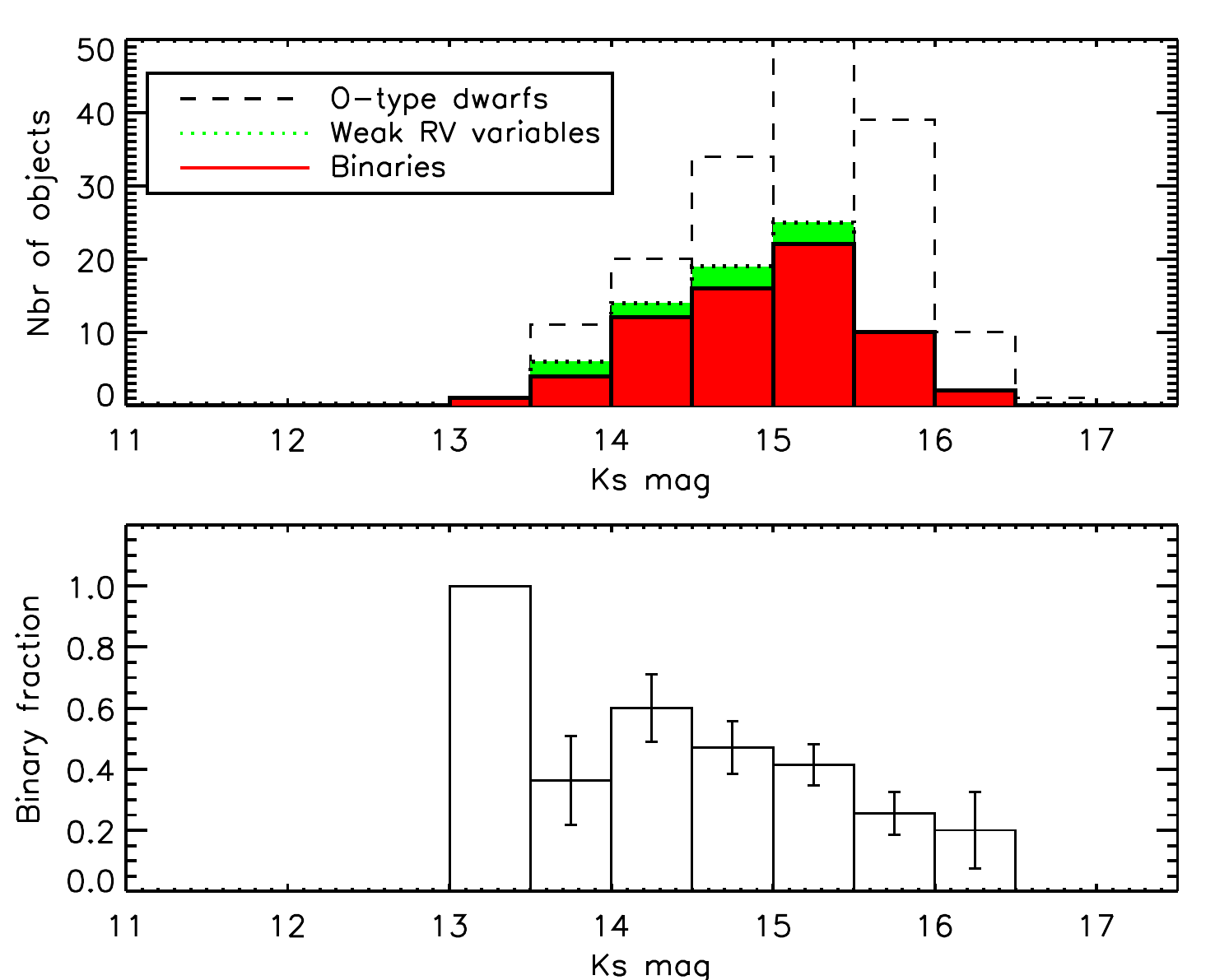}
\includegraphics[width=0.33\textwidth]{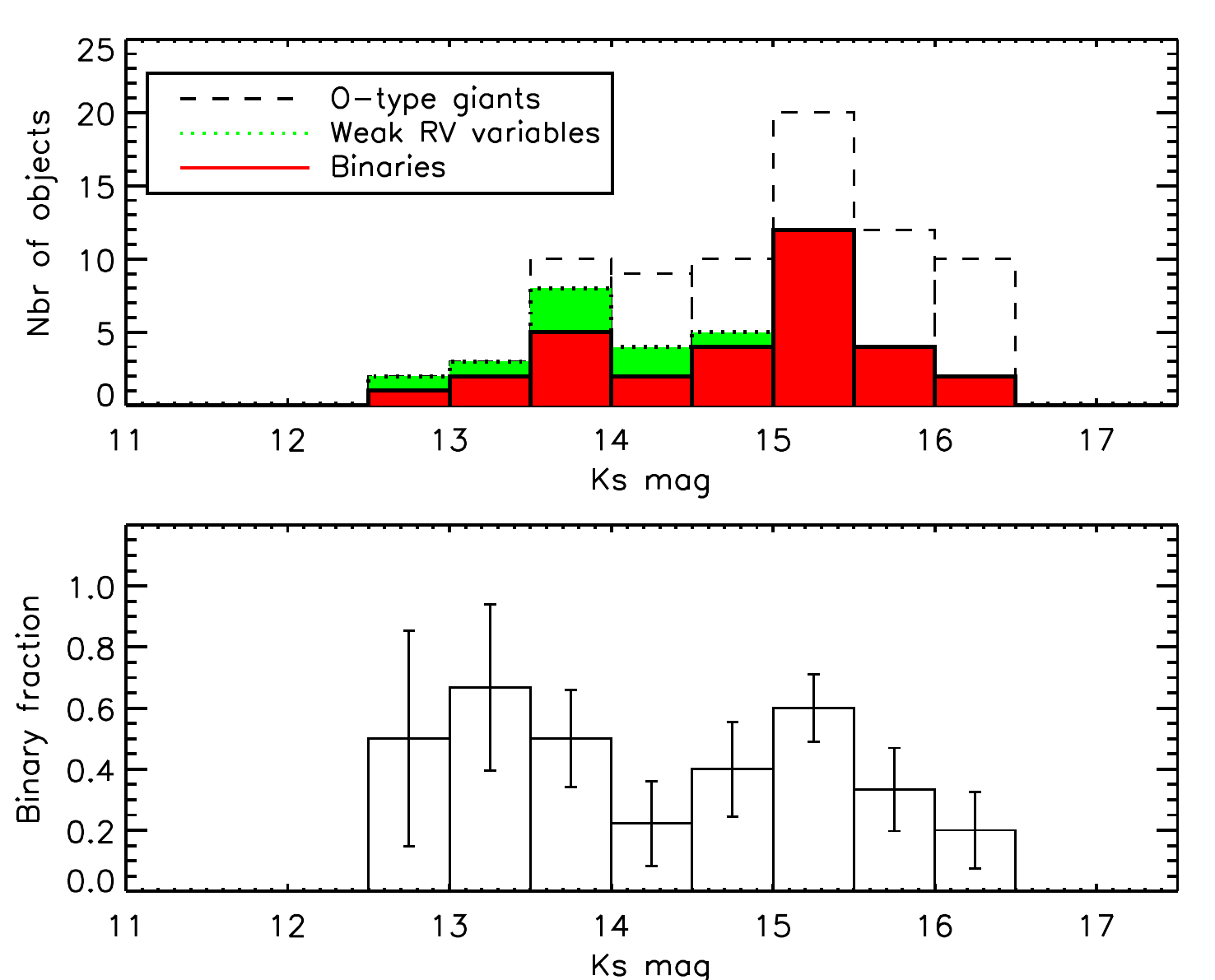}
\includegraphics[width=0.33\textwidth]{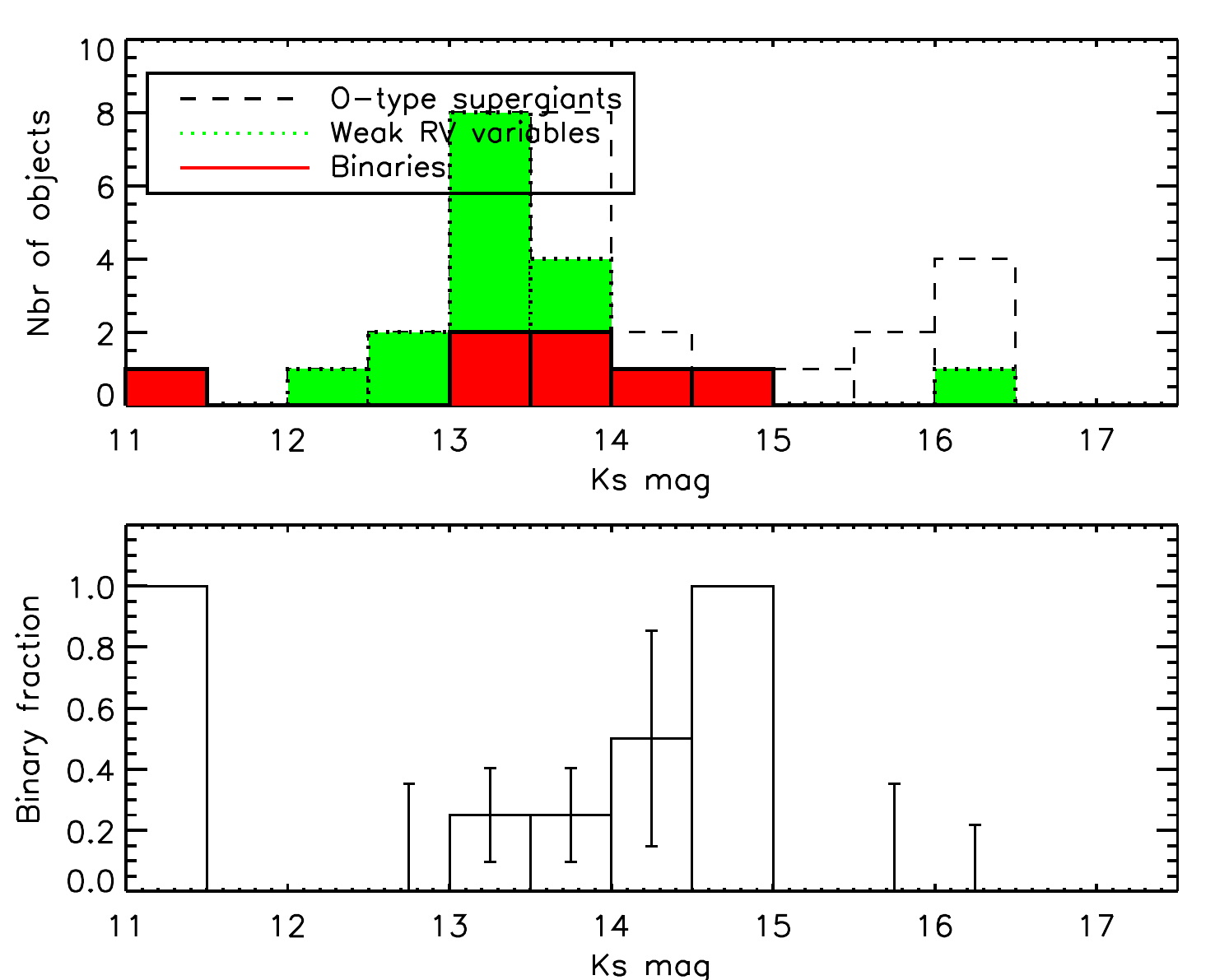}
\caption{Same as Fig.~\ref{fig: bf_Kmag} for the luminosity classes V-IV (left panels), III (middle panels) and II-I (right panels). } \label{fig: bf_Kmag_lc}
\end{figure*}

\begin{figure*}
\includegraphics[width=0.33\textwidth]{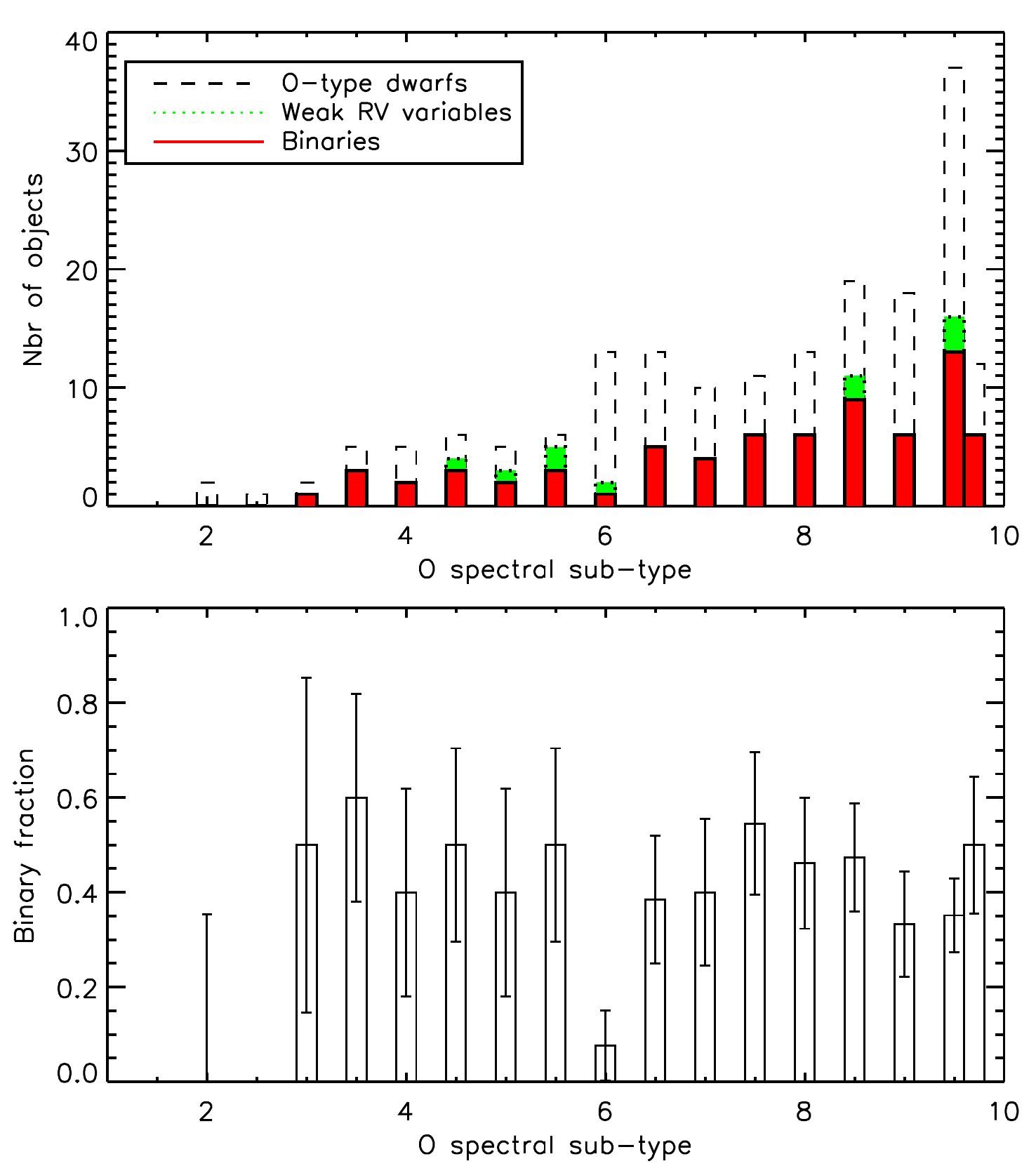}
\includegraphics[width=0.33\textwidth]{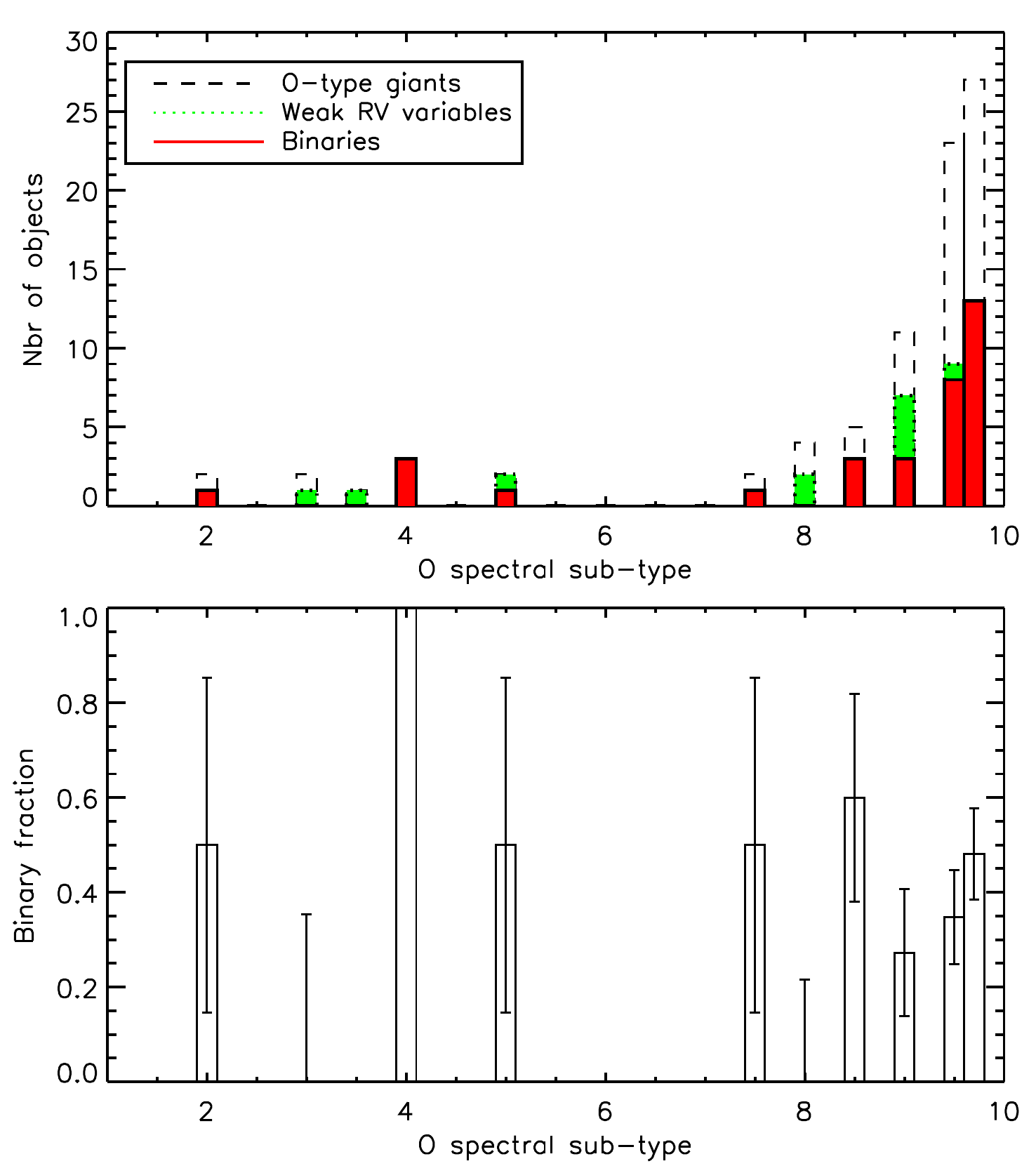}
\includegraphics[width=0.33\textwidth]{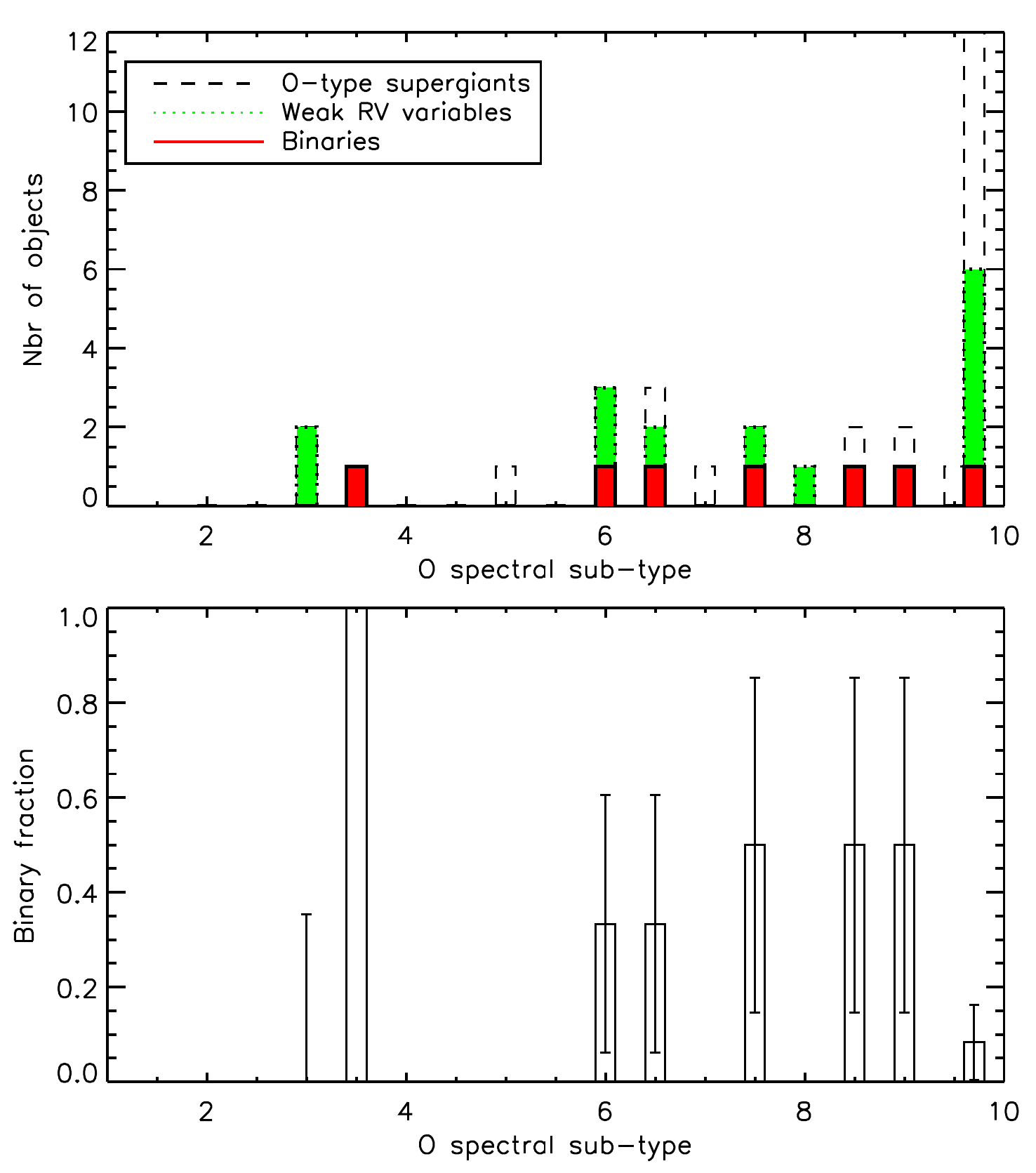}
\caption{Same as Fig.~\ref{fig: bf_spt} for the luminosity classes V-IV (left panels), III (middle panels) and II-I (right panels). } \label{fig: bf_spt_lc}
\end{figure*}

\end{appendix}
\end{document}